\documentclass[longauth]{aa}
\usepackage{txfonts}
\usepackage{natbib}
\usepackage{graphicx}
\usepackage{gensymb}
\usepackage{hyperref}
\usepackage{xspace}
\newcommand{\orcit}[1]{\protect\href{https://orcid.org/#1}{\protect\includegraphics[width=8pt]{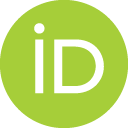}}}

\providecommand{\gaia}{\textit{Gaia}\xspace}
\providecommand{\gdr}[1]{\textit{Gaia}~DR{#1}\xspace}
\newcommand{\teff}{T_{\rm eff}}

\def\kms{\,{\rm km\,s^{-1}}}

\def\url#1{{\tt#1}}
\newcommand{\KI}{\rm K\,\uppercase\expandafter{\romannumeral1}}
\newcommand{\HI}{\rm H\,\uppercase\expandafter{\romannumeral1}}
\newcommand{\NI}{\rm N\,\uppercase\expandafter{\romannumeral1}}
\newcommand{\MgI}{\rm Mg\,\uppercase\expandafter{\romannumeral1}}
\newcommand{\NaI}{\rm Na\,\uppercase\expandafter{\romannumeral1}}
\newcommand{\FeI}{\rm Fe\,\uppercase\expandafter{\romannumeral1}}
\newcommand{\TiI}{\rm Ti\,\uppercase\expandafter{\romannumeral1}}
\newcommand{\CaI}{\rm Ca\,\uppercase\expandafter{\romannumeral1}}
\newcommand{\HeI}{\rm He\,\uppercase\expandafter{\romannumeral1}}
\newcommand{\HII}{\rm H\,\uppercase\expandafter{\romannumeral2}}
\newcommand{\MgII}{\rm Mg\,\uppercase\expandafter{\romannumeral2}}
\newcommand{\CaII}{\rm Ca\,\uppercase\expandafter{\romannumeral2}}
\newcommand{\HeII}{\rm He\,\uppercase\expandafter{\romannumeral2}}
\newcommand{\NIII}{\rm N\,\uppercase\expandafter{\romannumeral3}}

\newcommand{\Av}{A_{\rm V}}

\newcommand{\EBV}{E(B\,{-}\,V)}
\newcommand{\Vlsr}{V_{\rm LSR}}
\newcommand{\Usun}{U_{\odot}}

\newcommand{\Rgc}{R_{\rm GC}}

\newcommand{\xp}{x^{\prime}}

\defcitealias{SFD}{SFD}
\defcitealias{Surot20}{S20}
\defcitealias{Alvaro17}{R17}
\defcitealias{Zasowski2015c}{Z15}
\defcitealias{Elyajouri2019}{E19}
\defcitealias{Gonzalez12}{G12}

\makeatletter
\renewcommand*\maketitle{%
  \thispagestyle{firstpage}
\begingroup
    \if@wideboxfn
    \setlength\bibindent{1.4\parindent}
    \else
    \setlength\bibindent{\parindent}
    \fi
    \renewcommand*\thefootnote{\@fnsymbol\c@footnote}%
    \renewcommand\@makefntext[1]{%
    \ifaa@longfn\hsize\textwidth\fi
    \noindent
    \hb@xt@\bibindent{\hss\@makefnmark\enspace}##1}
  \ifaa@twocolumn
  \begingroup
    \begin{aa@strip}
          \aa@maketitle
    \end{aa@strip}
    \@thanks            
  \endgroup
  \else
    \begingroup
      \let\thanks\footnote
      \aa@maketitle
    \endgroup
  \fi
\endgroup
  \setcounter{footnote}{0}%
}
\makeatother
\usepackage[normalem]{ulem}

\begin{document}

\title{Gaia Data Release 3: Exploring and mapping the diffuse interstellar band at 862\,nm}

\author{
{\it Gaia} Collaboration
\and       M.~                    Schultheis\orcit{0000-0002-6590-1657}\inst{\ref{inst:0001}}
\and         H.~                          Zhao\orcit{0000-0003-2645-6869}\inst{\ref{inst:0001}}
\and         T.~                       Zwitter\orcit{0000-0002-2325-8763}\inst{\ref{inst:0003}}
\and       D.J.~                      Marshall\orcit{0000-0003-3956-3524}\inst{\ref{inst:0004}}
\and         R.~                       Drimmel\orcit{0000-0002-1777-5502}\inst{\ref{inst:0005}}
\and         Y.~                    Fr\'{e}mat\orcit{0000-0002-4645-6017}\inst{\ref{inst:0006}}
\and     C.A.L.~                  Bailer-Jones\inst{\ref{inst:0007}}
\and         A.~                  Recio-Blanco\orcit{0000-0002-6550-7377}\inst{\ref{inst:0001}}
\and         G.~                    Kordopatis\orcit{0000-0002-9035-3920}\inst{\ref{inst:0001}}
\and         P.~                    de Laverny\orcit{0000-0002-2817-4104}\inst{\ref{inst:0001}}
\and         R.~                        Andrae\orcit{0000-0001-8006-6365}\inst{\ref{inst:0007}}
\and       T.E.~                 Dharmawardena\orcit{0000-0002-9583-5216}\inst{\ref{inst:0007}}
\and         M.~                     Fouesneau\orcit{0000-0001-9256-5516}\inst{\ref{inst:0007}}
\and         R.~                         Sordo\orcit{0000-0003-4979-0659}\inst{\ref{inst:0014}}
\and     A.G.A.~                         Brown\orcit{0000-0002-7419-9679}\inst{\ref{inst:0015}}
\and         A.~                     Vallenari\orcit{0000-0003-0014-519X}\inst{\ref{inst:0014}}
\and         T.~                        Prusti\orcit{0000-0003-3120-7867}\inst{\ref{inst:0017}}
\and     J.H.J.~                    de Bruijne\orcit{0000-0001-6459-8599}\inst{\ref{inst:0017}}
\and         F.~                        Arenou\orcit{0000-0003-2837-3899}\inst{\ref{inst:0019}}
\and         C.~                     Babusiaux\orcit{0000-0002-7631-348X}\inst{\ref{inst:0020},\ref{inst:0019}}
\and         M.~                      Biermann\inst{\ref{inst:0022}}
\and       O.L.~                       Creevey\orcit{0000-0003-1853-6631}\inst{\ref{inst:0001}}
\and         C.~                     Ducourant\orcit{0000-0003-4843-8979}\inst{\ref{inst:0024}}
\and       D.W.~                         Evans\orcit{0000-0002-6685-5998}\inst{\ref{inst:0025}}
\and         L.~                          Eyer\orcit{0000-0002-0182-8040}\inst{\ref{inst:0026}}
\and         R.~                        Guerra\orcit{0000-0002-9850-8982}\inst{\ref{inst:0027}}
\and         A.~                        Hutton\inst{\ref{inst:0028}}
\and         C.~                         Jordi\orcit{0000-0001-5495-9602}\inst{\ref{inst:0029}}
\and       S.A.~                       Klioner\orcit{0000-0003-4682-7831}\inst{\ref{inst:0030}}
\and       U.L.~                       Lammers\orcit{0000-0001-8309-3801}\inst{\ref{inst:0027}}
\and         L.~                     Lindegren\orcit{0000-0002-5443-3026}\inst{\ref{inst:0032}}
\and         X.~                          Luri\orcit{0000-0001-5428-9397}\inst{\ref{inst:0029}}
\and         F.~                       Mignard\inst{\ref{inst:0001}}
\and         C.~                         Panem\inst{\ref{inst:0035}}
\and         D.~            Pourbaix$^\dagger$\orcit{0000-0002-3020-1837}\inst{\ref{inst:0036},\ref{inst:0037}}
\and         S.~                       Randich\orcit{0000-0003-2438-0899}\inst{\ref{inst:0038}}
\and         P.~                    Sartoretti\inst{\ref{inst:0019}}
\and         C.~                      Soubiran\orcit{0000-0003-3304-8134}\inst{\ref{inst:0024}}
\and         P.~                         Tanga\orcit{0000-0002-2718-997X}\inst{\ref{inst:0001}}
\and       N.A.~                        Walton\orcit{0000-0003-3983-8778}\inst{\ref{inst:0025}}
\and         U.~                       Bastian\orcit{0000-0002-8667-1715}\inst{\ref{inst:0022}}
\and         F.~                        Jansen\inst{\ref{inst:0044}}
\and         D.~                          Katz\orcit{0000-0001-7986-3164}\inst{\ref{inst:0019}}
\and       M.G.~                      Lattanzi\orcit{0000-0003-0429-7748}\inst{\ref{inst:0005},\ref{inst:0047}}
\and         F.~                   van Leeuwen\inst{\ref{inst:0025}}
\and         J.~                        Bakker\inst{\ref{inst:0027}}
\and         C.~                      Cacciari\orcit{0000-0001-5174-3179}\inst{\ref{inst:0050}}
\and         J.~                 Casta\~{n}eda\orcit{0000-0001-7820-946X}\inst{\ref{inst:0051}}
\and         F.~                     De Angeli\orcit{0000-0003-1879-0488}\inst{\ref{inst:0025}}
\and         C.~                     Fabricius\orcit{0000-0003-2639-1372}\inst{\ref{inst:0029}}
\and         L.~                     Galluccio\orcit{0000-0002-8541-0476}\inst{\ref{inst:0001}}
\and         A.~                      Guerrier\inst{\ref{inst:0035}}
\and         U.~                        Heiter\orcit{0000-0001-6825-1066}\inst{\ref{inst:0056}}
\and         E.~                        Masana\orcit{0000-0002-4819-329X}\inst{\ref{inst:0029}}
\and         R.~                      Messineo\inst{\ref{inst:0058}}
\and         N.~                       Mowlavi\orcit{0000-0003-1578-6993}\inst{\ref{inst:0026}}
\and         C.~                       Nicolas\inst{\ref{inst:0035}}
\and         K.~                  Nienartowicz\orcit{0000-0001-5415-0547}\inst{\ref{inst:0061},\ref{inst:0062}}
\and         F.~                       Pailler\orcit{0000-0002-4834-481X}\inst{\ref{inst:0035}}
\and         P.~                       Panuzzo\orcit{0000-0002-0016-8271}\inst{\ref{inst:0019}}
\and         F.~                        Riclet\inst{\ref{inst:0035}}
\and         W.~                          Roux\orcit{0000-0002-7816-1950}\inst{\ref{inst:0035}}
\and       G.M.~                      Seabroke\orcit{0000-0003-4072-9536}\inst{\ref{inst:0067}}
\and         F.~                  Th\'{e}venin\inst{\ref{inst:0001}}
\and         G.~                  Gracia-Abril\inst{\ref{inst:0069},\ref{inst:0022}}
\and         J.~                       Portell\orcit{0000-0002-8886-8925}\inst{\ref{inst:0029}}
\and         D.~                      Teyssier\orcit{0000-0002-6261-5292}\inst{\ref{inst:0072}}
\and         M.~                       Altmann\orcit{0000-0002-0530-0913}\inst{\ref{inst:0022},\ref{inst:0074}}
\and         M.~                        Audard\orcit{0000-0003-4721-034X}\inst{\ref{inst:0026},\ref{inst:0062}}
\and         I.~                Bellas-Velidis\inst{\ref{inst:0077}}
\and         K.~                        Benson\inst{\ref{inst:0067}}
\and         J.~                      Berthier\orcit{0000-0003-1846-6485}\inst{\ref{inst:0079}}
\and         R.~                        Blomme\orcit{0000-0002-2526-346X}\inst{\ref{inst:0006}}
\and       P.W.~                       Burgess\inst{\ref{inst:0025}}
\and         D.~                      Busonero\orcit{0000-0002-3903-7076}\inst{\ref{inst:0005}}
\and         G.~                         Busso\orcit{0000-0003-0937-9849}\inst{\ref{inst:0025}}
\and         H.~                   C\'{a}novas\orcit{0000-0001-7668-8022}\inst{\ref{inst:0072}}
\and         B.~                         Carry\orcit{0000-0001-5242-3089}\inst{\ref{inst:0001}}
\and         A.~                       Cellino\orcit{0000-0002-6645-334X}\inst{\ref{inst:0005}}
\and         N.~                         Cheek\inst{\ref{inst:0087}}
\and         G.~                    Clementini\orcit{0000-0001-9206-9723}\inst{\ref{inst:0050}}
\and         Y.~                      Damerdji\orcit{0000-0002-3107-4024}\inst{\ref{inst:0089},\ref{inst:0090}}
\and         M.~                      Davidson\inst{\ref{inst:0091}}
\and         P.~                    de Teodoro\inst{\ref{inst:0027}}
\and         M.~              Nu\~{n}ez Campos\inst{\ref{inst:0028}}
\and         L.~                    Delchambre\orcit{0000-0003-2559-408X}\inst{\ref{inst:0089}}
\and         A.~                      Dell'Oro\orcit{0000-0003-1561-9685}\inst{\ref{inst:0038}}
\and         P.~                        Esquej\orcit{0000-0001-8195-628X}\inst{\ref{inst:0096}}
\and         J.~   Fern\'{a}ndez-Hern\'{a}ndez\inst{\ref{inst:0097}}
\and         E.~                        Fraile\inst{\ref{inst:0096}}
\and         D.~                      Garabato\orcit{0000-0002-7133-6623}\inst{\ref{inst:0099}}
\and         P.~              Garc\'{i}a-Lario\orcit{0000-0003-4039-8212}\inst{\ref{inst:0027}}
\and         E.~                        Gosset\inst{\ref{inst:0089},\ref{inst:0037}}
\and         R.~                       Haigron\inst{\ref{inst:0019}}
\and      J.-L.~                     Halbwachs\orcit{0000-0003-2968-6395}\inst{\ref{inst:0104}}
\and       N.C.~                        Hambly\orcit{0000-0002-9901-9064}\inst{\ref{inst:0091}}
\and       D.L.~                      Harrison\orcit{0000-0001-8687-6588}\inst{\ref{inst:0025},\ref{inst:0107}}
\and         J.~                 Hern\'{a}ndez\orcit{0000-0002-0361-4994}\inst{\ref{inst:0027}}
\and         D.~                    Hestroffer\orcit{0000-0003-0472-9459}\inst{\ref{inst:0079}}
\and       S.T.~                       Hodgkin\orcit{0000-0002-5470-3962}\inst{\ref{inst:0025}}
\and         B.~                          Holl\orcit{0000-0001-6220-3266}\inst{\ref{inst:0026},\ref{inst:0062}}
\and         K.~                    Jan{\ss}en\orcit{0000-0002-8163-2493}\inst{\ref{inst:0113}}
\and         G.~          Jevardat de Fombelle\inst{\ref{inst:0026}}
\and         S.~                        Jordan\orcit{0000-0001-6316-6831}\inst{\ref{inst:0022}}
\and         A.~                 Krone-Martins\orcit{0000-0002-2308-6623}\inst{\ref{inst:0116},\ref{inst:0117}}
\and       A.C.~                     Lanzafame\orcit{0000-0002-2697-3607}\inst{\ref{inst:0118},\ref{inst:0119}}
\and         W.~                  L\"{ o}ffler\inst{\ref{inst:0022}}
\and         O.~                       Marchal\orcit{ 0000-0001-7461-892}\inst{\ref{inst:0104}}
\and       P.M.~                       Marrese\orcit{0000-0002-8162-3810}\inst{\ref{inst:0122},\ref{inst:0123}}
\and         A.~                      Moitinho\orcit{0000-0003-0822-5995}\inst{\ref{inst:0116}}
\and         K.~                      Muinonen\orcit{0000-0001-8058-2642}\inst{\ref{inst:0125},\ref{inst:0126}}
\and         P.~                       Osborne\inst{\ref{inst:0025}}
\and         E.~                       Pancino\orcit{0000-0003-0788-5879}\inst{\ref{inst:0038},\ref{inst:0123}}
\and         T.~                       Pauwels\inst{\ref{inst:0006}}
\and         C.~                     Reyl\'{e}\orcit{0000-0003-2258-2403}\inst{\ref{inst:0131}}
\and         M.~                        Riello\orcit{0000-0002-3134-0935}\inst{\ref{inst:0025}}
\and         L.~                     Rimoldini\orcit{0000-0002-0306-585X}\inst{\ref{inst:0062}}
\and         T.~                      Roegiers\orcit{0000-0002-1231-4440}\inst{\ref{inst:0134}}
\and         J.~                       Rybizki\orcit{0000-0002-0993-6089}\inst{\ref{inst:0007}}
\and       L.M.~                         Sarro\orcit{0000-0002-5622-5191}\inst{\ref{inst:0136}}
\and         C.~                        Siopis\orcit{0000-0002-6267-2924}\inst{\ref{inst:0036}}
\and         M.~                         Smith\inst{\ref{inst:0067}}
\and         A.~                      Sozzetti\orcit{0000-0002-7504-365X}\inst{\ref{inst:0005}}
\and         E.~                       Utrilla\inst{\ref{inst:0028}}
\and         M.~                   van Leeuwen\orcit{0000-0001-9698-2392}\inst{\ref{inst:0025}}
\and         U.~                         Abbas\orcit{0000-0002-5076-766X}\inst{\ref{inst:0005}}
\and         P.~               \'{A}brah\'{a}m\orcit{0000-0001-6015-646X}\inst{\ref{inst:0143},\ref{inst:0144}}
\and         A.~                Abreu Aramburu\inst{\ref{inst:0097}}
\and         C.~                         Aerts\orcit{0000-0003-1822-7126}\inst{\ref{inst:0146},\ref{inst:0147},\ref{inst:0007}}
\and       J.J.~                        Aguado\inst{\ref{inst:0136}}
\and         M.~                          Ajaj\inst{\ref{inst:0019}}
\and         F.~                 Aldea-Montero\inst{\ref{inst:0027}}
\and         G.~                     Altavilla\orcit{0000-0002-9934-1352}\inst{\ref{inst:0122},\ref{inst:0123}}
\and       M.A.~                   \'{A}lvarez\orcit{0000-0002-6786-2620}\inst{\ref{inst:0099}}
\and         J.~                         Alves\orcit{0000-0002-4355-0921}\inst{\ref{inst:0155}}
\and         F.~                        Anders\inst{\ref{inst:0029}}
\and       R.I.~                      Anderson\orcit{0000-0001-8089-4419}\inst{\ref{inst:0157}}
\and         E.~                Anglada Varela\orcit{0000-0001-7563-0689}\inst{\ref{inst:0097}}
\and         T.~                        Antoja\orcit{0000-0003-2595-5148}\inst{\ref{inst:0029}}
\and         D.~                        Baines\orcit{0000-0002-6923-3756}\inst{\ref{inst:0072}}
\and       S.G.~                         Baker\orcit{0000-0002-6436-1257}\inst{\ref{inst:0067}}
\and         L.~        Balaguer-N\'{u}\~{n}ez\orcit{0000-0001-9789-7069}\inst{\ref{inst:0029}}
\and         E.~                      Balbinot\orcit{0000-0002-1322-3153}\inst{\ref{inst:0163}}
\and         Z.~                         Balog\orcit{0000-0003-1748-2926}\inst{\ref{inst:0022},\ref{inst:0007}}
\and         C.~                       Barache\inst{\ref{inst:0074}}
\and         D.~                       Barbato\inst{\ref{inst:0026},\ref{inst:0005}}
\and         M.~                        Barros\orcit{0000-0002-9728-9618}\inst{\ref{inst:0116}}
\and       M.A.~                       Barstow\orcit{0000-0002-7116-3259}\inst{\ref{inst:0170}}
\and         S.~                 Bartolom\'{e}\orcit{0000-0002-6290-6030}\inst{\ref{inst:0029}}
\and      J.-L.~                     Bassilana\inst{\ref{inst:0172}}
\and         N.~                       Bauchet\inst{\ref{inst:0019}}
\and         U.~                      Becciani\orcit{0000-0002-4389-8688}\inst{\ref{inst:0118}}
\and         M.~                    Bellazzini\orcit{0000-0001-8200-810X}\inst{\ref{inst:0050}}
\and         A.~                     Berihuete\orcit{0000-0002-8589-4423}\inst{\ref{inst:0176}}
\and         M.~                        Bernet\orcit{0000-0001-7503-1010}\inst{\ref{inst:0029}}
\and         S.~                       Bertone\orcit{0000-0001-9885-8440}\inst{\ref{inst:0178},\ref{inst:0179},\ref{inst:0005}}
\and         L.~                       Bianchi\orcit{0000-0002-7999-4372}\inst{\ref{inst:0181}}
\and         A.~                    Binnenfeld\orcit{0000-0002-9319-3838}\inst{\ref{inst:0182}}
\and         S.~               Blanco-Cuaresma\orcit{0000-0002-1584-0171}\inst{\ref{inst:0183}}
\and         T.~                          Boch\orcit{0000-0001-5818-2781}\inst{\ref{inst:0104}}
\and         A.~                       Bombrun\inst{\ref{inst:0185}}
\and         D.~                       Bossini\orcit{0000-0002-9480-8400}\inst{\ref{inst:0186}}
\and         S.~                    Bouquillon\inst{\ref{inst:0074},\ref{inst:0188}}
\and         A.~                     Bragaglia\orcit{0000-0002-0338-7883}\inst{\ref{inst:0050}}
\and         L.~                      Bramante\inst{\ref{inst:0058}}
\and         E.~                        Breedt\orcit{0000-0001-6180-3438}\inst{\ref{inst:0025}}
\and         A.~                       Bressan\orcit{0000-0002-7922-8440}\inst{\ref{inst:0192}}
\and         N.~                     Brouillet\orcit{0000-0002-3274-7024}\inst{\ref{inst:0024}}
\and         E.~                    Brugaletta\orcit{0000-0003-2598-6737}\inst{\ref{inst:0118}}
\and         B.~                   Bucciarelli\orcit{0000-0002-5303-0268}\inst{\ref{inst:0005},\ref{inst:0047}}
\and         A.~                       Burlacu\inst{\ref{inst:0197}}
\and       A.G.~                     Butkevich\orcit{0000-0002-4098-3588}\inst{\ref{inst:0005}}
\and         R.~                         Buzzi\orcit{0000-0001-9389-5701}\inst{\ref{inst:0005}}
\and         E.~                        Caffau\orcit{0000-0001-6011-6134}\inst{\ref{inst:0019}}
\and         R.~                   Cancelliere\orcit{0000-0002-9120-3799}\inst{\ref{inst:0201}}
\and         T.~                 Cantat-Gaudin\orcit{0000-0001-8726-2588}\inst{\ref{inst:0029},\ref{inst:0007}}
\and         R.~                      Carballo\orcit{0000-0001-7412-2498}\inst{\ref{inst:0204}}
\and         T.~                      Carlucci\inst{\ref{inst:0074}}
\and       M.I.~                     Carnerero\orcit{0000-0001-5843-5515}\inst{\ref{inst:0005}}
\and       J.M.~                      Carrasco\orcit{0000-0002-3029-5853}\inst{\ref{inst:0029}}
\and         L.~                   Casamiquela\orcit{0000-0001-5238-8674}\inst{\ref{inst:0024},\ref{inst:0019}}
\and         M.~                    Castellani\orcit{0000-0002-7650-7428}\inst{\ref{inst:0122}}
\and         A.~                 Castro-Ginard\orcit{0000-0002-9419-3725}\inst{\ref{inst:0015}}
\and         L.~                        Chaoul\inst{\ref{inst:0035}}
\and         P.~                       Charlot\orcit{0000-0002-9142-716X}\inst{\ref{inst:0024}}
\and         L.~                        Chemin\orcit{0000-0002-3834-7937}\inst{\ref{inst:0214}}
\and         V.~                    Chiaramida\inst{\ref{inst:0058}}
\and         A.~                     Chiavassa\orcit{0000-0003-3891-7554}\inst{\ref{inst:0001}}
\and         N.~                       Chornay\orcit{0000-0002-8767-3907}\inst{\ref{inst:0025}}
\and         G.~                     Comoretto\inst{\ref{inst:0072},\ref{inst:0219}}
\and         G.~                      Contursi\orcit{0000-0001-5370-1511}\inst{\ref{inst:0001}}
\and       W.J.~                        Cooper\orcit{0000-0003-3501-8967}\inst{\ref{inst:0221},\ref{inst:0005}}
\and         T.~                        Cornez\inst{\ref{inst:0172}}
\and         S.~                        Cowell\inst{\ref{inst:0025}}
\and         F.~                         Crifo\inst{\ref{inst:0019}}
\and         M.~                       Cropper\orcit{0000-0003-4571-9468}\inst{\ref{inst:0067}}
\and         M.~                        Crosta\orcit{0000-0003-4369-3786}\inst{\ref{inst:0005},\ref{inst:0228}}
\and         C.~                       Crowley\inst{\ref{inst:0185}}
\and         C.~                       Dafonte\orcit{0000-0003-4693-7555}\inst{\ref{inst:0099}}
\and         A.~                    Dapergolas\inst{\ref{inst:0077}}
\and         P.~                         David\inst{\ref{inst:0079}}
\and         F.~                      De Luise\orcit{0000-0002-6570-8208}\inst{\ref{inst:0233}}
\and         R.~                      De March\orcit{0000-0003-0567-842X}\inst{\ref{inst:0058}}
\and         J.~                     De Ridder\orcit{0000-0001-6726-2863}\inst{\ref{inst:0146}}
\and         R.~                      de Souza\inst{\ref{inst:0236}}
\and         A.~                     de Torres\inst{\ref{inst:0185}}
\and       E.F.~                    del Peloso\inst{\ref{inst:0022}}
\and         E.~                      del Pozo\inst{\ref{inst:0028}}
\and         M.~                         Delbo\orcit{0000-0002-8963-2404}\inst{\ref{inst:0001}}
\and         A.~                       Delgado\inst{\ref{inst:0096}}
\and      J.-B.~                       Delisle\orcit{0000-0001-5844-9888}\inst{\ref{inst:0026}}
\and         C.~                      Demouchy\inst{\ref{inst:0243}}
\and         S.~                       Diakite\inst{\ref{inst:0244}}
\and         C.~                        Diener\inst{\ref{inst:0025}}
\and         E.~                     Distefano\orcit{0000-0002-2448-2513}\inst{\ref{inst:0118}}
\and         C.~                       Dolding\inst{\ref{inst:0067}}
\and         H.~                          Enke\orcit{0000-0002-2366-8316}\inst{\ref{inst:0113}}
\and         C.~                         Fabre\inst{\ref{inst:0249}}
\and         M.~                      Fabrizio\orcit{0000-0001-5829-111X}\inst{\ref{inst:0122},\ref{inst:0123}}
\and         S.~                       Faigler\orcit{0000-0002-8368-5724}\inst{\ref{inst:0252}}
\and         G.~                      Fedorets\orcit{0000-0002-8418-4809}\inst{\ref{inst:0125},\ref{inst:0254}}
\and         P.~                      Fernique\orcit{0000-0002-3304-2923}\inst{\ref{inst:0104},\ref{inst:0256}}
\and         F.~                      Figueras\orcit{0000-0002-3393-0007}\inst{\ref{inst:0029}}
\and         Y.~                      Fournier\orcit{0000-0002-6633-9088}\inst{\ref{inst:0113}}
\and         C.~                        Fouron\inst{\ref{inst:0197}}
\and         F.~                     Fragkoudi\orcit{0000-0002-0897-3013}\inst{\ref{inst:0260},\ref{inst:0261},\ref{inst:0262}}
\and         M.~                           Gai\orcit{0000-0001-9008-134X}\inst{\ref{inst:0005}}
\and         A.~              Garcia-Gutierrez\inst{\ref{inst:0029}}
\and         M.~              Garcia-Reinaldos\inst{\ref{inst:0027}}
\and         M.~             Garc\'{i}a-Torres\orcit{0000-0002-6867-7080}\inst{\ref{inst:0266}}
\and         A.~                      Garofalo\orcit{0000-0002-5907-0375}\inst{\ref{inst:0050}}
\and         A.~                         Gavel\orcit{0000-0002-2963-722X}\inst{\ref{inst:0056}}
\and         P.~                        Gavras\orcit{0000-0002-4383-4836}\inst{\ref{inst:0096}}
\and         E.~                       Gerlach\orcit{0000-0002-9533-2168}\inst{\ref{inst:0030}}
\and         R.~                         Geyer\orcit{0000-0001-6967-8707}\inst{\ref{inst:0030}}
\and         P.~                      Giacobbe\orcit{0000-0001-7034-7024}\inst{\ref{inst:0005}}
\and         G.~                       Gilmore\orcit{0000-0003-4632-0213}\inst{\ref{inst:0025}}
\and         S.~                        Girona\orcit{0000-0002-1975-1918}\inst{\ref{inst:0274}}
\and         G.~                     Giuffrida\inst{\ref{inst:0122}}
\and         R.~                         Gomel\inst{\ref{inst:0252}}
\and         A.~                         Gomez\orcit{0000-0002-3796-3690}\inst{\ref{inst:0099}}
\and         J.~    Gonz\'{a}lez-N\'{u}\~{n}ez\orcit{0000-0001-5311-5555}\inst{\ref{inst:0087},\ref{inst:0279}}
\and         I.~   Gonz\'{a}lez-Santamar\'{i}a\orcit{0000-0002-8537-9384}\inst{\ref{inst:0099}}
\and       J.J.~            Gonz\'{a}lez-Vidal\inst{\ref{inst:0029}}
\and         M.~                       Granvik\orcit{0000-0002-5624-1888}\inst{\ref{inst:0125},\ref{inst:0283}}
\and         P.~                      Guillout\inst{\ref{inst:0104}}
\and         J.~                       Guiraud\inst{\ref{inst:0035}}
\and         R.~     Guti\'{e}rrez-S\'{a}nchez\inst{\ref{inst:0072}}
\and       L.P.~                           Guy\orcit{0000-0003-0800-8755}\inst{\ref{inst:0062},\ref{inst:0288}}
\and         D.~                Hatzidimitriou\orcit{0000-0002-5415-0464}\inst{\ref{inst:0289},\ref{inst:0077}}
\and         M.~                        Hauser\inst{\ref{inst:0007},\ref{inst:0292}}
\and         M.~                       Haywood\orcit{0000-0003-0434-0400}\inst{\ref{inst:0019}}
\and         A.~                        Helmer\inst{\ref{inst:0172}}
\and         A.~                         Helmi\orcit{0000-0003-3937-7641}\inst{\ref{inst:0163}}
\and       M.H.~                     Sarmiento\orcit{0000-0003-4252-5115}\inst{\ref{inst:0028}}
\and       S.L.~                       Hidalgo\orcit{0000-0002-0002-9298}\inst{\ref{inst:0297},\ref{inst:0298}}
\and         N.~                   H\l{}adczuk\orcit{0000-0001-9163-4209}\inst{\ref{inst:0027},\ref{inst:0300}}
\and         D.~                         Hobbs\orcit{0000-0002-2696-1366}\inst{\ref{inst:0032}}
\and         G.~                       Holland\inst{\ref{inst:0025}}
\and       H.E.~                        Huckle\inst{\ref{inst:0067}}
\and         K.~                       Jardine\inst{\ref{inst:0304}}
\and         G.~                    Jasniewicz\inst{\ref{inst:0305}}
\and         A.~          Jean-Antoine Piccolo\orcit{0000-0001-8622-212X}\inst{\ref{inst:0035}}
\and     \'{O}.~            Jim\'{e}nez-Arranz\orcit{0000-0001-7434-5165}\inst{\ref{inst:0029}}
\and         J.~             Juaristi Campillo\inst{\ref{inst:0022}}
\and         F.~                         Julbe\inst{\ref{inst:0029}}
\and         L.~                     Karbevska\inst{\ref{inst:0062},\ref{inst:0311}}
\and         P.~                      Kervella\orcit{0000-0003-0626-1749}\inst{\ref{inst:0312}}
\and         S.~                        Khanna\orcit{0000-0002-2604-4277}\inst{\ref{inst:0163},\ref{inst:0005}}
\and       A.J.~                          Korn\orcit{0000-0002-3881-6756}\inst{\ref{inst:0056}}
\and      \'{A}~                K\'{o}sp\'{a}l\orcit{'{u}t 15-17, 1121 B}\inst{\ref{inst:0143},\ref{inst:0007},\ref{inst:0144}}
\and         Z.~           Kostrzewa-Rutkowska\inst{\ref{inst:0015},\ref{inst:0320}}
\and         K.~                Kruszy\'{n}ska\orcit{0000-0002-2729-5369}\inst{\ref{inst:0321}}
\and         M.~                           Kun\orcit{0000-0002-7538-5166}\inst{\ref{inst:0143}}
\and         P.~                       Laizeau\inst{\ref{inst:0323}}
\and         S.~                       Lambert\orcit{0000-0001-6759-5502}\inst{\ref{inst:0074}}
\and       A.F.~                         Lanza\orcit{0000-0001-5928-7251}\inst{\ref{inst:0118}}
\and         Y.~                         Lasne\inst{\ref{inst:0172}}
\and      J.-F.~                    Le Campion\inst{\ref{inst:0024}}
\and         Y.~                      Lebreton\orcit{0000-0002-4834-2144}\inst{\ref{inst:0312},\ref{inst:0329}}
\and         T.~                     Lebzelter\orcit{0000-0002-0702-7551}\inst{\ref{inst:0155}}
\and         S.~                        Leccia\orcit{0000-0001-5685-6930}\inst{\ref{inst:0331}}
\and         N.~                       Leclerc\inst{\ref{inst:0019}}
\and         I.~                 Lecoeur-Taibi\orcit{0000-0003-0029-8575}\inst{\ref{inst:0062}}
\and         S.~                          Liao\orcit{0000-0002-9346-0211}\inst{\ref{inst:0334},\ref{inst:0005},\ref{inst:0336}}
\and       E.L.~                        Licata\orcit{0000-0002-5203-0135}\inst{\ref{inst:0005}}
\and     H.E.P.~                  Lindstr{\o}m\inst{\ref{inst:0005},\ref{inst:0339},\ref{inst:0340}}
\and       T.A.~                        Lister\orcit{0000-0002-3818-7769}\inst{\ref{inst:0341}}
\and         E.~                       Livanou\orcit{0000-0003-0628-2347}\inst{\ref{inst:0289}}
\and         A.~                         Lobel\orcit{0000-0001-5030-019X}\inst{\ref{inst:0006}}
\and         A.~                         Lorca\inst{\ref{inst:0028}}
\and         C.~                          Loup\inst{\ref{inst:0104}}
\and         P.~                 Madrero Pardo\inst{\ref{inst:0029}}
\and         A.~               Magdaleno Romeo\inst{\ref{inst:0197}}
\and         S.~                       Managau\inst{\ref{inst:0172}}
\and       R.G.~                          Mann\orcit{0000-0002-0194-325X}\inst{\ref{inst:0091}}
\and         M.~                      Manteiga\orcit{0000-0002-7711-5581}\inst{\ref{inst:0350}}
\and       J.M.~                      Marchant\orcit{0000-0002-3678-3145}\inst{\ref{inst:0351}}
\and         M.~                       Marconi\orcit{0000-0002-1330-2927}\inst{\ref{inst:0331}}
\and         J.~                        Marcos\inst{\ref{inst:0072}}
\and     M.M.S.~                 Marcos Santos\inst{\ref{inst:0087}}
\and         D.~                Mar\'{i}n Pina\orcit{0000-0001-6482-1842}\inst{\ref{inst:0029}}
\and         S.~                      Marinoni\orcit{0000-0001-7990-6849}\inst{\ref{inst:0122},\ref{inst:0123}}
\and         F.~                       Marocco\orcit{0000-0001-7519-1700}\inst{\ref{inst:0358}}
\and         L.~                   Martin Polo\inst{\ref{inst:0087}}
\and       J.M.~            Mart\'{i}n-Fleitas\orcit{0000-0002-8594-569X}\inst{\ref{inst:0028}}
\and         G.~                        Marton\orcit{0000-0002-1326-1686}\inst{\ref{inst:0143}}
\and         N.~                          Mary\inst{\ref{inst:0172}}
\and         A.~                         Masip\orcit{0000-0003-1419-0020}\inst{\ref{inst:0029}}
\and         D.~                       Massari\orcit{0000-0001-8892-4301}\inst{\ref{inst:0050}}
\and         A.~          Mastrobuono-Battisti\orcit{0000-0002-2386-9142}\inst{\ref{inst:0019}}
\and         T.~                         Mazeh\orcit{0000-0002-3569-3391}\inst{\ref{inst:0252}}
\and       P.J.~                      McMillan\orcit{0000-0002-8861-2620}\inst{\ref{inst:0032}}
\and         S.~                       Messina\orcit{0000-0002-2851-2468}\inst{\ref{inst:0118}}
\and         D.~                      Michalik\orcit{0000-0002-7618-6556}\inst{\ref{inst:0017}}
\and       N.R.~                        Millar\inst{\ref{inst:0025}}
\and         A.~                         Mints\orcit{0000-0002-8440-1455}\inst{\ref{inst:0113}}
\and         D.~                        Molina\orcit{0000-0003-4814-0275}\inst{\ref{inst:0029}}
\and         R.~                      Molinaro\orcit{0000-0003-3055-6002}\inst{\ref{inst:0331}}
\and         L.~                    Moln\'{a}r\orcit{0000-0002-8159-1599}\inst{\ref{inst:0143},\ref{inst:0375},\ref{inst:0144}}
\and         G.~                        Monari\orcit{0000-0002-6863-0661}\inst{\ref{inst:0104}}
\and         M.~                   Mongui\'{o}\orcit{0000-0002-4519-6700}\inst{\ref{inst:0029}}
\and         P.~                   Montegriffo\orcit{0000-0001-5013-5948}\inst{\ref{inst:0050}}
\and         A.~                       Montero\inst{\ref{inst:0028}}
\and         R.~                           Mor\orcit{0000-0002-8179-6527}\inst{\ref{inst:0029}}
\and         A.~                          Mora\inst{\ref{inst:0028}}
\and         R.~                    Morbidelli\orcit{0000-0001-7627-4946}\inst{\ref{inst:0005}}
\and         T.~                         Morel\orcit{0000-0002-8176-4816}\inst{\ref{inst:0089}}
\and         D.~                        Morris\inst{\ref{inst:0091}}
\and         T.~                      Muraveva\orcit{0000-0002-0969-1915}\inst{\ref{inst:0050}}
\and       C.P.~                        Murphy\inst{\ref{inst:0027}}
\and         I.~                       Musella\orcit{0000-0001-5909-6615}\inst{\ref{inst:0331}}
\and         Z.~                          Nagy\orcit{0000-0002-3632-1194}\inst{\ref{inst:0143}}
\and         L.~                         Noval\inst{\ref{inst:0172}}
\and         F.~                     Oca\~{n}a\inst{\ref{inst:0072},\ref{inst:0391}}
\and         A.~                         Ogden\inst{\ref{inst:0025}}
\and         C.~                     Ordenovic\inst{\ref{inst:0001}}
\and       J.O.~                        Osinde\inst{\ref{inst:0096}}
\and         C.~                        Pagani\orcit{0000-0001-5477-4720}\inst{\ref{inst:0170}}
\and         I.~                        Pagano\orcit{0000-0001-9573-4928}\inst{\ref{inst:0118}}
\and         L.~                     Palaversa\orcit{0000-0003-3710-0331}\inst{\ref{inst:0397},\ref{inst:0025}}
\and       P.A.~                       Palicio\orcit{0000-0002-7432-8709}\inst{\ref{inst:0001}}
\and         L.~               Pallas-Quintela\orcit{0000-0001-9296-3100}\inst{\ref{inst:0099}}
\and         A.~                        Panahi\orcit{0000-0001-5850-4373}\inst{\ref{inst:0252}}
\and         S.~               Payne-Wardenaar\inst{\ref{inst:0022}}
\and         X.~         Pe\~{n}alosa Esteller\inst{\ref{inst:0029}}
\and         A.~                 Penttil\"{ a}\orcit{0000-0001-7403-1721}\inst{\ref{inst:0125}}
\and         B.~                        Pichon\orcit{0000 0000 0062 1449}\inst{\ref{inst:0001}}
\and       A.M.~                    Piersimoni\orcit{0000-0002-8019-3708}\inst{\ref{inst:0233}}
\and      F.-X.~                        Pineau\orcit{0000-0002-2335-4499}\inst{\ref{inst:0104}}
\and         E.~                        Plachy\orcit{0000-0002-5481-3352}\inst{\ref{inst:0143},\ref{inst:0375},\ref{inst:0144}}
\and         G.~                          Plum\inst{\ref{inst:0019}}
\and         E.~                        Poggio\orcit{0000-0003-3793-8505}\inst{\ref{inst:0001},\ref{inst:0005}}
\and         A.~                      Pr\v{s}a\orcit{0000-0002-1913-0281}\inst{\ref{inst:0413}}
\and         L.~                        Pulone\orcit{0000-0002-5285-998X}\inst{\ref{inst:0122}}
\and         E.~                        Racero\orcit{0000-0002-6101-9050}\inst{\ref{inst:0087},\ref{inst:0391}}
\and         S.~                       Ragaini\inst{\ref{inst:0050}}
\and         M.~                        Rainer\orcit{0000-0002-8786-2572}\inst{\ref{inst:0038},\ref{inst:0419}}
\and       C.M.~                       Raiteri\orcit{0000-0003-1784-2784}\inst{\ref{inst:0005}}
\and         P.~                         Ramos\orcit{0000-0002-5080-7027}\inst{\ref{inst:0029},\ref{inst:0104}}
\and         M.~                  Ramos-Lerate\inst{\ref{inst:0072}}
\and         P.~                  Re Fiorentin\orcit{0000-0002-4995-0475}\inst{\ref{inst:0005}}
\and         S.~                        Regibo\inst{\ref{inst:0146}}
\and       P.J.~                      Richards\inst{\ref{inst:0426}}
\and         C.~                     Rios Diaz\inst{\ref{inst:0096}}
\and         V.~                        Ripepi\orcit{0000-0003-1801-426X}\inst{\ref{inst:0331}}
\and         A.~                          Riva\orcit{0000-0002-6928-8589}\inst{\ref{inst:0005}}
\and      H.-W.~                           Rix\orcit{0000-0003-4996-9069}\inst{\ref{inst:0007}}
\and         G.~                         Rixon\orcit{0000-0003-4399-6568}\inst{\ref{inst:0025}}
\and         N.~                      Robichon\orcit{0000-0003-4545-7517}\inst{\ref{inst:0019}}
\and       A.C.~                         Robin\orcit{0000-0001-8654-9499}\inst{\ref{inst:0131}}
\and         C.~                         Robin\inst{\ref{inst:0172}}
\and         M.~                       Roelens\orcit{0000-0003-0876-4673}\inst{\ref{inst:0026}}
\and     H.R.O.~                        Rogues\inst{\ref{inst:0243}}
\and         L.~                    Rohrbasser\inst{\ref{inst:0062}}
\and         M.~              Romero-G\'{o}mez\orcit{0000-0003-3936-1025}\inst{\ref{inst:0029}}
\and         N.~                        Rowell\orcit{0000-0003-3809-1895}\inst{\ref{inst:0091}}
\and         F.~                         Royer\orcit{0000-0002-9374-8645}\inst{\ref{inst:0019}}
\and         D.~                    Ruz Mieres\orcit{0000-0002-9455-157X}\inst{\ref{inst:0025}}
\and       K.A.~                       Rybicki\orcit{0000-0002-9326-9329}\inst{\ref{inst:0321}}
\and         G.~                      Sadowski\orcit{0000-0002-3411-1003}\inst{\ref{inst:0036}}
\and         A.~        S\'{a}ez N\'{u}\~{n}ez\inst{\ref{inst:0029}}
\and         A.~       Sagrist\`{a} Sell\'{e}s\orcit{0000-0001-6191-2028}\inst{\ref{inst:0022}}
\and         J.~                      Sahlmann\orcit{0000-0001-9525-3673}\inst{\ref{inst:0096}}
\and         E.~                      Salguero\inst{\ref{inst:0097}}
\and         N.~                       Samaras\orcit{0000-0001-8375-6652}\inst{\ref{inst:0006},\ref{inst:0449}}
\and         V.~               Sanchez Gimenez\orcit{0000-0003-1797-3557}\inst{\ref{inst:0029}}
\and         N.~                         Sanna\orcit{0000-0001-9275-9492}\inst{\ref{inst:0038}}
\and         R.~                 Santove\~{n}a\orcit{0000-0002-9257-2131}\inst{\ref{inst:0099}}
\and         M.~                       Sarasso\orcit{0000-0001-5121-0727}\inst{\ref{inst:0005}}
\and         E.~                       Sciacca\orcit{0000-0002-5574-2787}\inst{\ref{inst:0118}}
\and         M.~                         Segol\inst{\ref{inst:0243}}
\and       J.C.~                       Segovia\inst{\ref{inst:0087}}
\and         D.~                 S\'{e}gransan\orcit{0000-0003-2355-8034}\inst{\ref{inst:0026}}
\and         D.~                        Semeux\inst{\ref{inst:0249}}
\and         S.~                        Shahaf\orcit{0000-0001-9298-8068}\inst{\ref{inst:0459}}
\and       H.I.~                      Siddiqui\orcit{0000-0003-1853-6033}\inst{\ref{inst:0460}}
\and         A.~                       Siebert\orcit{0000-0001-8059-2840}\inst{\ref{inst:0104},\ref{inst:0256}}
\and         L.~                       Siltala\orcit{0000-0002-6938-794X}\inst{\ref{inst:0125}}
\and         A.~                       Silvelo\orcit{0000-0002-5126-6365}\inst{\ref{inst:0099}}
\and         E.~                        Slezak\inst{\ref{inst:0001}}
\and         I.~                        Slezak\inst{\ref{inst:0001}}
\and       R.L.~                         Smart\orcit{0000-0002-4424-4766}\inst{\ref{inst:0005}}
\and       O.N.~                        Snaith\inst{\ref{inst:0019}}
\and         E.~                        Solano\inst{\ref{inst:0469}}
\and         F.~                       Solitro\inst{\ref{inst:0058}}
\and         D.~                        Souami\orcit{0000-0003-4058-0815}\inst{\ref{inst:0312},\ref{inst:0472}}
\and         J.~                       Souchay\inst{\ref{inst:0074}}
\and         A.~                        Spagna\orcit{0000-0003-1732-2412}\inst{\ref{inst:0005}}
\and         L.~                         Spina\orcit{0000-0002-9760-6249}\inst{\ref{inst:0014}}
\and         F.~                         Spoto\orcit{0000-0001-7319-5847}\inst{\ref{inst:0183}}
\and       I.A.~                        Steele\orcit{0000-0001-8397-5759}\inst{\ref{inst:0351}}
\and         H.~            Steidelm\"{ u}ller\inst{\ref{inst:0030}}
\and       C.A.~                    Stephenson\inst{\ref{inst:0072},\ref{inst:0480}}
\and         M.~                  S\"{ u}veges\orcit{0000-0003-3017-5322}\inst{\ref{inst:0481}}
\and         J.~                        Surdej\orcit{0000-0002-7005-1976}\inst{\ref{inst:0089},\ref{inst:0483}}
\and         L.~                      Szabados\orcit{0000-0002-2046-4131}\inst{\ref{inst:0143}}
\and         E.~                  Szegedi-Elek\orcit{0000-0001-7807-6644}\inst{\ref{inst:0143}}
\and         F.~                         Taris\inst{\ref{inst:0074}}
\and       M.B.~                        Taylor\orcit{0000-0002-4209-1479}\inst{\ref{inst:0487}}
\and         R.~                      Teixeira\orcit{0000-0002-6806-6626}\inst{\ref{inst:0236}}
\and         L.~                       Tolomei\orcit{0000-0002-3541-3230}\inst{\ref{inst:0058}}
\and         N.~                       Tonello\orcit{0000-0003-0550-1667}\inst{\ref{inst:0274}}
\and         F.~                         Torra\orcit{0000-0002-8429-299X}\inst{\ref{inst:0051}}
\and         J.~               Torra$^\dagger$\inst{\ref{inst:0029}}
\and         G.~                Torralba Elipe\orcit{0000-0001-8738-194X}\inst{\ref{inst:0099}}
\and         M.~                     Trabucchi\orcit{0000-0002-1429-2388}\inst{\ref{inst:0494},\ref{inst:0026}}
\and       A.T.~                       Tsounis\inst{\ref{inst:0496}}
\and         C.~                         Turon\orcit{0000-0003-1236-5157}\inst{\ref{inst:0019}}
\and         A.~                          Ulla\orcit{0000-0001-6424-5005}\inst{\ref{inst:0498}}
\and         N.~                         Unger\orcit{0000-0003-3993-7127}\inst{\ref{inst:0026}}
\and       M.V.~                      Vaillant\inst{\ref{inst:0172}}
\and         E.~                    van Dillen\inst{\ref{inst:0243}}
\and         W.~                    van Reeven\inst{\ref{inst:0502}}
\and         O.~                         Vanel\orcit{0000-0002-7898-0454}\inst{\ref{inst:0019}}
\and         A.~                     Vecchiato\orcit{0000-0003-1399-5556}\inst{\ref{inst:0005}}
\and         Y.~                         Viala\inst{\ref{inst:0019}}
\and         D.~                       Vicente\orcit{0000-0002-1584-1182}\inst{\ref{inst:0274}}
\and         S.~                     Voutsinas\inst{\ref{inst:0091}}
\and         M.~                        Weiler\inst{\ref{inst:0029}}
\and         T.~                        Wevers\orcit{0000-0002-4043-9400}\inst{\ref{inst:0025},\ref{inst:0510}}
\and      \L{}.~                   Wyrzykowski\orcit{0000-0002-9658-6151}\inst{\ref{inst:0321}}
\and         A.~                        Yoldas\inst{\ref{inst:0025}}
\and         P.~                         Yvard\inst{\ref{inst:0243}}
\and         J.~                         Zorec\inst{\ref{inst:0514}}
\and         S.~                        Zucker\orcit{0000-0003-3173-3138}\inst{\ref{inst:0182}}
}
\institute{
     Universit\'{e} C\^{o}te d'Azur, Observatoire de la C\^{o}te d'Azur, CNRS, Laboratoire Lagrange, Bd de l'Observatoire, CS 34229, 06304 Nice Cedex 4, France\relax                                                                                                                                                                                              \label{inst:0001}
\and Faculty of Mathematics and Physics, University of Ljubljana, Jadranska ulica 19, 1000 Ljubljana, Slovenia\relax                                                                                                                                                                                                                                               \label{inst:0003}\vfill
\and IRAP, Universit\'{e} de Toulouse, CNRS, UPS, CNES, 9 Av. colonel Roche, BP 44346, 31028 Toulouse Cedex 4, France\relax                                                                                                                                                                                                                                        \label{inst:0004}\vfill
\and INAF - Osservatorio Astrofisico di Torino, via Osservatorio 20, 10025 Pino Torinese (TO), Italy\relax                                                                                                                                                                                                                                                         \label{inst:0005}\vfill
\and Royal Observatory of Belgium, Ringlaan 3, 1180 Brussels, Belgium\relax                                                                                                                                                                                                                                                                                        \label{inst:0006}\vfill
\and Max Planck Institute for Astronomy, K\"{ o}nigstuhl 17, 69117 Heidelberg, Germany\relax                                                                                                                                                                                                                                                                       \label{inst:0007}\vfill
\and INAF - Osservatorio astronomico di Padova, Vicolo Osservatorio 5, 35122 Padova, Italy\relax                                                                                                                                                                                                                                                                   \label{inst:0014}\vfill
\and Leiden Observatory, Leiden University, Niels Bohrweg 2, 2333 CA Leiden, The Netherlands\relax                                                                                                                                                                                                                                                                 \label{inst:0015}\vfill
\and European Space Agency (ESA), European Space Research and Technology Centre (ESTEC), Keplerlaan 1, 2201AZ, Noordwijk, The Netherlands\relax                                                                                                                                                                                                                    \label{inst:0017}\vfill
\and GEPI, Observatoire de Paris, Universit\'{e} PSL, CNRS, 5 Place Jules Janssen, 92190 Meudon, France\relax                                                                                                                                                                                                                                                      \label{inst:0019}\vfill
\and Univ. Grenoble Alpes, CNRS, IPAG, 38000 Grenoble, France\relax                                                                                                                                                                                                                                                                                                \label{inst:0020}\vfill
\and Astronomisches Rechen-Institut, Zentrum f\"{ u}r Astronomie der Universit\"{ a}t Heidelberg, M\"{ o}nchhofstr. 12-14, 69120 Heidelberg, Germany\relax                                                                                                                                                                                                         \label{inst:0022}\vfill
\and Laboratoire d'astrophysique de Bordeaux, Univ. Bordeaux, CNRS, B18N, all{\'e}e Geoffroy Saint-Hilaire, 33615 Pessac, France\relax                                                                                                                                                                                                                             \label{inst:0024}\vfill
\and Institute of Astronomy, University of Cambridge, Madingley Road, Cambridge CB3 0HA, United Kingdom\relax                                                                                                                                                                                                                                                      \label{inst:0025}\vfill
\and Department of Astronomy, University of Geneva, Chemin Pegasi 51, 1290 Versoix, Switzerland\relax                                                                                                                                                                                                                                                              \label{inst:0026}\vfill
\and European Space Agency (ESA), European Space Astronomy Centre (ESAC), Camino bajo del Castillo, s/n, Urbanizacion Villafranca del Castillo, Villanueva de la Ca\~{n}ada, 28692 Madrid, Spain\relax                                                                                                                                                             \label{inst:0027}\vfill
\and Aurora Technology for European Space Agency (ESA), Camino bajo del Castillo, s/n, Urbanizacion Villafranca del Castillo, Villanueva de la Ca\~{n}ada, 28692 Madrid, Spain\relax                                                                                                                                                                               \label{inst:0028}\vfill
\and Institut de Ci\`{e}ncies del Cosmos (ICCUB), Universitat  de  Barcelona  (IEEC-UB), Mart\'{i} i  Franqu\`{e}s  1, 08028 Barcelona, Spain\relax                                                                                                                                                                                                                \label{inst:0029}\vfill
\and Lohrmann Observatory, Technische Universit\"{ a}t Dresden, Mommsenstra{\ss}e 13, 01062 Dresden, Germany\relax                                                                                                                                                                                                                                                 \label{inst:0030}\vfill
\and Lund Observatory, Department of Astronomy and Theoretical Physics, Lund University, Box 43, 22100 Lund, Sweden\relax                                                                                                                                                                                                                                          \label{inst:0032}\vfill
\and CNES Centre Spatial de Toulouse, 18 avenue Edouard Belin, 31401 Toulouse Cedex 9, France\relax                                                                                                                                                                                                                                                                \label{inst:0035}\vfill
\and Institut d'Astronomie et d'Astrophysique, Universit\'{e} Libre de Bruxelles CP 226, Boulevard du Triomphe, 1050 Brussels, Belgium\relax                                                                                                                                                                                                                       \label{inst:0036}\vfill
\and F.R.S.-FNRS, Rue d'Egmont 5, 1000 Brussels, Belgium\relax                                                                                                                                                                                                                                                                                                     \label{inst:0037}\vfill
\and INAF - Osservatorio Astrofisico di Arcetri, Largo Enrico Fermi 5, 50125 Firenze, Italy\relax                                                                                                                                                                                                                                                                  \label{inst:0038}\vfill
\and European Space Agency (ESA, retired)\relax                                                                                                                                                                                                                                                                                                                    \label{inst:0044}\vfill
\and University of Turin, Department of Physics, Via Pietro Giuria 1, 10125 Torino, Italy\relax                                                                                                                                                                                                                                                                    \label{inst:0047}\vfill
\and INAF - Osservatorio di Astrofisica e Scienza dello Spazio di Bologna, via Piero Gobetti 93/3, 40129 Bologna, Italy\relax                                                                                                                                                                                                                                      \label{inst:0050}\vfill
\and DAPCOM for Institut de Ci\`{e}ncies del Cosmos (ICCUB), Universitat  de  Barcelona  (IEEC-UB), Mart\'{i} i  Franqu\`{e}s  1, 08028 Barcelona, Spain\relax                                                                                                                                                                                                     \label{inst:0051}\vfill
\and Observational Astrophysics, Division of Astronomy and Space Physics, Department of Physics and Astronomy, Uppsala University, Box 516, 751 20 Uppsala, Sweden\relax                                                                                                                                                                                           \label{inst:0056}\vfill
\and ALTEC S.p.a, Corso Marche, 79,10146 Torino, Italy\relax                                                                                                                                                                                                                                                                                                       \label{inst:0058}\vfill
\and S\`{a}rl, Geneva, Switzerland\relax                                                                                                                                                                                                                                                                                                                           \label{inst:0061}\vfill
\and Department of Astronomy, University of Geneva, Chemin d'Ecogia 16, 1290 Versoix, Switzerland\relax                                                                                                                                                                                                                                                            \label{inst:0062}\vfill
\and Mullard Space Science Laboratory, University College London, Holmbury St Mary, Dorking, Surrey RH5 6NT, United Kingdom\relax                                                                                                                                                                                                                                  \label{inst:0067}\vfill
\and Gaia DPAC Project Office, ESAC, Camino bajo del Castillo, s/n, Urbanizacion Villafranca del Castillo, Villanueva de la Ca\~{n}ada, 28692 Madrid, Spain\relax                                                                                                                                                                                                  \label{inst:0069}\vfill
\and Telespazio UK S.L. for European Space Agency (ESA), Camino bajo del Castillo, s/n, Urbanizacion Villafranca del Castillo, Villanueva de la Ca\~{n}ada, 28692 Madrid, Spain\relax                                                                                                                                                                              \label{inst:0072}\vfill
\and SYRTE, Observatoire de Paris, Universit\'{e} PSL, CNRS,  Sorbonne Universit\'{e}, LNE, 61 avenue de l'Observatoire 75014 Paris, France\relax                                                                                                                                                                                                                  \label{inst:0074}\vfill
\and National Observatory of Athens, I. Metaxa and Vas. Pavlou, Palaia Penteli, 15236 Athens, Greece\relax                                                                                                                                                                                                                                                         \label{inst:0077}\vfill
\and IMCCE, Observatoire de Paris, Universit\'{e} PSL, CNRS, Sorbonne Universit{\'e}, Univ. Lille, 77 av. Denfert-Rochereau, 75014 Paris, France\relax                                                                                                                                                                                                             \label{inst:0079}\vfill
\and Serco Gesti\'{o}n de Negocios for European Space Agency (ESA), Camino bajo del Castillo, s/n, Urbanizacion Villafranca del Castillo, Villanueva de la Ca\~{n}ada, 28692 Madrid, Spain\relax                                                                                                                                                                   \label{inst:0087}\vfill
\and Institut d'Astrophysique et de G\'{e}ophysique, Universit\'{e} de Li\`{e}ge, 19c, All\'{e}e du 6 Ao\^{u}t, B-4000 Li\`{e}ge, Belgium\relax                                                                                                                                                                                                                    \label{inst:0089}\vfill
\and CRAAG - Centre de Recherche en Astronomie, Astrophysique et G\'{e}ophysique, Route de l'Observatoire Bp 63 Bouzareah 16340 Algiers, Algeria\relax                                                                                                                                                                                                             \label{inst:0090}\vfill
\and Institute for Astronomy, University of Edinburgh, Royal Observatory, Blackford Hill, Edinburgh EH9 3HJ, United Kingdom\relax                                                                                                                                                                                                                                  \label{inst:0091}\vfill
\and RHEA for European Space Agency (ESA), Camino bajo del Castillo, s/n, Urbanizacion Villafranca del Castillo, Villanueva de la Ca\~{n}ada, 28692 Madrid, Spain\relax                                                                                                                                                                                            \label{inst:0096}\vfill
\and ATG Europe for European Space Agency (ESA), Camino bajo del Castillo, s/n, Urbanizacion Villafranca del Castillo, Villanueva de la Ca\~{n}ada, 28692 Madrid, Spain\relax                                                                                                                                                                                      \label{inst:0097}\vfill
\and CIGUS CITIC - Department of Computer Science and Information Technologies, University of A Coru\~{n}a, Campus de Elvi\~{n}a s/n, A Coru\~{n}a, 15071, Spain\relax                                                                                                                                                                                             \label{inst:0099}\vfill
\and Universit\'{e} de Strasbourg, CNRS, Observatoire astronomique de Strasbourg, UMR 7550,  11 rue de l'Universit\'{e}, 67000 Strasbourg, France\relax                                                                                                                                                                                                            \label{inst:0104}\vfill
\and Kavli Institute for Cosmology Cambridge, Institute of Astronomy, Madingley Road, Cambridge, CB3 0HA\relax                                                                                                                                                                                                                                                     \label{inst:0107}\vfill
\and Leibniz Institute for Astrophysics Potsdam (AIP), An der Sternwarte 16, 14482 Potsdam, Germany\relax                                                                                                                                                                                                                                                          \label{inst:0113}\vfill
\and CENTRA, Faculdade de Ci\^{e}ncias, Universidade de Lisboa, Edif. C8, Campo Grande, 1749-016 Lisboa, Portugal\relax                                                                                                                                                                                                                                            \label{inst:0116}\vfill
\and Department of Informatics, Donald Bren School of Information and Computer Sciences, University of California, Irvine, 5226 Donald Bren Hall, 92697-3440 CA Irvine, United States\relax                                                                                                                                                                        \label{inst:0117}\vfill
\and INAF - Osservatorio Astrofisico di Catania, via S. Sofia 78, 95123 Catania, Italy\relax                                                                                                                                                                                                                                                                       \label{inst:0118}\vfill
\and Dipartimento di Fisica e Astronomia ""Ettore Majorana"", Universit\`{a} di Catania, Via S. Sofia 64, 95123 Catania, Italy\relax                                                                                                                                                                                                                               \label{inst:0119}\vfill
\and INAF - Osservatorio Astronomico di Roma, Via Frascati 33, 00078 Monte Porzio Catone (Roma), Italy\relax                                                                                                                                                                                                                                                       \label{inst:0122}\vfill
\and Space Science Data Center - ASI, Via del Politecnico SNC, 00133 Roma, Italy\relax                                                                                                                                                                                                                                                                             \label{inst:0123}\vfill
\and Department of Physics, University of Helsinki, P.O. Box 64, 00014 Helsinki, Finland\relax                                                                                                                                                                                                                                                                     \label{inst:0125}\vfill
\and Finnish Geospatial Research Institute FGI, Geodeetinrinne 2, 02430 Masala, Finland\relax                                                                                                                                                                                                                                                                      \label{inst:0126}\vfill
\and Institut UTINAM CNRS UMR6213, Universit\'{e} Bourgogne Franche-Comt\'{e}, OSU THETA Franche-Comt\'{e} Bourgogne, Observatoire de Besan\c{c}on, BP1615, 25010 Besan\c{c}on Cedex, France\relax                                                                                                                                                                 \label{inst:0131}\vfill
\and HE Space Operations BV for European Space Agency (ESA), Keplerlaan 1, 2201AZ, Noordwijk, The Netherlands\relax                                                                                                                                                                                                                                                \label{inst:0134}\vfill
\and Dpto. de Inteligencia Artificial, UNED, c/ Juan del Rosal 16, 28040 Madrid, Spain\relax                                                                                                                                                                                                                                                                       \label{inst:0136}\vfill
\and Konkoly Observatory, Research Centre for Astronomy and Earth Sciences, E\"{ o}tv\"{ o}s Lor{\'a}nd Research Network (ELKH), MTA Centre of Excellence, Konkoly Thege Mikl\'{o}s \'{u}t 15-17, 1121 Budapest, Hungary\relax                                                                                                                                     \label{inst:0143}\vfill
\and ELTE E\"{ o}tv\"{ o}s Lor\'{a}nd University, Institute of Physics, 1117, P\'{a}zm\'{a}ny P\'{e}ter s\'{e}t\'{a}ny 1A, Budapest, Hungary\relax                                                                                                                                                                                                                 \label{inst:0144}\vfill
\and Instituut voor Sterrenkunde, KU Leuven, Celestijnenlaan 200D, 3001 Leuven, Belgium\relax                                                                                                                                                                                                                                                                      \label{inst:0146}\vfill
\and Department of Astrophysics/IMAPP, Radboud University, P.O.Box 9010, 6500 GL Nijmegen, The Netherlands\relax                                                                                                                                                                                                                                                   \label{inst:0147}\vfill
\and University of Vienna, Department of Astrophysics, T\"{ u}rkenschanzstra{\ss}e 17, A1180 Vienna, Austria\relax                                                                                                                                                                                                                                                 \label{inst:0155}\vfill
\and Institute of Physics, Laboratory of Astrophysics, Ecole Polytechnique F\'ed\'erale de Lausanne (EPFL), Observatoire de Sauverny, 1290 Versoix, Switzerland\relax                                                                                                                                                                                              \label{inst:0157}\vfill
\and Kapteyn Astronomical Institute, University of Groningen, Landleven 12, 9747 AD Groningen, The Netherlands\relax                                                                                                                                                                                                                                               \label{inst:0163}\vfill
\and School of Physics and Astronomy / Space Park Leicester, University of Leicester, University Road, Leicester LE1 7RH, United Kingdom\relax                                                                                                                                                                                                                     \label{inst:0170}\vfill
\and Thales Services for CNES Centre Spatial de Toulouse, 18 avenue Edouard Belin, 31401 Toulouse Cedex 9, France\relax                                                                                                                                                                                                                                            \label{inst:0172}\vfill
\and Depto. Estad\'istica e Investigaci\'on Operativa. Universidad de C\'adiz, Avda. Rep\'ublica Saharaui s/n, 11510 Puerto Real, C\'adiz, Spain\relax                                                                                                                                                                                                             \label{inst:0176}\vfill
\and Center for Research and Exploration in Space Science and Technology, University of Maryland Baltimore County, 1000 Hilltop Circle, Baltimore MD, USA\relax                                                                                                                                                                                                    \label{inst:0178}\vfill
\and GSFC - Goddard Space Flight Center, Code 698, 8800 Greenbelt Rd, 20771 MD Greenbelt, United States\relax                                                                                                                                                                                                                                                      \label{inst:0179}\vfill
\and EURIX S.r.l., Corso Vittorio Emanuele II 61, 10128, Torino, Italy\relax                                                                                                                                                                                                                                                                                       \label{inst:0181}\vfill
\and Porter School of the Environment and Earth Sciences, Tel Aviv University, Tel Aviv 6997801, Israel\relax                                                                                                                                                                                                                                                      \label{inst:0182}\vfill
\and Harvard-Smithsonian Center for Astrophysics, 60 Garden St., MS 15, Cambridge, MA 02138, USA\relax                                                                                                                                                                                                                                                             \label{inst:0183}\vfill
\and HE Space Operations BV for European Space Agency (ESA), Camino bajo del Castillo, s/n, Urbanizacion Villafranca del Castillo, Villanueva de la Ca\~{n}ada, 28692 Madrid, Spain\relax                                                                                                                                                                          \label{inst:0185}\vfill
\and Instituto de Astrof\'{i}sica e Ci\^{e}ncias do Espa\c{c}o, Universidade do Porto, CAUP, Rua das Estrelas, PT4150-762 Porto, Portugal\relax                                                                                                                                                                                                                    \label{inst:0186}\vfill
\and LFCA/DAS,Universidad de Chile,CNRS,Casilla 36-D, Santiago, Chile\relax                                                                                                                                                                                                                                                                                        \label{inst:0188}\vfill
\and SISSA - Scuola Internazionale Superiore di Studi Avanzati, via Bonomea 265, 34136 Trieste, Italy\relax                                                                                                                                                                                                                                                        \label{inst:0192}\vfill
\and Telespazio for CNES Centre Spatial de Toulouse, 18 avenue Edouard Belin, 31401 Toulouse Cedex 9, France\relax                                                                                                                                                                                                                                                 \label{inst:0197}\vfill
\and University of Turin, Department of Computer Sciences, Corso Svizzera 185, 10149 Torino, Italy\relax                                                                                                                                                                                                                                                           \label{inst:0201}\vfill
\and Dpto. de Matem\'{a}tica Aplicada y Ciencias de la Computaci\'{o}n, Univ. de Cantabria, ETS Ingenieros de Caminos, Canales y Puertos, Avda. de los Castros s/n, 39005 Santander, Spain\relax                                                                                                                                                                   \label{inst:0204}\vfill
\and Centro de Astronom\'{i}a - CITEVA, Universidad de Antofagasta, Avenida Angamos 601, Antofagasta 1270300, Chile\relax                                                                                                                                                                                                                                          \label{inst:0214}\vfill
\and DLR Gesellschaft f\"{ u}r Raumfahrtanwendungen (GfR) mbH M\"{ u}nchener Stra{\ss}e 20 , 82234 We{\ss}ling\relax                                                                                                                                                                                                                                               \label{inst:0219}\vfill
\and Centre for Astrophysics Research, University of Hertfordshire, College Lane, AL10 9AB, Hatfield, United Kingdom\relax                                                                                                                                                                                                                                         \label{inst:0221}\vfill
\and University of Turin, Mathematical Department ""G.Peano"", Via Carlo Alberto 10, 10123 Torino, Italy\relax                                                                                                                                                                                                                                                     \label{inst:0228}\vfill
\and INAF - Osservatorio Astronomico d'Abruzzo, Via Mentore Maggini, 64100 Teramo, Italy\relax                                                                                                                                                                                                                                                                     \label{inst:0233}\vfill
\and Instituto de Astronomia, Geof\`{i}sica e Ci\^{e}ncias Atmosf\'{e}ricas, Universidade de S\~{a}o Paulo, Rua do Mat\~{a}o, 1226, Cidade Universitaria, 05508-900 S\~{a}o Paulo, SP, Brazil\relax                                                                                                                                                                \label{inst:0236}\vfill
\and APAVE SUDEUROPE SAS for CNES Centre Spatial de Toulouse, 18 avenue Edouard Belin, 31401 Toulouse Cedex 9, France\relax                                                                                                                                                                                                                                        \label{inst:0243}\vfill
\and M\'{e}socentre de calcul de Franche-Comt\'{e}, Universit\'{e} de Franche-Comt\'{e}, 16 route de Gray, 25030 Besan\c{c}on Cedex, France\relax                                                                                                                                                                                                                  \label{inst:0244}\vfill
\and ATOS for CNES Centre Spatial de Toulouse, 18 avenue Edouard Belin, 31401 Toulouse Cedex 9, France\relax                                                                                                                                                                                                                                                       \label{inst:0249}\vfill
\and School of Physics and Astronomy, Tel Aviv University, Tel Aviv 6997801, Israel\relax                                                                                                                                                                                                                                                                          \label{inst:0252}\vfill
\and Astrophysics Research Centre, School of Mathematics and Physics, Queen's University Belfast, Belfast BT7 1NN, UK\relax                                                                                                                                                                                                                                        \label{inst:0254}\vfill
\and Centre de Donn\'{e}es Astronomique de Strasbourg, Strasbourg, France\relax                                                                                                                                                                                                                                                                                    \label{inst:0256}\vfill
\and Institute for Computational Cosmology, Department of Physics, Durham University, Durham DH1 3LE, UK\relax                                                                                                                                                                                                                                                     \label{inst:0260}\vfill
\and European Southern Observatory, Karl-Schwarzschild-Str. 2, 85748 Garching, Germany\relax                                                                                                                                                                                                                                                                       \label{inst:0261}\vfill
\and Max-Planck-Institut f\"{ u}r Astrophysik, Karl-Schwarzschild-Stra{\ss}e 1, 85748 Garching, Germany\relax                                                                                                                                                                                                                                                      \label{inst:0262}\vfill
\and Data Science and Big Data Lab, Pablo de Olavide University, 41013, Seville, Spain\relax                                                                                                                                                                                                                                                                       \label{inst:0266}\vfill
\and Barcelona Supercomputing Center (BSC), Pla\c{c}a Eusebi G\"{ u}ell 1-3, 08034-Barcelona, Spain\relax                                                                                                                                                                                                                                                          \label{inst:0274}\vfill
\and ETSE Telecomunicaci\'{o}n, Universidade de Vigo, Campus Lagoas-Marcosende, 36310 Vigo, Galicia, Spain\relax                                                                                                                                                                                                                                                   \label{inst:0279}\vfill
\and Asteroid Engineering Laboratory, Space Systems, Lule\aa{} University of Technology, Box 848, S-981 28 Kiruna, Sweden\relax                                                                                                                                                                                                                                    \label{inst:0283}\vfill
\and Vera C Rubin Observatory,  950 N. Cherry Avenue, Tucson, AZ 85719, USA\relax                                                                                                                                                                                                                                                                                  \label{inst:0288}\vfill
\and Department of Astrophysics, Astronomy and Mechanics, National and Kapodistrian University of Athens, Panepistimiopolis, Zografos, 15783 Athens, Greece\relax                                                                                                                                                                                                  \label{inst:0289}\vfill
\and TRUMPF Photonic Components GmbH, Lise-Meitner-Stra{\ss}e 13,  89081 Ulm, Germany\relax                                                                                                                                                                                                                                                                        \label{inst:0292}\vfill
\and IAC - Instituto de Astrofisica de Canarias, Via L\'{a}ctea s/n, 38200 La Laguna S.C., Tenerife, Spain\relax                                                                                                                                                                                                                                                   \label{inst:0297}\vfill
\and Department of Astrophysics, University of La Laguna, Via L\'{a}ctea s/n, 38200 La Laguna S.C., Tenerife, Spain\relax                                                                                                                                                                                                                                          \label{inst:0298}\vfill
\and Faculty of Aerospace Engineering, Delft University of Technology, Kluyverweg 1, 2629 HS Delft, The Netherlands\relax                                                                                                                                                                                                                                          \label{inst:0300}\vfill
\and Radagast Solutions\relax                                                                                                                                                                                                                                                                                                                                      \label{inst:0304}\vfill
\and Laboratoire Univers et Particules de Montpellier, CNRS Universit\'{e} Montpellier, Place Eug\`{e}ne Bataillon, CC72, 34095 Montpellier Cedex 05, France\relax                                                                                                                                                                                                 \label{inst:0305}\vfill
\and Universit\'{e} de Caen Normandie, C\^{o}te de Nacre Boulevard Mar\'{e}chal Juin, 14032 Caen, France\relax                                                                                                                                                                                                                                                     \label{inst:0311}\vfill
\and LESIA, Observatoire de Paris, Universit\'{e} PSL, CNRS, Sorbonne Universit\'{e}, Universit\'{e} de Paris, 5 Place Jules Janssen, 92190 Meudon, France\relax                                                                                                                                                                                                   \label{inst:0312}\vfill
\and SRON Netherlands Institute for Space Research, Niels Bohrweg 4, 2333 CA Leiden, The Netherlands\relax                                                                                                                                                                                                                                                         \label{inst:0320}\vfill
\and Astronomical Observatory, University of Warsaw,  Al. Ujazdowskie 4, 00-478 Warszawa, Poland\relax                                                                                                                                                                                                                                                             \label{inst:0321}\vfill
\and Scalian for CNES Centre Spatial de Toulouse, 18 avenue Edouard Belin, 31401 Toulouse Cedex 9, France\relax                                                                                                                                                                                                                                                    \label{inst:0323}\vfill
\and Universit\'{e} Rennes, CNRS, IPR (Institut de Physique de Rennes) - UMR 6251, 35000 Rennes, France\relax                                                                                                                                                                                                                                                      \label{inst:0329}\vfill
\and INAF - Osservatorio Astronomico di Capodimonte, Via Moiariello 16, 80131, Napoli, Italy\relax                                                                                                                                                                                                                                                                 \label{inst:0331}\vfill
\and Shanghai Astronomical Observatory, Chinese Academy of Sciences, 80 Nandan Road, Shanghai 200030, People's Republic of China\relax                                                                                                                                                                                                                             \label{inst:0334}\vfill
\and University of Chinese Academy of Sciences, No.19(A) Yuquan Road, Shijingshan District, Beijing 100049, People's Republic of China\relax                                                                                                                                                                                                                       \label{inst:0336}\vfill
\and Niels Bohr Institute, University of Copenhagen, Juliane Maries Vej 30, 2100 Copenhagen {\O}, Denmark\relax                                                                                                                                                                                                                                                    \label{inst:0339}\vfill
\and DXC Technology, Retortvej 8, 2500 Valby, Denmark\relax                                                                                                                                                                                                                                                                                                        \label{inst:0340}\vfill
\and Las Cumbres Observatory, 6740 Cortona Drive Suite 102, Goleta, CA 93117, USA\relax                                                                                                                                                                                                                                                                            \label{inst:0341}\vfill
\and CIGUS CITIC, Department of Nautical Sciences and Marine Engineering, University of A Coru\~{n}a, Paseo de Ronda 51, 15071, A Coru\~{n}a, Spain\relax                                                                                                                                                                                                          \label{inst:0350}\vfill
\and Astrophysics Research Institute, Liverpool John Moores University, 146 Brownlow Hill, Liverpool L3 5RF, United Kingdom\relax                                                                                                                                                                                                                                  \label{inst:0351}\vfill
\and IPAC, Mail Code 100-22, California Institute of Technology, 1200 E. California Blvd., Pasadena, CA 91125, USA\relax                                                                                                                                                                                                                                           \label{inst:0358}\vfill
\and MTA CSFK Lend\"{ u}let Near-Field Cosmology Research Group, Konkoly Observatory, MTA Research Centre for Astronomy and Earth Sciences, Konkoly Thege Mikl\'{o}s \'{u}t 15-17, 1121 Budapest, Hungary\relax                                                                                                                                                    \label{inst:0375}\vfill
\and Departmento de F\'{i}sica de la Tierra y Astrof\'{i}sica, Universidad Complutense de Madrid, 28040 Madrid, Spain\relax                                                                                                                                                                                                                                        \label{inst:0391}\vfill
\and Ru{\dj}er Bo\v{s}kovi\'{c} Institute, Bijeni\v{c}ka cesta 54, 10000 Zagreb, Croatia\relax                                                                                                                                                                                                                                                                     \label{inst:0397}\vfill
\and Villanova University, Department of Astrophysics and Planetary Science, 800 E Lancaster Avenue, Villanova PA 19085, USA\relax                                                                                                                                                                                                                                 \label{inst:0413}\vfill
\and INAF - Osservatorio Astronomico di Brera, via E. Bianchi, 46, 23807 Merate (LC), Italy\relax                                                                                                                                                                                                                                                                  \label{inst:0419}\vfill
\and STFC, Rutherford Appleton Laboratory, Harwell, Didcot, OX11 0QX, United Kingdom\relax                                                                                                                                                                                                                                                                         \label{inst:0426}\vfill
\and Charles University, Faculty of Mathematics and Physics, Astronomical Institute of Charles University, V Holesovickach 2, 18000 Prague, Czech Republic\relax                                                                                                                                                                                                   \label{inst:0449}\vfill
\and Department of Particle Physics and Astrophysics, Weizmann Institute of Science, Rehovot 7610001, Israel\relax                                                                                                                                                                                                                                                 \label{inst:0459}\vfill
\and Department of Astrophysical Sciences, 4 Ivy Lane, Princeton University, Princeton NJ 08544, USA\relax                                                                                                                                                                                                                                                         \label{inst:0460}\vfill
\and Departamento de Astrof\'{i}sica, Centro de Astrobiolog\'{i}a (CSIC-INTA), ESA-ESAC. Camino Bajo del Castillo s/n. 28692 Villanueva de la Ca\~{n}ada, Madrid, Spain\relax                                                                                                                                                                                      \label{inst:0469}\vfill
\and naXys, University of Namur, Rempart de la Vierge, 5000 Namur, Belgium\relax                                                                                                                                                                                                                                                                                   \label{inst:0472}\vfill
\and CGI Deutschland B.V. \& Co. KG, Mornewegstr. 30, 64293 Darmstadt, Germany\relax                                                                                                                                                                                                                                                                               \label{inst:0480}\vfill
\and Institute of Global Health, University of Geneva\relax                                                                                                                                                                                                                                                                                                        \label{inst:0481}\vfill
\and Astronomical Observatory Institute, Faculty of Physics, Adam Mickiewicz University, Pozna\'{n}, Poland\relax                                                                                                                                                                                                                                                  \label{inst:0483}\vfill
\and H H Wills Physics Laboratory, University of Bristol, Tyndall Avenue, Bristol BS8 1TL, United Kingdom\relax                                                                                                                                                                                                                                                    \label{inst:0487}\vfill
\and Department of Physics and Astronomy G. Galilei, University of Padova, Vicolo dell'Osservatorio 3, 35122, Padova, Italy\relax                                                                                                                                                                                                                                  \label{inst:0494}\vfill
\and CERN, Geneva, Switzerland\relax                                                                                                                                                                                                                                                                                                                               \label{inst:0496}\vfill
\and Applied Physics Department, Universidade de Vigo, 36310 Vigo, Spain\relax                                                                                                                                                                                                                                                                                     \label{inst:0498}\vfill
\and Association of Universities for Research in Astronomy, 1331 Pennsylvania Ave. NW, Washington, DC 20004, USA\relax                                                                                                                                                                                                                                             \label{inst:0502}\vfill
\and European Southern Observatory, Alonso de C\'ordova 3107, Casilla 19, Santiago, Chile\relax                                                                                                                                                                                                                                                                    \label{inst:0510}\vfill
\and Sorbonne Universit\'{e}, CNRS, UMR7095, Institut d'Astrophysique de Paris, 98bis bd. Arago, 75014 Paris, France\relax                                                                                                                                                                                                                                         \label{inst:0514}\vfill
}

%\author{\textit{Gaia Collaboration}
%  \and \\
%  M. Schultheis\inst{1}%\thanks{Corresponding author: %Mathias Schultheis, Mathias.Schultheis@oca.eu}
%  \and H. Zhao\inst{1} 
%  \and T. Zwitter\inst{2}
%  \and D.J.\ Marshall\inst{3}
%  \and R. Drimmel\inst{4}
%  \and Y. Fr\'emat\inst{5} 
%  \and C.A.L.\ Bailer-Jones\inst{6}
%  \and A. Recio-Blanco\inst{1}
%  \and G. Kordopatis\inst{1}
%  \and P. de Laverny\inst{1}
%  \and R. Andrae\inst{6}
%  \and T.E.\ Dharmawardena \inst{6}
%  \and M. Fouesneau\inst{6}
%  \and R. Sordo \inst{7}
%and
%\\
%\textit{rest of DPAC}
%}

%\institute{Universit\'e C\^ote d'Azur, Observatoire de la C\^ote d'Azur, CNRS,  Laboratoire Lagrange, Nice, \\
% e-mail: mathias.schultheis@oca.eu
% \and
% Faculty of Mathematics and Physics, University of Ljubljana, Jadranska 19, 1000 Ljubljana, Slovenia
% \and
%IRAP, Universit\'{e} de Toulouse, CNRS, UPS, CNES, 9 Av. colonel Roche, BP 44346, 31028 Toulouse Cedex 4, France\label{inst:irap}
%\and
% INAF - Osservatorio Astrofisico di Torino, Istituto Nazionale di Astrofisica, I-10025 Pino Torinese, Italy
% \and
% Royal Observatory of Belgium, 3 avenue circulaire, 1180 Brussels, Belgium
% \and
% Max Planck Institute for Astronomy, K\"onigstuhl 17, D-69117 Heidelberg, Germany
% \and
% INAF - Osservatorio Astronomico di Padova, Vicolo Osservatorio 5, 35122 Padova, Italy
% }

\date{Received ; accepted}

\abstract {Diffuse interstellar bands (DIBs) are common interstellar absorption features in spectroscopic observations  but their origins remain unclear.  DIBs play an important role in the life cycle of the interstellar medium (ISM) and can also be used to trace Galactic structure. }
{Here, we demonstrate the capacity of the Gaia-Radial Velocity Spectrometer (RVS) in \gdr{3} to reveal the spatial distribution of the unknown molecular species responsible for the most prominent DIB at 862\,nm in the RVS passband, exploring  the Galactic ISM within  a few kiloparsecs from the Sun. }
{The DIBs are measured within the GSP-Spec module using a Gaussian profile fit for cool stars and a Gaussian process for hot stars. In addition to the equivalent widths and their uncertainties, \gdr{3} provides their characteristic central  wavelength, width, and quality flags.}
{We present an extensive  sample of 476\,117 individual DIB measurements obtained in a homogeneous way covering the entire sky. We compare spatial distributions of the DIB carrier with  interstellar reddening and find evidence that DIB carriers are present in a local bubble around the Sun which contains nearly no dust. We characterised the DIB equivalent width with a  local density of $0.19 \pm 0.04$~\AA/kpc and a scale height of $\rm 98.60_{-8.46}^{+11.10}$\,pc. The latter is smaller than the dust scale height, indicating that DIBs are more concentrated towards the Galactic plane. We determine the rest-frame wavelength with unprecedented precision ($\rm \lambda_{0} = 8620.86\, \pm 0.019$\,{\AA} in air)   and reveal  a remarkable correspondence between the DIB velocities and the CO gas velocities, suggesting that the 862\,nm DIB carrier is related to macro-molecules.}
{We demonstrate the unique capacity of Gaia to trace the spatial structure of the Galactic ISM using the 862\,nm DIB.}
 \maketitle
 
\titlerunning{The 862\,nm DIB in \gdr{3}}
\authorrunning{Gaia Collaboration}

\section{Introduction}
Diffuse  interstellar  bands  (DIBs)  are  interstellar  absorption  features  that primarily  exist  in  the optical and near-infrared (NIR) wavelength range, the physical origin of which  is still debated. The name was formally given by \citet{Merrill1930}, where `diffuse' refers to the fact that  their profiles are broader than those of interstellar atomic lines (e.g. NaI lines). DIBs presumably originate from molecular absorption, which is supported by the fact that their central wavelength does not match any known atomic transition lines.
The fine structure observed in some DIBs also suggests that the molecular carriers are probably in the gas phase. For reviews on  DIBs, see \citet{Leger1984}, \citet{Herbig1995}, \citet{Sarre2006}, and \cite{Snow2006}.

Nowadays, molecules are strongly suggested to be associated with the DIB carrier, because DIB profiles
are usually much broader than atomic lines and contain substructures even through single-cloud
sight lines (e.g. \citealt{Sarre1995}, \citealt{Cami1997}, \citealt{Kerr1998}, \citealt{Galazutdinov2008a}).
Carbon-bearing molecules are the most favoured species in this respect as carbon can form many  stable compounds and is
 relatively  abundant in the Universe (\citealt{Puget1989})

The DIB at 862\,nm (hereafter referred to as DIB\,$\lambda$862) is a  strong band, but was not identified until 1975 \citep{Geary1975}, 
more than 50 years after the discovery of the first DIBs, because the wavelength range beyond 8600\,{\AA} was not covered by earlier work.  
The DIB\,$\lambda$862 was confirmed by \citet{Sanner1978}, who further reported $\lambda_0\,{=}\,8620.7 \pm 0.3$\,{\AA} and a tight linear 
correlation between the DIB equivalent width ($\rm EW_{862}$) and the colour excess,
that is $\EBV = 2.85 \pm 0.11 \times {\rm EW_{862}}$ (coefficient calculated by \citealt{Kos2013}). 
%As DIB\,$\lambda$862 is falling within the spectral region of the {\it Gaia}$-$RVS spectra (\citealt{Wilkinson2005}, \citealt{Katz2019}, \citealt{Recio-Blanco2016}), 
 \citet{Munari1999} and \citet{Munari2000} made preliminary studies of the relation between the $\rm EW_{862}$ of DIB\,$\lambda$862 and interstellar 
extinction. This author found a surprisingly tight correlation with 
$\EBV/{\rm EW_{862}}\,{=}\,2.63$ (\citealt{Munari1999}) and $2.69 \pm 0.03$ (\citealt{Munari2000}), respectively. Therefore, the DIB\,$\lambda$862 was suggested to be a tracer of Galactic extinction 
in the context of the {\it Gaia} mission, while \citet{Krelowski2018} and \citet{Krelowski2019b} argued that $\EBV/{\rm EW_{862}}$ can vary depending on the line of sight. \citet{Munari2008} measured the DIB\,$\lambda$862 in the spectra of 68 early-type stars observed by the RAdial Velocity Experiment (RAVE; \citealt{Steinmetz2006}) and derived a very good correlation between $\rm EW_{862}$  and $\EBV$ with $\EBV/{\rm EW_{862}}\,{=}\,2.72\pm0.03$. 
These results, as well as those of 
\citet{Munari1999} and \citet{Munari2000}, were all consistent with each other, but none agreed with those of \citet{Wallerstein2007}, who derived a much higher ratio of $\EBV/{\rm EW_{862}}$.

\begin{figure*}[!htbp]
    \centering
    \includegraphics[width=1\textwidth]{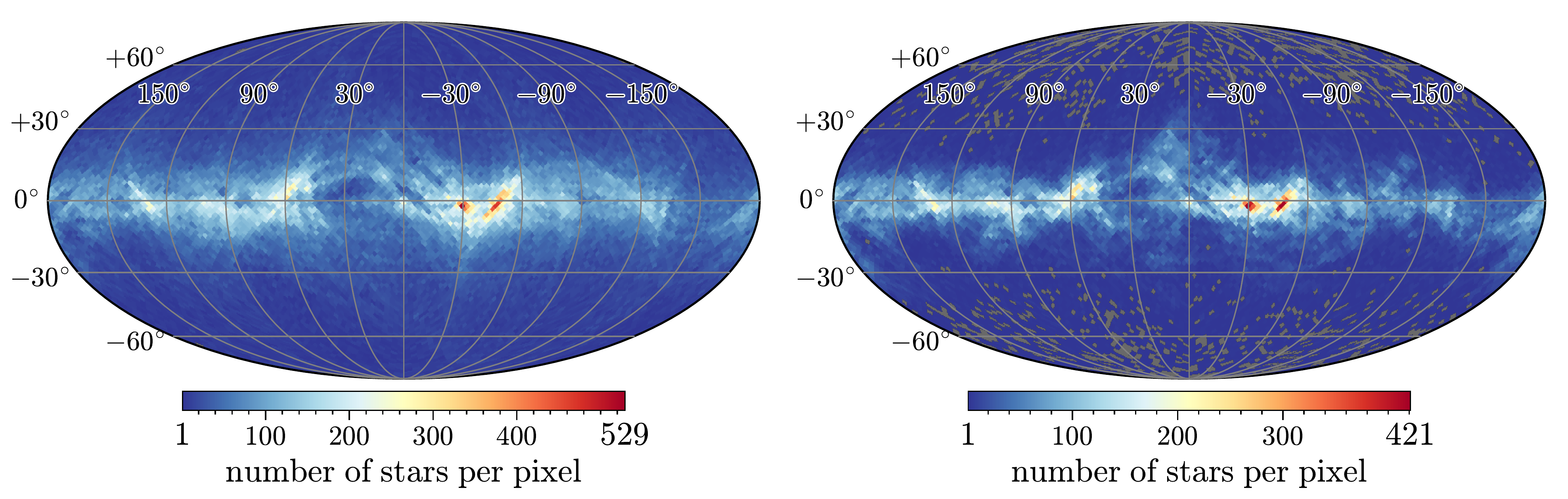}
    \caption{(Left panel:) Galactic distribution of the  476\,117 DIBs\,$\lambda$862  in \gdr{3}. The spatial resolution is 1.8\degree per HEALpixel (level 5). The colour scale indicates the number of measurements per pixel.  (Right panel:) The subset of 236\,836 sources with high-quality measurements  ($\rm QF \leqslant 2$, see Sect.~\ref{HQ}).}
    \label{lbdistribution}
\end{figure*}

\citet{Munari2008} determined the rest-frame 
wavelength of DIB\,$\lambda$862 as $\lambda_0\,{=}\,8620.4 \pm 0.1$\,{\AA} based on the assumption that the average velocity of their 
carriers towards the Galactic center is approximately zero, as derived from the interstellar-medium (ISM) radial-velocity map of \citet{BB1993}.

To make use of the vast number of cool-star ($3500 \leqslant \teff \leqslant \,7000\,K$) spectra in RAVE, \citet{Kos2013} implemented a data-driven method to derive the $\rm EW_{862}$ of interstellar spectra 
using real spectra at high Galactic latitudes ($b\,{<}\,{-}65^{\circ}$) and furthermore stacked spectra in small spatial volumes to increase 
the final signal-to-noise ratio (S/N) and measure $\rm EW_{862}$  with high precision. In this way, they confirmed the linear $\rm EW_{862}$ ${-}\EBV$ correlation in a statistical way. 

Based on measurements 
with a large number of RAVE spectra, \citet{Kos2014} built the first projected DIB\,$\lambda$862 intensity map, mainly within 3\,kpc from the Sun, where for the first time the large-scale structure of the distribution of the 
DIB\,$\lambda$862 carrier was shown. The findings of these authors further suggested an exponential  
distribution of $\rm EW_{862}$ in the direction perpendicular to the Galactic plane with a scale height of $209.0 \pm 11.9$\,pc, 
larger than the scale height of $117.7 \pm 4.7$\,pc for the dust derived by their $\Av$ map. \citet{Puspitarini2015} measured 
the DIB\,$\lambda$862  in the spectra of 64 late-type stars from the Gaia$-$ESO (GES) survey \citep{Gilmore2012} towards a 
Galactic anticentre region at $(\ell,\,b)\,{=}\,(212.9^{\circ},\,{-}2.0^{\circ})$. \citet{Puspitarini2015} fitted the observed spectra with synthetic 
spectra containing stellar components, telluric transmissions, and a DIB empirical profile. For DIB\,$\lambda$862, they obtained the 
empirical model by averaging the profiles detected in several spectra based on the data analysis reported by \citet{Chen2013}. 

Similar to \citet{Puspitarini2015}, \citet{Krelowski2019b} also argued that a simple Gaussian 
fit was not enough to describe the irregular profile of the DIB\,$\lambda$862. They therefore used the observation towards $\rm BD\,{+}\,40\ 4220$, 
a heavily reddened and rapidly rotating star, as a template for the profile of $\lambda$862. Measurements of other targets were 
obtained by rescaling the depth of the template to match the observed band profiles. 

Using this method, \citet{Krelowski2019b} measured 56 high-resolution spectra ($R\,{>}\,30\,000$) and derived a ratio of 
$\EBV/{\rm EW_{862}}\,{=}\,2.03 \pm 0.15$ with an offset of 0.22, which was close to the result of \citet{Puspitarini2015}. 
\citet{Maiz-Apellaniz2015a} showed a linear relation between $\rm EW_{862}$ and the colour excess $E(4405\,{-}\,5495)$ up to 
$\Av\,{\sim}\,6$\,mag with a Pearson coefficient of $r_p\,{=}\,0.878$. 
All previous studies suggested a linear relation between 
$\rm EW_{862}$ and extinction except \citet{Damineli2016}, who reported a quadratic relation based on the observations of 12 
bright field stars and 11 members of Westerlund 1 cluster. Their relation is in good agreement with those found by \citet{Wallerstein2007} and 
\citet{Munari2008} for $\rm EW_{862}\,{<}\,0.8$\,{\AA}.

In this paper, we discuss the DIB\,$\lambda$862 measurements of nearly half a million DIBs measured by the RVS spectrometer. This is, by one order of magnitude, the largest sample of individual DIB measurements with full sky-coverage to be obtained so far.

In Sect. \ref{sample}, we discuss the DIB\,$\lambda$862 sample. In Sect.~\ref{HQ} we define our high-quality sample and in Sect.~\ref{HR} we validate the DIB\,$\lambda$862 measurements in the HR diagram. In Sect.~\ref{dust} we show the correlation with the dust extinction and in Sect.~\ref{Spatial} we present our analysis of the spatial distribution of the DIBs\,$\lambda$862. In Sect.~\ref{sect:lambda0} we describe how we determined the rest-frame wavelength of DIB\,$\lambda$862, and in Sect.~\ref{Kinematics} we look briefly at an application to kinematic studies.  We conclude in Sect.~\ref{Discussions}.

\section{Description of the  sample of diffuse  interstellar  bands} \label{sample}

This work makes use of the DIB\,$\lambda$862 parameterisation derived from the Gaia RVS spectra using the General Stellar Parameteriser spectroscopy (GSP-Spec, \citealt{GSPspecDR3}) module and made available through the astrophysical\_parameters table of the Gaia third data release (DR3). We note that the RVS wavelength range is [845, 870] nm (\citealt{Sartoretti2018A}), and its medium resolving power is $R  = \lambda / \Delta \lambda \sim 11500$ (\citealt{Cropper2018}). In addition to the DIB\,$\lambda$862 parameterisation, GSP-Spec estimates the main atmospheric parameters and the individual abundances of 12 different chemical elements from Gaia RVS spectra of single stars. When necessary (e.g. stars with $\teff\,{<}\,7000$\,K), the DIB\,$\lambda$862 spectral parameterisation is based on the MatisseGauguin GSP-Spec workflow. More details on the DIB\,$\lambda$862 measurement algorithms can be found in \citet{paperI}.
A GSP-Spec catalogue flag was implemented (\citealt{GSPspecDR3})
during the post-processing with a chain of 41 digits including all the adopted failure criteria and uncertainty sources considered during the post-processing. In this chain, value `0' is the best, and `9' is the worst, generally implying the parameter masking. For our purposes, we use only the first 13 characters  (see Sect.\ref{HQ}, Table.~\ref{GSPspectable})

\begin{figure}[!htbp]
    \centering
    
    \includegraphics[width=0.5\textwidth]{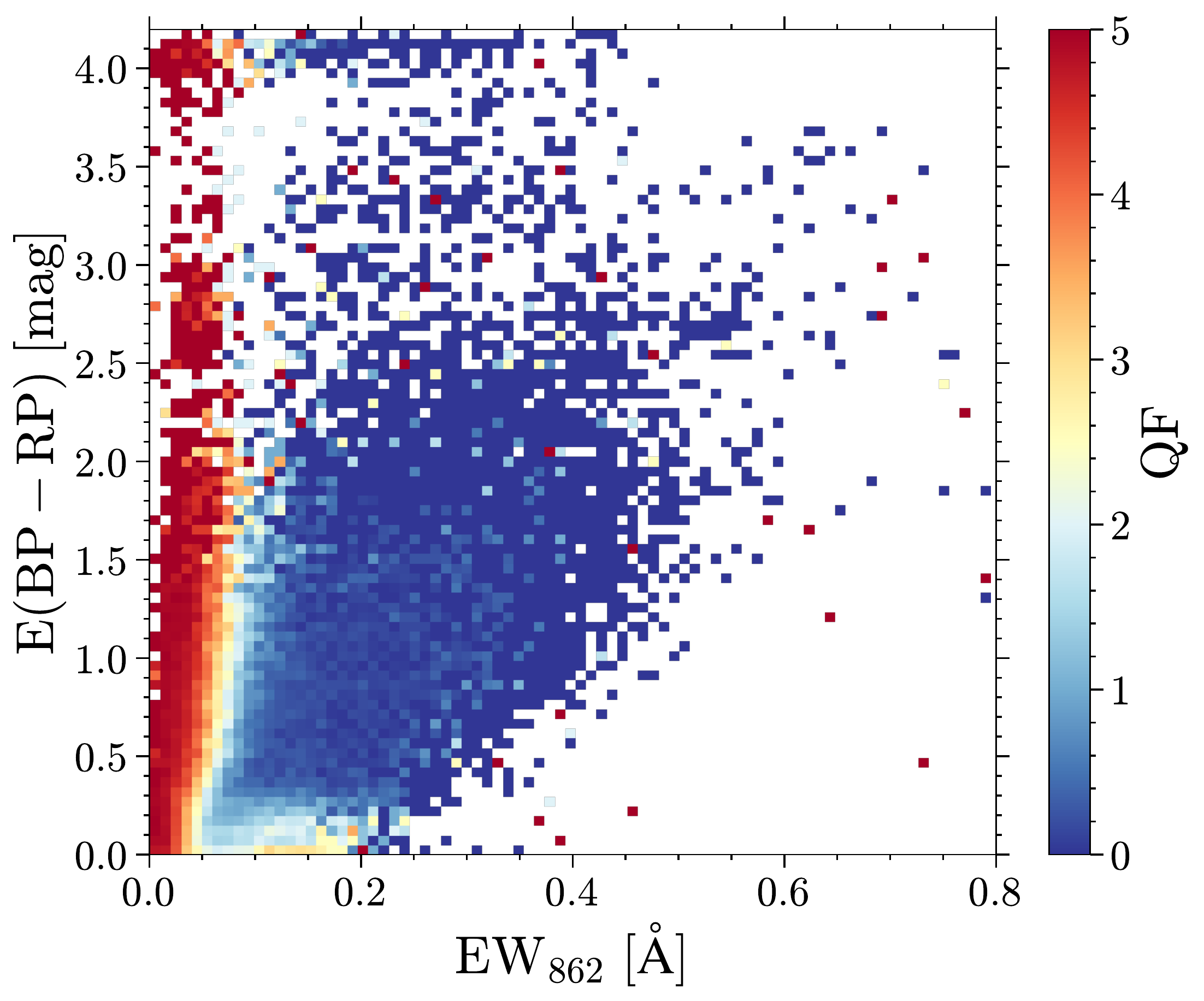}
    \caption{Equivalent width vs. $\rm E(BP-RP)$ for the DIB\,$\lambda$862 sample coloured by the 
    mean QF calculated in 0.01\,{\AA} $\times$ 0.05\,mag bins.} 
    \label{QFdib}
\end{figure}

We performed a  local renormalisation of the spectrum  around the DIB\,$\lambda$862 feature (35\,{\AA} wide around its central wavelength) for each Gaia-RVS spectrum. We carried out a preliminary fit using a preliminary detection of the DIB\,$\lambda$862 profile  and  sources where noise
is at the level of or exceeds the depth of the DIB\,$\lambda$862 feature were eliminated. Only detections above the 3\,$\rm \sigma$-level are considered as true detections. In order to perform the main fitting process of the DIB\,$\lambda$862, our sample is separated  into cool ($3500\,{<}\,\teff\,{\leqslant}\,7000$\,K) and hot ($\teff\,{>}\,7000$\,K) stars.  For cool stars, we divided the observed spectrum  by the best matching synthetic spectrum from GSP-Spec (corresponding to the derived atmospheric parameters), and fitted the DIB\,$\lambda$862 profile  with a Gaussian function and a constant that accounts for the continuum:

\begin{equation} \label{eq:Gauss-fit}
    f_{\Theta}(\lambda;p_0,p_1,p_2) = p_0 \times 
    {\rm exp}\left(-\frac{(\lambda-p_1)^2}{2 p_2^2}\right) + C,
\end{equation}

\noindent where $p_0$ and $p_2$ are the depth and width of the DIB profile, $p_1$ is the measured central wavelength, $C$ is 
the constant continuum, and $\lambda$ is the spectral wavelength.

For hot stars, we applied a Gaussian process similar to  \citet{Kos2017}  in which the DIB\,$\lambda$862 profile is fitted by  a Gaussian process regression (\citealt{GB12}). In order to extract the information of the DIB feature, we applied a Gaussian mean function (Eq. \ref{eq:Gauss-fit}) with $C\,{\equiv}\,1$.  For the kernels, we followed the 
strategy of \citet{Kos2017} and used exponential-squared kernel models  for the  stellar absorption lines:

\begin{equation}
   k_{se}(x,\xp) = a~{\rm exp}\left(-\frac{||x-\xp||^2}{2l^2}\right),
\end{equation}

 \noindent and a Mat\'ern 3/2 kernel model for the correlated noise:

\begin{equation}
    k_{m3/2}(x,\xp) = a\left(1+\frac{\sqrt{3}||x-\xp||}{l}\right)
    {\rm exp}\left(-\frac{\sqrt{3}||x-\xp||}{l}\right),
\end{equation}

\noindent where $a$ scales the kernels, and $l$ is the characteristic width of each
kernel.  We refer to \citet{paperI} for  a more detailed description of this process.

For each of the sources, the $\rm EW_{862}$,
depth ($p_0$), central wavelength ($p_1$), and width ($p_2$)  together with their uncertainties are determined  with ${\rm EW_{862}} = \sqrt{2\pi} \times |p_0| \times p_2 / C$  where C is the continuum level 
and $p_2 =  {\rm FWHM}/(2\sqrt{2{\rm ln}(2)}),$ where FWHM is the full width at half maximum of the DIB\,$\lambda$862 profile.

We consider two main uncertainties on the derived EW: the random noise error ($\rm \sigma_{noise}$), which is related to the signal-to-noise ratio (S/N) of the spectrum, and the mismatch between the observed spectrum and the synthetic one ($\rm \sigma_{spect}$). $\rm \sigma_{noise}$ was estimated for different DIB profiles using a random-noise simulation (see Sect. 2.6 in \citealt{paperI} for more details). The total uncertainty of the EW is considered to be $\rm \sigma^{2}_{EW} = \sigma_{noise}^{2} + \sigma_{spect}^{2}$. We refer to \citet{paperI} for a more detailed description of the derived uncertainties.

Quality flags (QFs) ranging from $\rm QF\,{=}\,0$ (highest 
quality) to $\rm QF\,{=}\,5$ (lowest quality) are generated. The defined values of the QF depend on the parameters $p_0$, $p_1$, and $p_2$, but also on the global noise level $R_A$ defined by the standard deviation of the data--model residuals between 8605 and 8640\,{\AA} as well as the local noise level $R_B$ within the DIB\,$\lambda$862 profile. Table~\ref{QF} shows the definition of the QF values. For a more detailed description of QF, we refer to \citet{paperI} and \citet{GSPspecDR3}. In this paper, we concentrate on  a high-quality sample ($\rm QF\,{\leq}\,2$, see Sect.~\ref{HQ}) but we stress that the full DIB\,$\lambda$862 sample should be scientifically exploited; for example,\ weak DIBs\,$\lambda$862 in low extinction areas.

\begin{figure}[!htbp]
    \centering
    
    \includegraphics[width=0.48\textwidth]{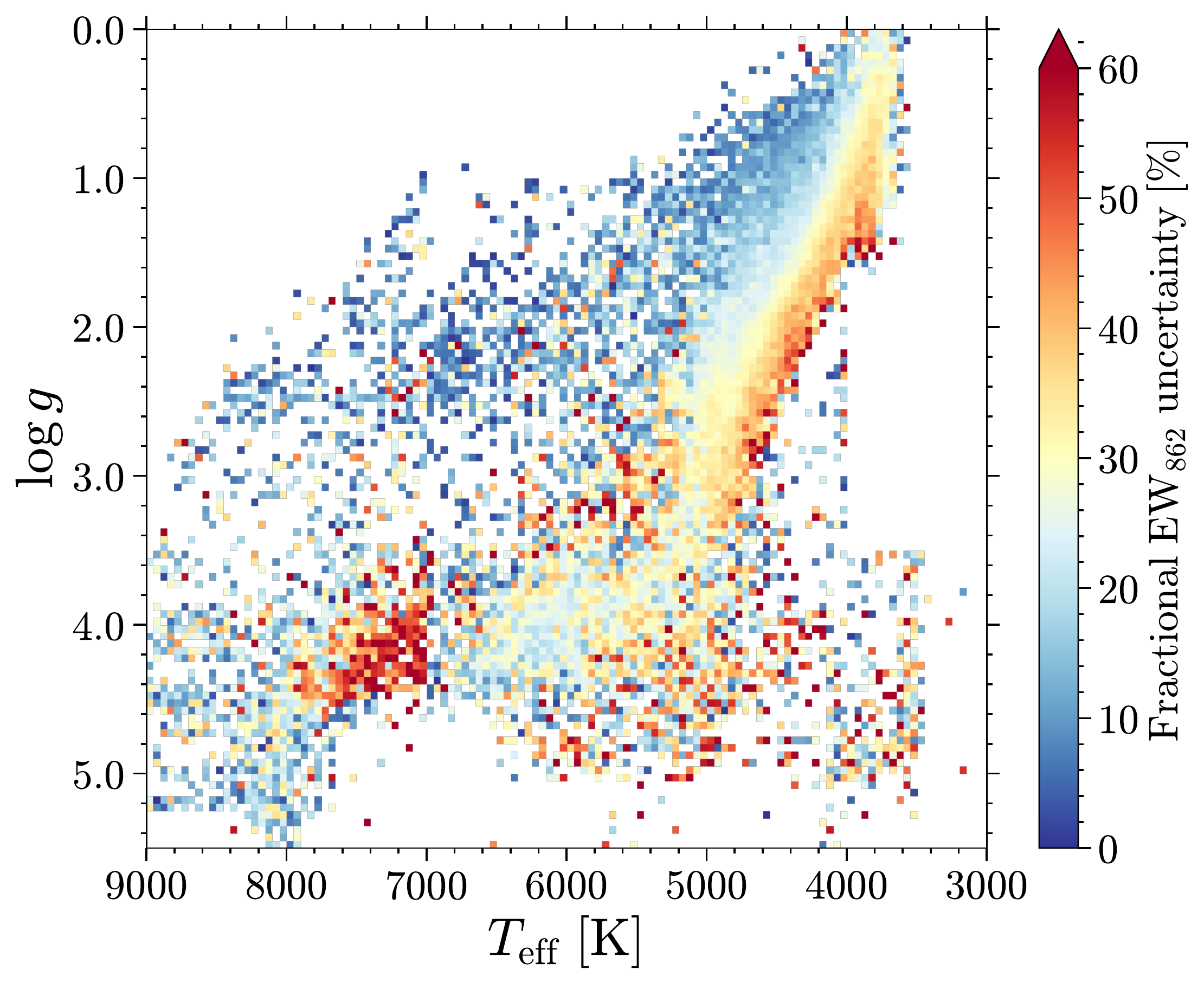}
    \caption{Kiel diagram as a function of the  fractional $\rm EW_{862}$ uncertainty (err($\rm EW_{862}$)/$\rm EW_{862}$) for a subsample with $\rm QF < 5$. The mean $\rm EW_{862}$ uncertainty is calculated in 50\,K $\times$ 0.05\,dex bins. }
    \label{HRvserror}
\end{figure}

\begin{figure}[!htbp]
    \centering
    
    \includegraphics[width=0.48\textwidth]{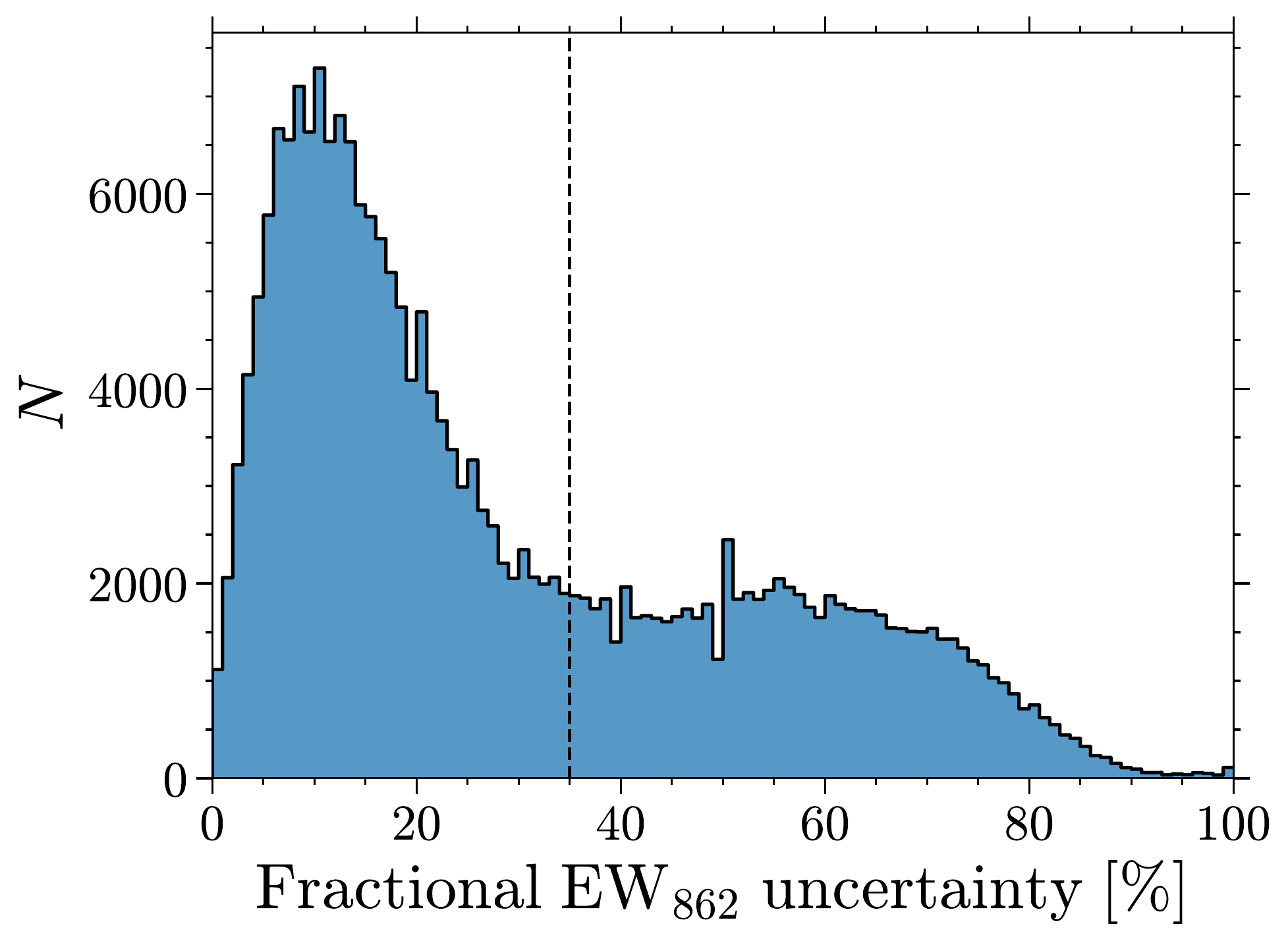}
    \caption{Histogram of the fractional uncertainties err($\rm EW_{862}$)/$\rm EW_{862}$ for targets with $\rm QF < 5$. The dashed line shows the cut-off in the uncertainties at 35\%.} 
    \label{fractional}
\end{figure}

\begin{figure*}[!htbp]
    \centering
    
    \includegraphics[width=1\textwidth]{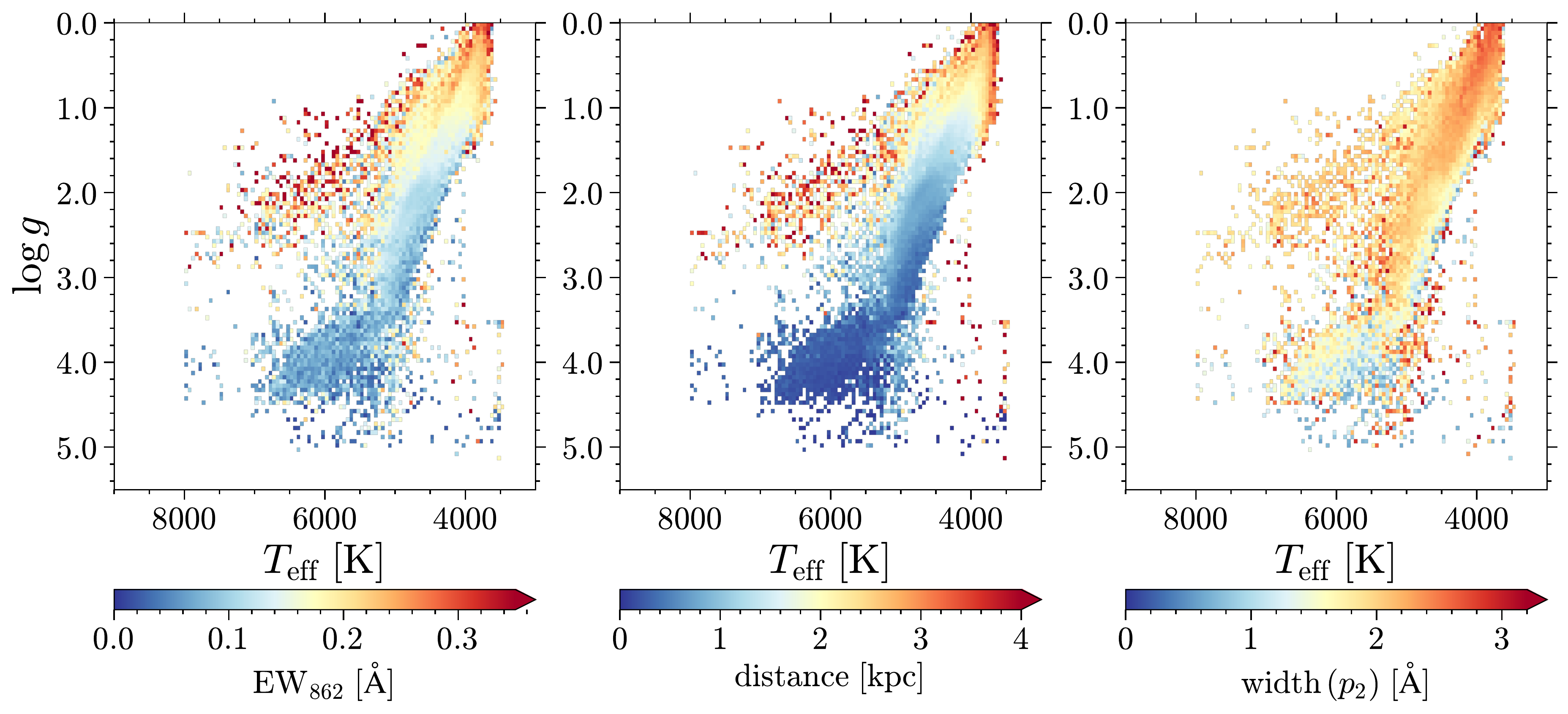}
    
    \caption{Mean $\rm EW_{862}$ ({\it left panel}), heliocentric photogeometric distance from \citet{Bailer-Jones2021} ({\it middle panel}), and the width ($p_2$, {\it right panel}), calculated in 50\,K $\times$ 0.05\,dex bins, as a function of the Kiel diagram, respectively.}
    \label{HRvsEW}
\end{figure*}

The full GSP-Spec sample contains 5\,591\,594 sources. Of these, 
476\,117  have a valid DIB\,$\lambda$862 measurement ($\sim$8.5\%). The number of sources for each QF is specified in Tab.~\ref{QF}.

\begin{table}[!htbp]
    \centering
    \begin{tabular}{cccr}
    QF&$p_0$& $p_2$ (\AA)&N\\
    \hline
     0 & $<$ 0.15  AND $>$ max($R_A$,$R_B$)           & 1.2--3.2 & 180\,879  \\
     1 & $<$ 0.15  AND $<$ max($R_A$,$R_B$) \& $>R_B$ & 1.2--3.2 & 1149\\
     2 & $<$ 0.15  AND $>$ max($R_A$,$R_B$)           & 0.6--1.2 & 54\,808 \\
     3 & $<$ 0.15  AND $<$ max($R_A$,$R_B$) \& $>R_B$ & 0.6--1.2 & 679\\
     4 & $<$ 0.15  AND $<$ max($R_A$,$R_B$) \& $<R_B$ & 0.6--1.2 & 2843      \\
     5 & $<$ 0.15                                    & --       & 235\,759\\
    \end{tabular}
    \caption{Definition of the Quality flags. $R_A$ is the fitting residual between the observed and synthetic spectrum for the global RVS spectrum, and $R_B$ is the region close to the DIB\,$\lambda$862 feature. Targets with a central wavelength beyond 816.6--8628.1\,{\AA} in vacuum are all labelled as $\rm QF=5$.
    The last column gives the number of sources for each QF.}
    \label{QF}
\end{table}

Figure~\ref{lbdistribution} shows the distribution on the sky of the DIB\,$\lambda$862 measurements at a resolution of $1.8^{\circ}$ (HEALPix map with level 5). 
As expected, the DIBs\,$\lambda$862 are concentrated towards the Galactic plane which is even more pronounced for the high-quality DIBs\,$\lambda$862 (right panel).  

Figure~\ref{QFdib} displays the relation between $\rm EW_{862}$ and the $\rm E(BP-RP)$ interstellar reddening measure from GSP-Phot (\citealt{DR3-DPACP-156}). We  see that  DIBs\,$\lambda$862 with low QFs ($\rm QF > 2$) show very small $\rm EW_{862}$ but a large range of $\rm E(BP-RP)$ which is not the case for the high quality (HQ) DIB\,$\lambda$862 measurements ($\rm QF  \leqslant 2$, see Sect.~\ref{HQ}).

\section{Definition of the high-quality sample} \label{HQ}

Figure~\ref{HRvserror} displays the GSP-Spec Kiel diagram of a subsample with $\rm QF<5$ as a function of the fractional uncertainty of the $\rm EW_{862}$. The vast majority of our sources show typical uncertainties below 20\%. However, on the red giant branch (RGB) sequence, the cooler stars (which are in general metal-richer) show larger uncertainties compared to the hotter ones.
This can be explained by the fact that for cooler metal-rich  stars, in general, we see a poorer agreement between the observed and the synthetic spectra due to the presence of molecular bands. This is also revealed by the larger $\rm log\, \chi^{2}$ values from GSP-spec. %(\citealt{GSPspec}).

We also notice higher uncertainties for hot dwarf stars in the range $7000\,{<}\,\teff\,{<}\,8000$\,K. 
The majority of those stars are classified as very metal-poor  with $\rm [M/H]\,{<}\,-3$\,dex by GSP-Spec. They further exhibit very large $v{\rm sin}i$ values from ESP-HS (Extended Stellar Parametrizer for Hot Stars; see Sect.~\ref{hotstars}). In addition to the parameter degeneracy between  $\teff$ and $\rm [M/H]$ for high-temperature stars, these objects present large $v{\rm sin}i$ values, which are not taken into account in the present GSP-Spec parameterisation, inducing parameter biases (c.f. \citealt{GSPspecDR3}).  Applying the specifically defined GSP-Spec flags  (see. Tab.~\ref{GSPspectable}) removes the majority of these stars.

Figure~\ref{fractional} shows the distribution of the fractional uncertainties (err($\rm EW_{862}$)/$\rm EW_{862}$) with $\rm QF < 5$. A clear bimodal distribution is apparent that is related to cool stars ($\teff\,{<}\,4500$\,K) with relatively weak DIBs\,$\lambda$862 ($<$0.2\,{\AA}) and a mismatch between the observed and the synthetic spectrum. 
We decided to reject sources with uncertainties larger than 35\%.  In addition, we  decided to neglect DIB\,$\lambda$862 measurements outside the wavelength interval $8620 < C_{\rm obs} < 8626\,{\AA}$ ---where $C_{\rm obs}$ is the measured central wavelength in the heliocentric frame with 
$C_{\rm obs} = p_1 + v_{\rm rad}\times p_1 / c$ where  $v_{\rm rad}$ is the stellar radial velocity and c the velocity of light---  because the majority of those are weak DIBs\,$\lambda$862, where the determination of the $p_1$ parameter could be corrupted and lead to high, unrealistic  velocities.  We stress that $p_1$ and $C_{\rm obs}$ are reported in the vacuum.

Our HQ sample is defined based on the criteria specified in Tab.~\ref{GSPspectable} which comprises 141\,103 objects. For a detailed explanation of the GSP-spec flag we refer here to \citet{GSPspecDR3}.

\begin{table}[!htbp]
\begin{center}
\begin{tabular}{c|c}
%\hline
  QF& $\rm \leqslant 2$  \\
  $\rm err( EW_{862})/ EW_{862}$ & $\leqslant 0.35$ \\
  $C_{\rm obs}$& 8620 -- 8626 \AA\\
  GSP-Specflag(vbroadT)& $\leqslant 1$\\
  GSP-Specflag(vbroadG)& $\leqslant 1$\\
  GSP-Specflag(vbroadM)& $\leqslant 1$\\
  GSP-Specflag(vradT)& $\leqslant 1$\\
  GSP-Specflag(vradG)& $\leqslant 1$\\
  GSP-Specflag(vradM)& $\leqslant 1$\\
  GSP-Specflag(fluxNoise) & $\leqslant 1$\\
  GSP-Specflag(extrapol) &  $\leqslant 1$\\
  GSP-Specflag(negFlux) &  $\leqslant 1$\\
  GSP-Specflag(nanFlux) &  $\leqslant 1$\\
  GSP-Specflag(emission) &  $\leqslant 1$\\
  GSP-Specflag(nullFluxErr)& $\leqslant 1$\\
  GSP-Specflag(KMgiantPar) & $\leqslant 1$\\
  \hline
  
\end{tabular}
\end{center}
\caption {Definiton of our high-quality sample}
\label{GSPspectable}
\end{table}

\begin{figure*}[!htbp]
    \centering
    
    \includegraphics[width=0.9\textwidth]{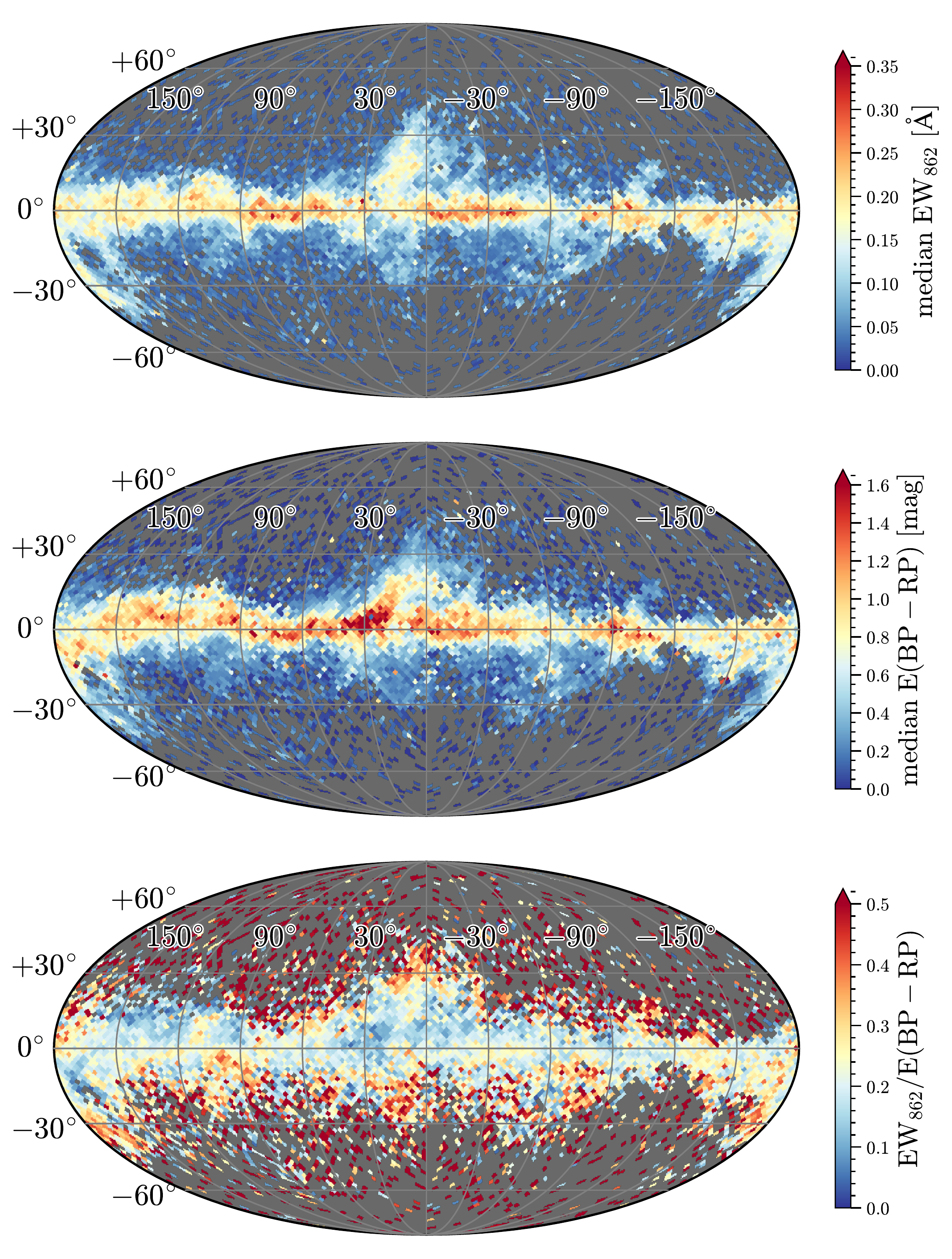}
    \caption{Comparison between the median $\rm EW_{862}$ of the DIB\,$\lambda$862 (upper panel), the median 
    $\rm E(BP-RP)$ (middle panel), and the ratio $\rm EW_{862}$/$\rm E(BP-RP)$ (lower panel) at HEALPix level 5 in the Mollweide projection.}
    \label{comparison}
\end{figure*}

\begin{figure}[!htbp]
    \centering
    
    \includegraphics[width=0.48\textwidth]{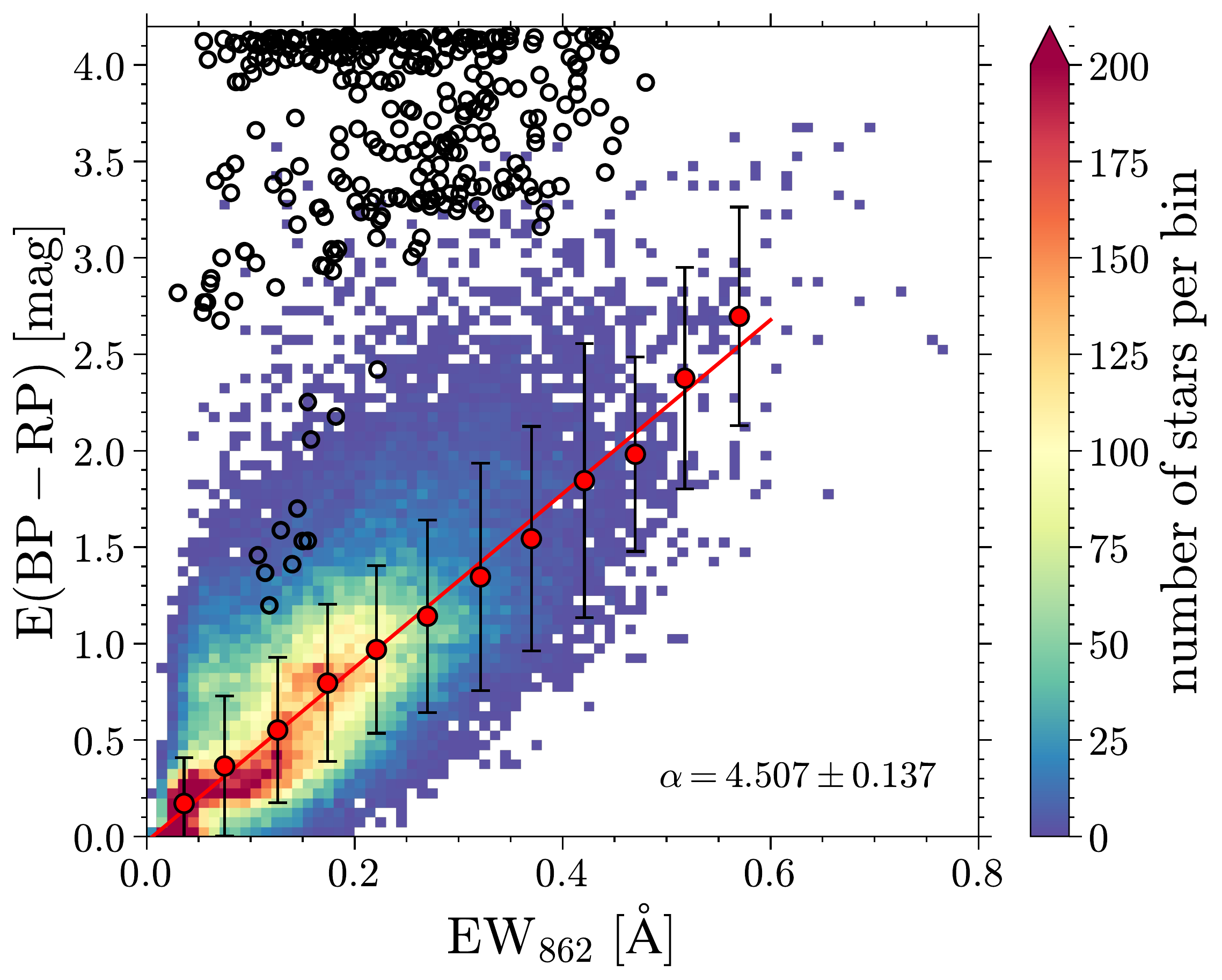}
    \caption{Correlation between $\rm EW_{862}$ and $\rm E(BP-RP)$ for 55 557 measurements in the
    high-quality sample with $\rm E(BP-RP)$ values. The colour scale  shows the number of stars per 0.01\,{\AA}$\times$0.05\,mag bin. The red dots are the median values taken in $\rm EW_{862}$ bins from 0 to 0.6\,{\AA} with a step of 0.05\,{\AA}. 
    The red line is the linear fit to the red dots. The fitting gradient and its uncertainty are also indicated.
    The open black circles (305 in total) are sources with a temperature difference (GSP-Phot -- GSP-Spec) larger than 5 000\,K.}

    \label{EWvsEBPRPvsTeff}
\end{figure}

\section{The Kiel diagram} \label{HR}

Figure~\ref{HRvsEW} shows the Kiel diagram colour-coded as a function of the  $\rm EW_{862}$ (left panel), the corresponding Gaia distances from Gaia EDR3 (middle panel, \citealt{Bailer-Jones2021}), and the DIB\,$\lambda$862 width ($p_2$). The very similar trend in these diagrams is striking, and indicates a clear relation between the $\rm EW_{862}$ of the DIB\,$\lambda$862 carrier and its distance, that is\ stars with larger distances show larger $\rm EW_{862}$. This is to be expected: As an interstellar feature, the DIB\,$\lambda$862 profile measured in the spectrum of a background star is the result of an integration of the DIB\,$\lambda$862 carrier between the observer and the star. DIB\,$\lambda$862 strength and dust extinction increase along the line of sight, and so both of them correlate with the distance and therefore also with each other. Also, we note that the distance of the background star is only an upper limit to the true distance of the DIB\,$\lambda$862  carrier clouds along the line of sight \citep{Zasowski2015c}. As shown by \citet{paperII}, direct measurements of the DIB\,$\lambda$862 carrier clouds can be obtained using kinematic  distances. This method will be further investigated in another paper. 

The right panel of Fig.~\ref{HRvsEW} shows how the measured width of the DIB\,$\lambda$862 (the value of the parameter $p_2$) increases with decreasing surface gravity; that is, we see that widths in giants are generally larger than in dwarfs. One may also conclude that the widths of DIB\,$\lambda$862 absorptions increase with distance, and explain this as a consequence of a superposition of an increasing number of clouds at slightly different radial velocities which accumulate along the line of sight. However, we also see DIB\,$\lambda$862 with large widths for close-by stars with $\teff< 5000$\,K and $\rm log\,g > 3$. This could be a consequence of spectral mismatches between observed spectra and the templates we use. These systematic trends will be investigated in a future work, but for now we stress that the measured widths of the DIB\,$\lambda$862 should be interpreted with caution.

From Fig.~\ref{HRvsEW} we see that stars with $5000 < \teff < 7000$\,K  and $\rm log\,g < 2.5$ have strong DIBs\,$\lambda$862. These massive stars  lie at distances of between 2 and 4 kpc and most of them are located in the  closest spiral arms (e.g. Sagittarius/Carina, Local and Perseus arms). This is in perfect agreement with the findings of \citet{GSPspecDR3}, who clearly identified those objects in their GSP-Spec Kiel diagram as massive stars that are  tracers of  the spiral arm structure, in agreement with the  spatial maps derived from \citet{Poggio2921}. The DIB\,$\lambda$862 measurements can therefore be considered as an excellent tracer of spiral arm  structures.

In contrast, our HQ sample lacks hot dwarf stars in the temperature range $7000 < \teff < 8000$\,K and $\rm 4.0 < log\,g < 4.5$ because their $\rm EW_{862}$ uncertainties are too high due to their high $v{\rm sin}i$ and  therefore large uncertainties in their stellar parameters (see Sect.~\ref{HQ}). A specific treatment of those stars is necessary but is beyond the scope of this work.

\section{Correlation with dust extinction} \label{dust}

As mentioned in Sect.~1, the  DIB\,$\lambda$862 shows a strong correlation with measurements of interstellar reddening such as $\rm E(B-V)$ (e.g. \citealt{Munari2008}, \citealt{Wallerstein2007}, \citealt{Kos2013}). Here, we use the interstellar reddening  $\rm E(BP-RP)$ derived from GSP-Phot as our main dust extinction tracer for individual objects. GSP-Phot  provides a detailed characterisation of single stars based on their BP/RP spectra, including stellar parameters ($\teff$, log\,g, [M/H]) and extinction $\rm A_{0}$. We refer to \citet{DR3-DPACP-156} 
for a detailed description of the GSP-Phot module. Due to the extensive filtering in GSP-Phot, only 66\,144 stars in our sample have $\rm E(BP-RP)$ measurements from GSP-Phot. Figure~\ref{comparison} compares the distribution on the sky of the median $\rm EW_{862}$ of the DIB\,$\lambda$862 with the median $\rm E(BP-RP)$. Overall, we see similarities between these two maps, with both showing larger values in the Galactic plane. Nevertheless, we also see some differences: (i) The DIBs\,$\lambda$862 seem to be generally  more concentrated towards the galactic plane compared to the interstellar dust (see also Sect.~\ref{Sec_scale_height}).  (ii) In the inner Galaxy ($|\ell| < 30^\circ$), DIBs\,$\lambda$862 show a larger scale height compared to the interstellar dust.   (iii) We notice at around   $\ell \sim 30^\circ$  a  low average $\rm EW_{862}$ of the DIB\,$\lambda$862 compared to the high amount of dust. This region covers several highly massive star forming regions which  were recently surveyed by the GLOSTAR Galactic plane survey in the frequency range between 4 and 8 GHz (\citealt{Brunthaler2021}). (iv) In the Galactic anticentre region ($|\ell| > 160^\circ$), some specific asymmetric tails of the DIB\,$\lambda$862 carrier are  visible (see third panel of Fig.~\ref{comparison}), reaching large Galactic latitudes ($b < -30^\circ$), which, interestingly,  are absent in the northern hemisphere. A detailed comparison between the correlation of interstellar dust and the DIB\,$\lambda$862 carrier along certain lines of sight is now possible thanks to the full sky coverage of Gaia  together with the distances; this should be further investigated.

\subsection{$\rm EW_{862}$ versus $\rm E(BP-RP)$} 
Figure~\ref{EWvsEBPRPvsTeff} shows the correlation between $\rm E(BP-RP)$ and $\rm EW_{862}$. We see the expected trend between $\rm EW_{862}$ and $\rm E(BP-RP)$ with a Pearson correlation coefficient (PCC) of 0.68 (the red circles with the uncertainty bars show the corresponding median values and their standard deviation). A linear fit through the median points (indicated by the red line in Fig.~\ref{EWvsEBPRPvsTeff}) is given by
\begin{equation}
\rm \mbox E(BP-RP) = 4.507(\pm0.137)\times EW_{862} - 0.026 (\pm0.047)
.\end{equation}

However, stars that were classified as hot stars by GSP-Phot but as cool stars by GSP-Spec deviate from this relation ---as indicated by the black open circles in Fig.~\ref{EWvsEBPRPvsTeff}--- in the sense that $\rm E(BP-RP)$ is too high compared to the measured DIB $\rm EW_{862}$. Due to the degeneracy between temperature and  extinction (see \citealt{DR3-DPACP-156}), the temperatures of those stars are overestimated by GSP-Phot,  leading to overestimation of $\rm E(BP-RP)$.  The DIB $\rm EW_{862}$ can therefore be used to find outliers of $\rm E(BP-RP)$ measurements.

For highly extincted regions, the  $\rm EW_{862}$ of the DIB\,$\lambda$862  should become smaller with increasing interstellar reddening and thus depart from a linear relation. \citet{Lan2015} attributed this behaviour to the `skin effect', noting that\ the DIB strength per unit reddening depends on cloud opacity.
\citet{Adamson1991} suggested that the DIB carriers must concentrate in the  surface layers (`skin') of the clouds and that the carrier depletion might be related to the reduction of the radiation field in the cloud interiors. \citet{Adamson1994} observed this effect with the NIR DIB, something that was later confirmed by \citet{Elyajouri2019} for the APOGEE DIB in the dense cores of the Taurus, Orion, and Cepheus clouds. We do not see this effect in our sample, which could be due to a selection effect in the sense that the Gaia RVS selection function  does not trace the most extincted regions.

\begin{figure*}[!htbp]
    \centering
    \includegraphics[width=0.8\textwidth]{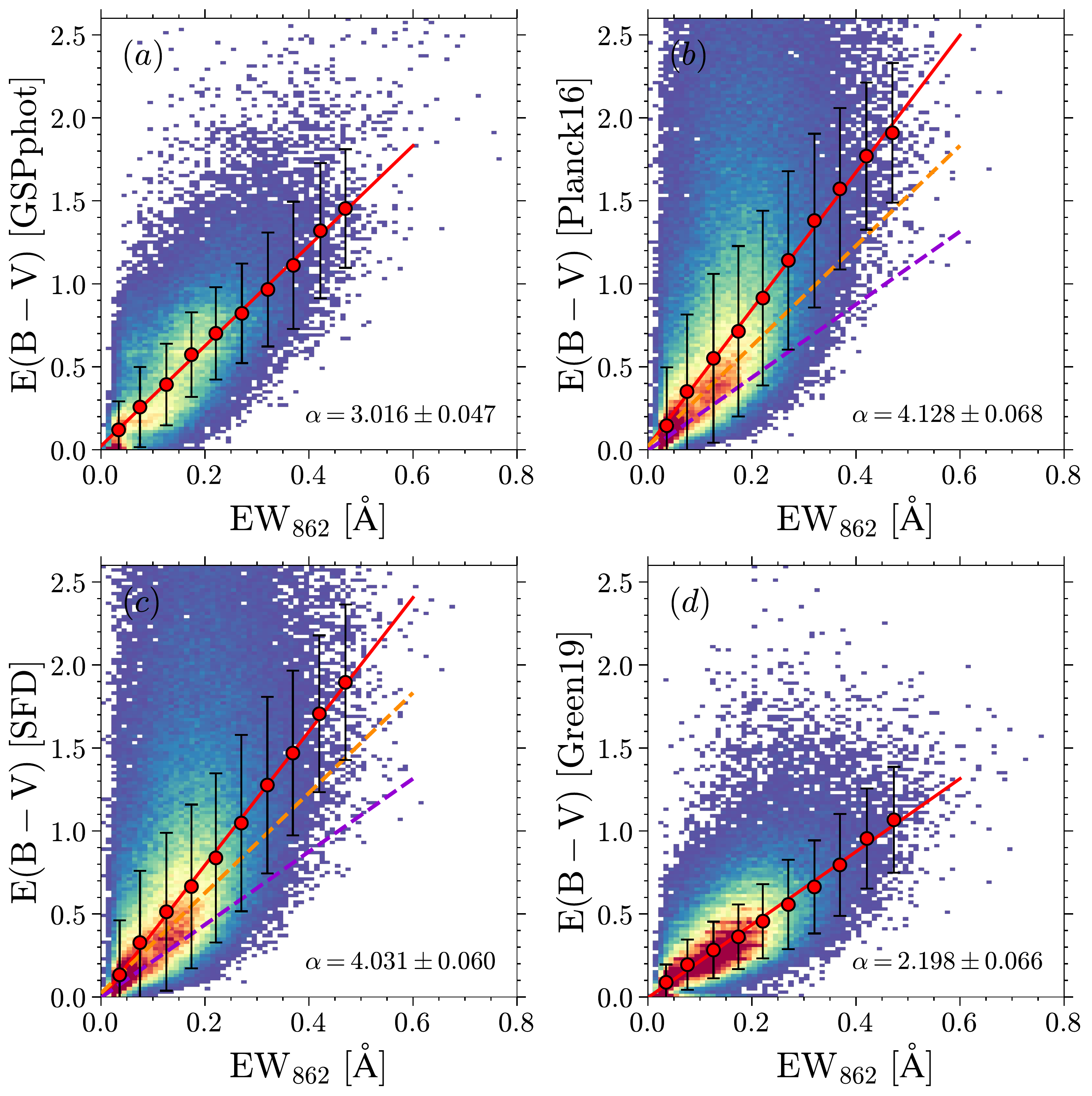}
    \caption{Correlations between $\rm EW_{862}$ and $\EBV$ derived from different extinction maps: (a) GSP-Phot,
    (b) \citet{Planck2016dust}, (c) \citet{SFD}, and (d) \citet{Green2019}. 
    The colours in each panel show the target number per 0.01\,{\AA}$\times$0.02\,mag bin. The colour bar is the same as in Fig. \ref{EWvsEBPRPvsTeff}. The red circles are the median values taken in $\rm EW_{862}$ bins from 0 to 0.5\,{\AA} with a step of 0.05\,{\AA}. The red lines are linear fits to the red dots in each panel, respectively. The fitting gradients ($\alpha$) and their uncertainties are indicated. They are also listed in Table \ref{literature}.
    The orange and violet dashed lines in (b) and (c) are the fit results to GSP-Phot and \citet{Green2019}, respectively.} 
    \label{EWvsEBV}
\end{figure*}

\subsection{$\rm EW_{862}$ versus $\EBV$}

$\EBV$ is the most frequently used reddening indicator to study the correlation with DIB strength, especially in early works. To compare our DIB--extinction relation to literature values, 
we derived the $\EBV/{\rm EW}_{862}$ coefficients from three dust extinction maps:  \citet{Planck2016dust}, \citet{SFD}, and \citet{Green2019}. We calculated  $\EBV$ from the  three maps  
using the Python package {\it dustmap} \citep{Green2018python}.

\citet{Planck2016dust} produced a full-sky two-dimensional extinction map using a generalised wavelet method to separate out Galactic dust emission from cosmic infrared background anisotropies. 
Such $\EBV$ values are asymptotic values and therefore represent overestimations for many of our objects (see Fig. \ref{EWvsEBV} (b)). This also applies to \citet{SFD} (Fig. \ref{EWvsEBV} (c)). 
Nonetheless, $\EBV$ derived from both of these maps for our objects present linear relations with $\rm EW_{862}$ with very high Pearson coefficients. For both \citet{Planck2016dust} and 
\citet{SFD}, we limit their $\EBV$ to values smaller than 2.6\,mag and get 121\,627 and 123\,175 individual measurements, respectively. We make use of 55\,252 available $\EBV$ values from GSP-Phot with 
a temperature difference between GSP-Spec and GSP-Phot of smaller than 5000\,K. Limited by the sky coverage, only 93\,247 objects have $\EBV$ from \citet{Green2019}, a three-dimensional dust reddening 
map inferred from 800 million stars with Pan--STARRS1 and 2MASS photometry. Based on \citet{SF11}, we apply a recalibration factor of 0.884 for $\EBV$ from \citet{SFD}. We also use this factor to 
convert the reddening unit of \citet{Green2019} to $\EBV$. We note
that the three-dimensional nature of the dust reddening maps from
GSP-Phot (Fig.\ \ref{EWvsEBV} a) and from  \citet{Green2019}  (Fig.\ \ref{EWvsEBV} d) negates the problem of overestimated $\EBV$ values. Table \ref{literature} lists the $\EBV/{\rm EW_{862}}$ coefficients and intercepts derived in this work together with  values from the literature.

Figure \ref{EWvsEBV} shows the correlation between  $\rm EW_{862}$ and $\EBV$ as well as their corresponding linear fits. We notice a large variation in the derived $\EBV/{\rm EW_{862}}$, which is due to 
the use of different methods for extinction calculation, with a very high value of $4.128 \pm 0.062$ from \citet{Planck2016dust} and a low value of $2.198 \pm 0.066$ from \citet{Green2019}. 
% The values from GSP-Phot show similar values and agree with the recent work of \citet{paperII} based on ground-based spectroscopic measurements toward the Galactic bulge. 
It is not surprising that different works report different values for the ratio of $\EBV/{\rm EW_{862}}$, depending on the sightlines  studied and the techniques  applied for DIB and extinction measurements.
The high coefficients with $\EBV$ from \citet{SFD} and \citet{Planck2016dust} imply that extinction measured from infrared emission is not only overestimated in some regions but presents
systematic differences (larger values) compared to the values calculated using other methods.

\begin{table}
        \begin{center}
        \small
        \caption{Coefficients and intercepts of the linear relations between DIB\,$\lambda$862 and $\EBV$
        derived in the literature and this work. 
        \label{literature}}
                \begin{tabular}{l c c}
                        \hline\hline
                        Works & $\EBV/{\rm EW_{862}}$  & intercept \\
                              & (mag\,{\AA}$^{-1}$)     &             \\ [0.5ex]
                        \hline
                        This work               &\ \ $3.016 \pm 0.047$\tablefootmark{a} &\ \ $0.023 \pm 0.013$ \\
                                                &\ \ $4.128 \pm 0.068$\tablefootmark{b} &\ \ $0.021  \pm 0.019$ \\
                                                &\ \ $4.031 \pm 0.060$\tablefootmark{c} &$-0.013  \pm 0.017$ \\
                                                &\ \ $2.198 \pm 0.066$\tablefootmark{d} & $-0.004 \pm 0.019$ \\
                        \citet{Sanner1978}      &\ \ $2.85 \pm 0.11$\tablefootmark{e} &\ \ -- \\
                        \citet{Munari2008}      & $2.72 \pm 0.03$ &\ \ -- \\
                        \citet{Wallerstein2007} & $4.61 \pm 0.56$ &\ \ -- \\
                        \citet{Kos2013}         & $2.49 \pm 0.23$ &\ \ $0.028 \pm 0.002$ \\
                        \citet{Puspitarini2015} &\ \ 2.12\tablefootmark{f} &\ \ -- \\
                        \citet{Krelowski2019b}  & $2.03 \pm 0.15$ &\ \ $0.22 \pm 0.05$ \\
                        \citet{paperII}         & $3.460 \pm 0.313$ & $-0.015 \pm 0.060$ \\ [0.5ex]
                        \hline
                \end{tabular}
        \end{center}
\tablefoot{ \\
\tablefoottext{a}{$\EBV$ from GSP-Phot} \\
\tablefoottext{b}{$\EBV$ from \citet{Planck2016dust}} \\
\tablefoottext{c}{$\EBV$ from \citet{SFD}} \\
\tablefoottext{d}{$\EBV$ from \citet{Green2019}} \\
\tablefoottext{e}{Calculated by \citet{Kos2013}} \\
\tablefoottext{f}{Estimated by their Fig. 7.}}
\end{table}

\begin{figure}[!htbp]
    \centering
    
    \includegraphics[width=1\linewidth]{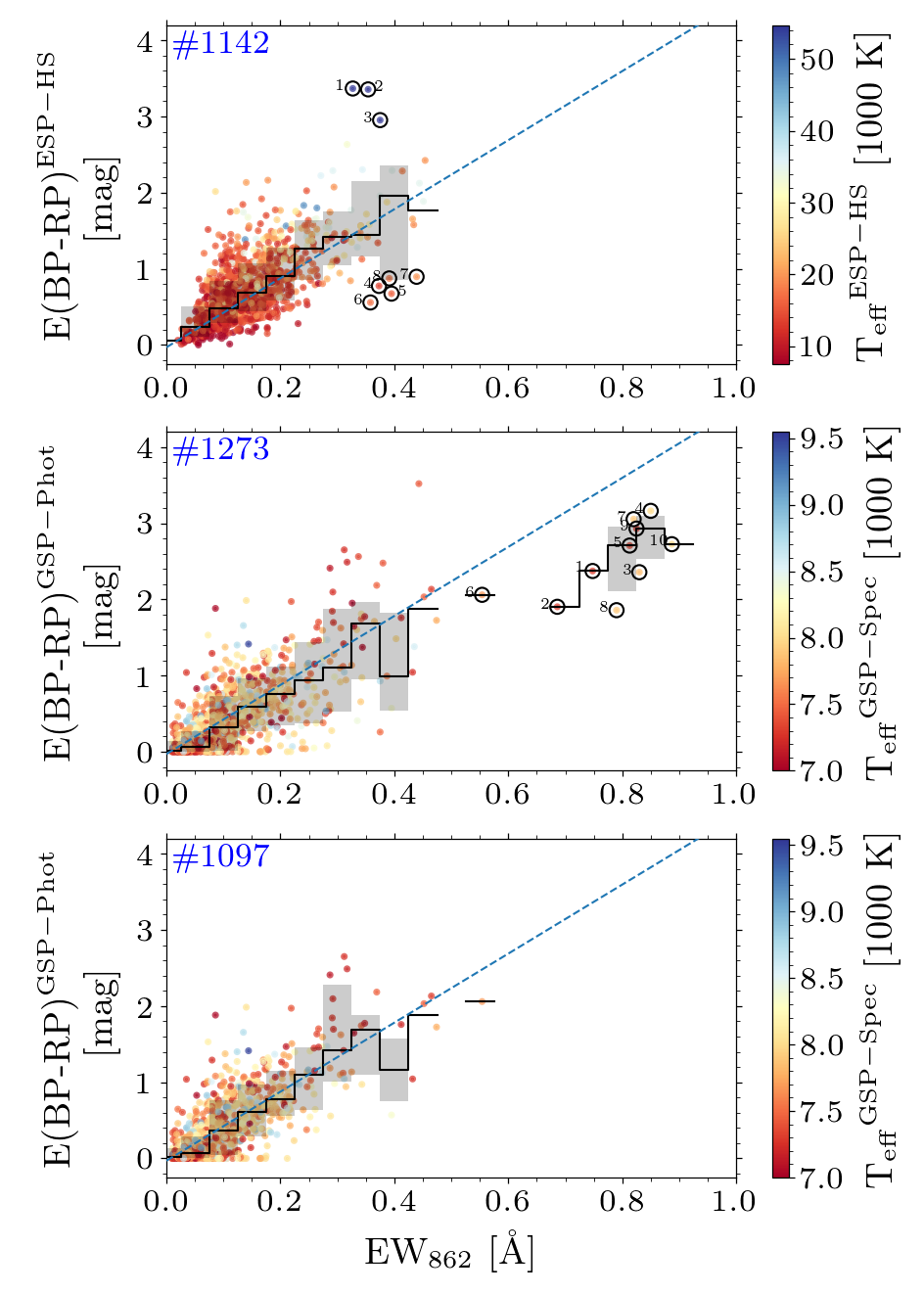}
    \caption{$\rm E(BP-RP)$ vs. $\rm EW_{862}$ of the DIB\,$\lambda$862 derived for the HQ sample by GSP-Spec for {\it hot} stars. The colour code follows the effective temperature derived by ESP-HS or GSP-Spec. The running median and interquantile (15 to 85 \%) are represented by a black step curve and the shaded area, respectively. The relation derived for the cooler stars is shown by the broken blue line. Upper panel: Reddening derived using the ESP-HS module for stars hotter than 7\,500~K. The outliers are identified with black circles and numbers. Middle panel: $\rm E(BP-RP)$ from GSP-Phot for targets hotter than 7\,000~K according to GSP-Spec only. Lower panel: $\rm E(BP-RP)$ from GSP-Phot, and hotter than 7\,000~K according GSP-Spec and GSP-Phot. Numbered black circles denote the outliers which are discussed in the main text, with their parameters listed in Table~\ref{tab:hot.outliers}. }

    \label{EW_ESPHS}
\end{figure}

\subsection{Hot stars} \label{hotstars}

In addition to the results obtained by GSP-Phot and GSP-Spec, the Apsis pipeline also contains the ESP-HS (Extended Stellar Parametrizer for Hot Stars) which specifically processes the BP/RP and RVS data for stars hotter than 7500\,K \citep{onlinedocdr3}. The module provides the astrophysical parameters  of O-, B-, and A-type stars, including an estimate of the interstellar extinction (A$_\mathrm{0}$, A$_\mathrm{G}$), and reddening $\rm E(BP-RP)$.
The target overlap between GSP-Phot, GSP-Spec, and ESP-HS is small due to the the post-processing filtering and quality assessment of the module, their $\teff$ validity domain (e.g. main valid AP domain of GSP-Spec is $\teff\,{<}\,8000$\,K), and/or parameter degeneracy. Keeping this in mind, there are 2929 ESP-HS hot stars with an estimate of the DIB $\rm EW_{862}$, and only 1142 that belong to the HQ sample. In the upper panel of Fig.\,\ref{EW_ESPHS}, we plot the interstellar reddening against the DIB $\rm EW_{862}$ for the latter sample, which provides a Pearson correlation coefficient (PCC)~of~$+$0.69. Eight outliers were identified. A brief description of these is provided in Table~\ref{tab:hot.outliers} (8 upper rows).

The hottest stars (labelled 1-3 in Fig.\,\ref{EW_ESPHS}) are targets cooler than 7500~K (according to GSP-Spec), and those that were treated with non-adapted synthetic spectra by ESP-HS. Outlier `7' is known from Simbad \citep{2000A&AS..143....9W} to exhibit emission. On the other hand, the H$\alpha$ pseudo-EW provided by the ESP-ELS module is positive (i.e. no significant emission is found in H$\alpha$ from the BP/RP spectrum), and its RVS spectrum appears normal. It therefore remains unclear as to why the derived APs (which include the extinction) do not provide a correct fit to the data. Outlier `6' has a very peculiar RVS spectrum belonging to an extreme He star (FQ Aqr). Outliers `4', `5',  and `8' show good agreement between observed and RVS fitted spectra.

\begin{figure*}[!htbp]
    \centering
    \includegraphics[clip,trim=0 0.8cm 0.8cm 0,width=0.49\textwidth]{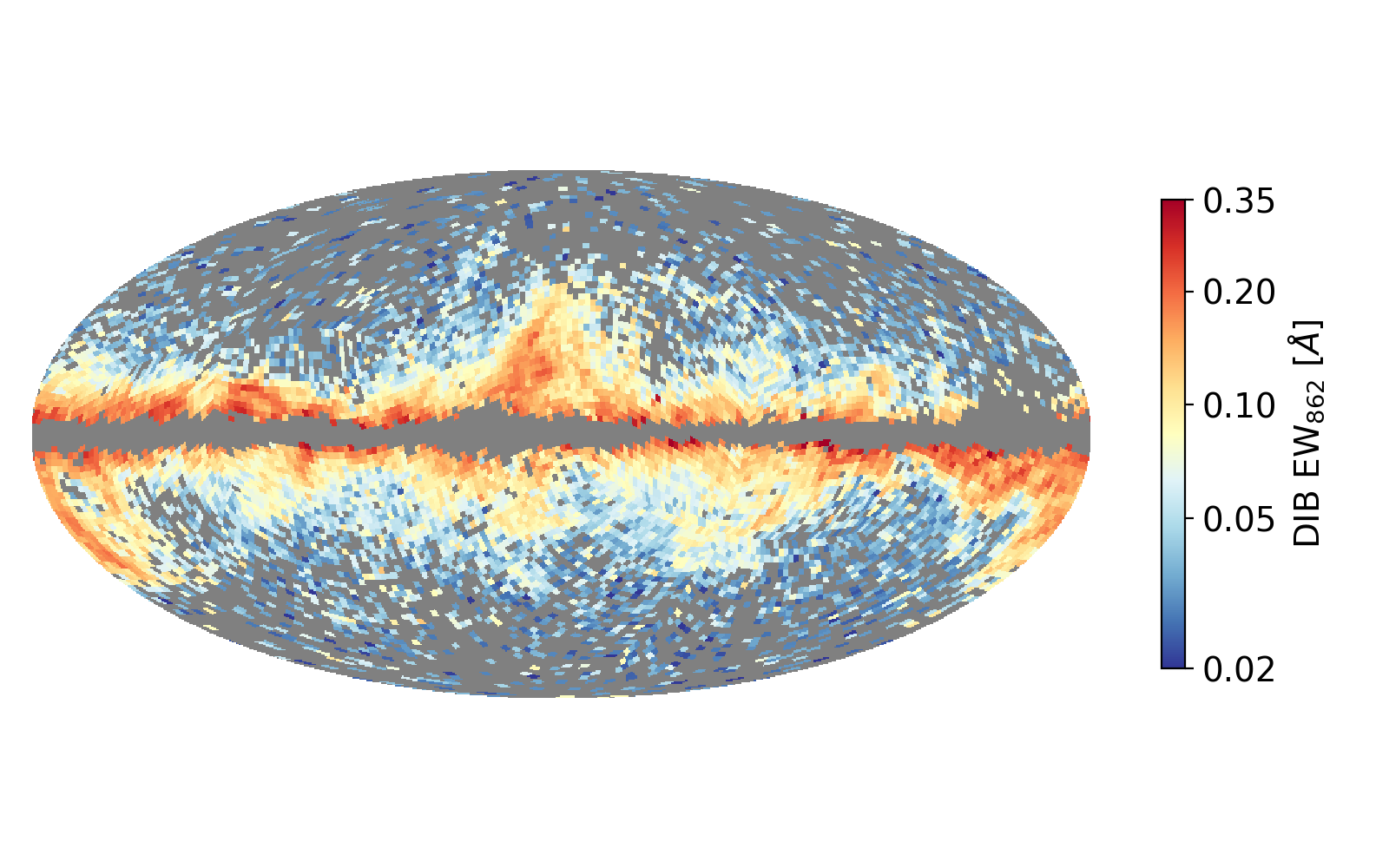}
    \includegraphics[clip,trim=0 0.8cm 0.8cm 0,width=0.49\textwidth]{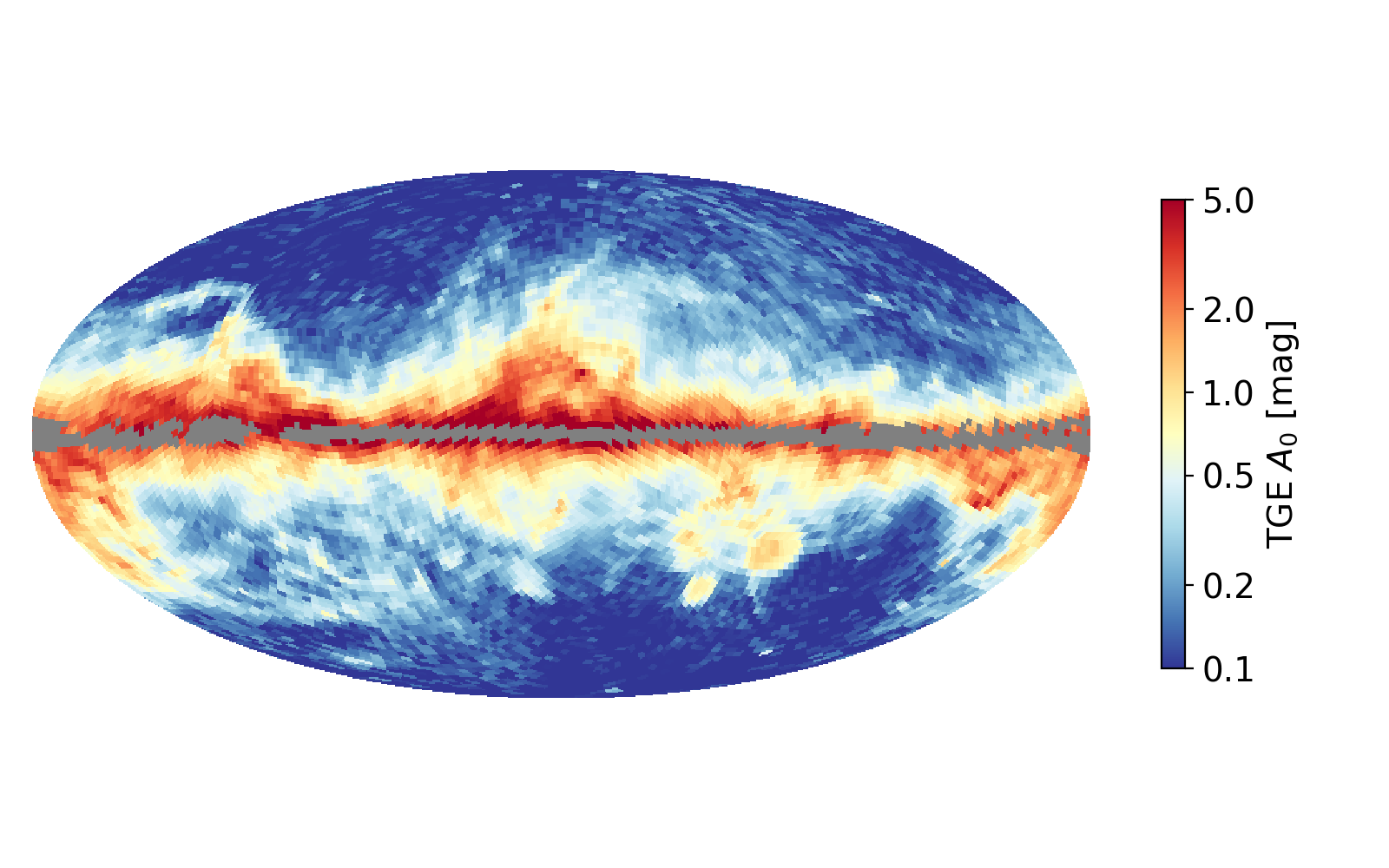}
    \includegraphics[clip,trim=0 0.8cm 0.8cm 0,width=0.49\textwidth]{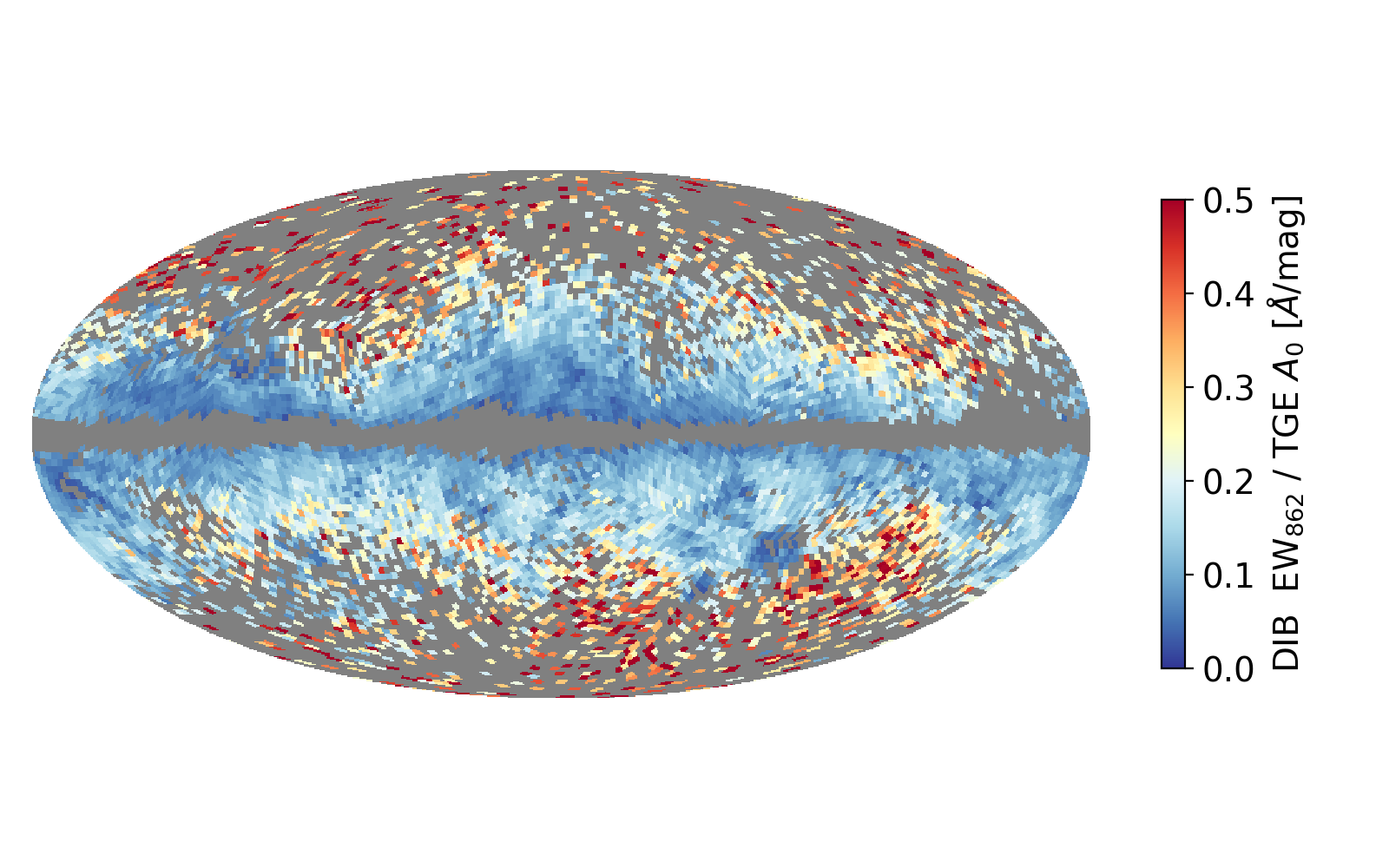}
       \includegraphics[width=0.49\textwidth]{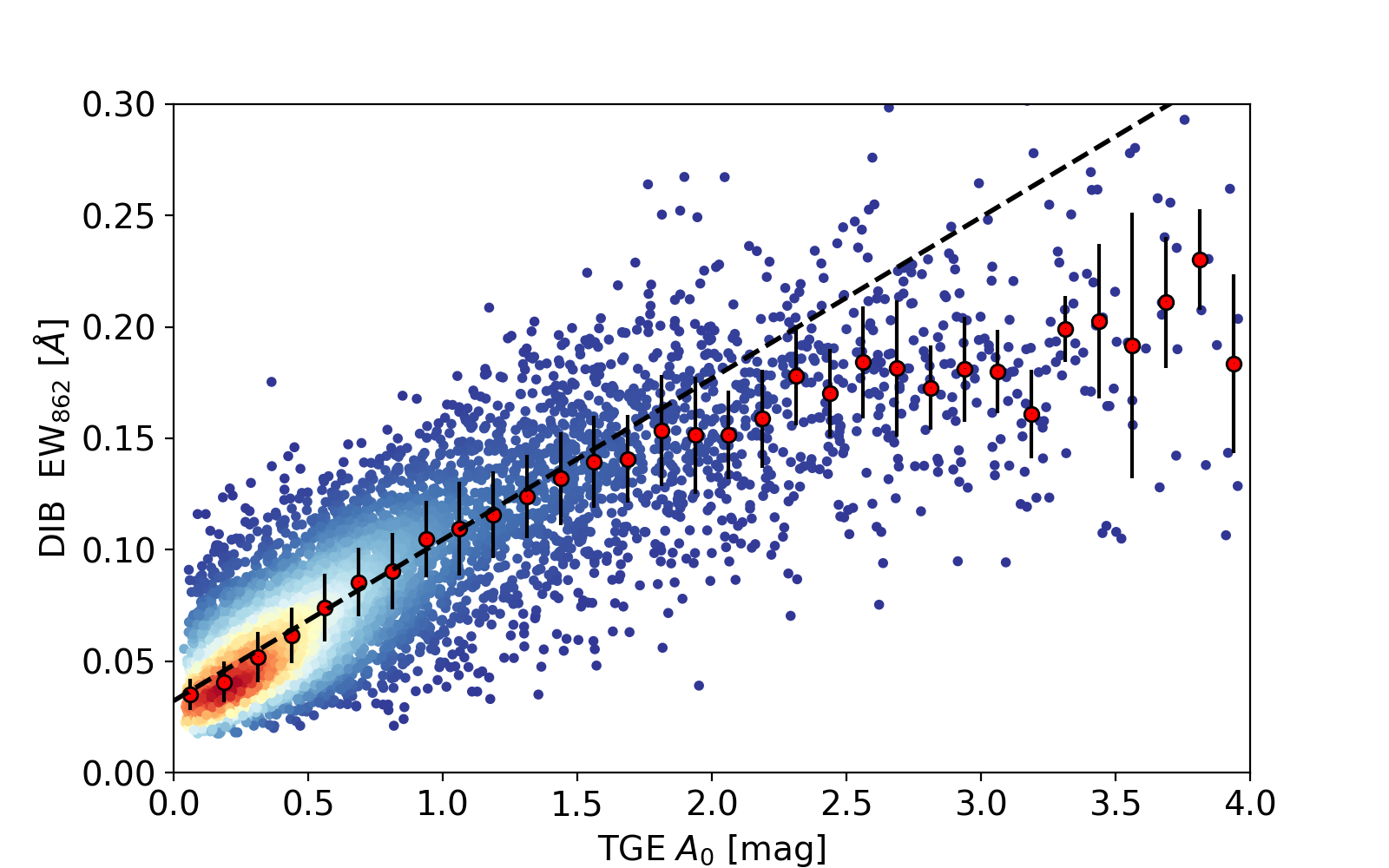}
    \caption{Top left: $\rm EW_{862}$ of the HQ sample for stars beyond the Galactic disk ($|z|>300$pc), averaged in each level-5 HEALPix. Grey pixels indicate no data, where there are fewer than two DIB\,$\lambda$862 measurements in the level-5 HEALPix. Top right: TGE $A_0$ at HEALPix level 5, again where grey signifies no data (i.e. where there are insufficient extinction tracers). Bottom left: $\rm EW_{862}$ vs. TGE over the sky. Bottom right:  Density plot of $\rm EW_{862}$ vs. TGE. The median $\rm EW_{862}$ in regular TGE bins is shown as red points. The uncertainty bars are derived using the average absolute deviation around the median.}
    \label{fig:tge_comp_level5}
\end{figure*}

A similar trend is observed in the GSP-Phot vs. GSP-Spec data, and plotted in the two lower panels of Fig.\,\ref{EW_ESPHS}. In the middle panel, the selection is solely based on the effective temperature provided by GSP-Spec. Targets with a DIB 
$\rm EW_{862}$ of greater than 0.5 \AA\ are identified and numbered (Table~\ref{tab:hot.outliers}). With the exception of the star labelled `6', which shows an RVS spectrum typical for an early-B or late-O star, all the stars have spectral features usually seen in M or late-K-type stars (which is confirmed by Simbad in two cases; in the other ones no additional information was found). Therefore, these are confirmed outliers, and to consistently (e.g. between the two GSP modules) remove those points, we performed a second selection based on the $\teff$ derived by both modules ($\teff\,{>}\,7000$\,K).  This last selection is plotted in the lower panel of Fig.\,\ref{EW_ESPHS}, and provides a PCC~=~$+$0.77. The first selection attempt (middle panel) provides a median $\rm E(BP-RP)$ versus $\rm EW_{862}$ that is slightly lower than the relation obtained for the cooler stars (represented by the broken blue line), while the first and third ones are in fair agreement with this latter.
The sample combination of the ESP-HS and GSP-Phot/GSP-Spec (Fig.\,\ref{EW_ESPHS}, lower panel) selections provides 1\,804 hot stars.

\subsection{Comparison with the TGE dust map}

The total galactic extinction (TGE)  map is a full-sky 2D representation of the foreground extinction from the Milky Way towards extragalactic sources, which is constructed from selected sources at large distances beyond the Galactic disk. To derive this map, distant giants were selected in order to obtain a set of stars situated beyond the dust layer of the disk of the Galaxy. The median of extinctions derived by GSP-Phot was then used to assign an extinction value for each HEALPix at different levels. For further details on the TGE maps, see \citet{DR3-DPACP-158}. 

In the following, we use the HQ DIB sample as defined in Sect. \ref{HQ}. In order to compare the $\rm EW_{862}$ of the DIB\,$\lambda$862 to the TGE map, it is first necessary to construct a HEALPix map of the $\rm EW_{862}$ in the same way as for the TGE map. We selected the DIB\,$\lambda$862 DIB $\rm EW_{862}$ measurements based on their Galactic altitude ($|z|>300$ pc) and then calculated the median $\rm EW_{862}$ in each HEALPix. Only HEALPixels with more than one DIB\,$\lambda$862 measurement were retained.

\begin{figure*}[!htbp]
    \centering
    \includegraphics[width=0.80\textwidth]{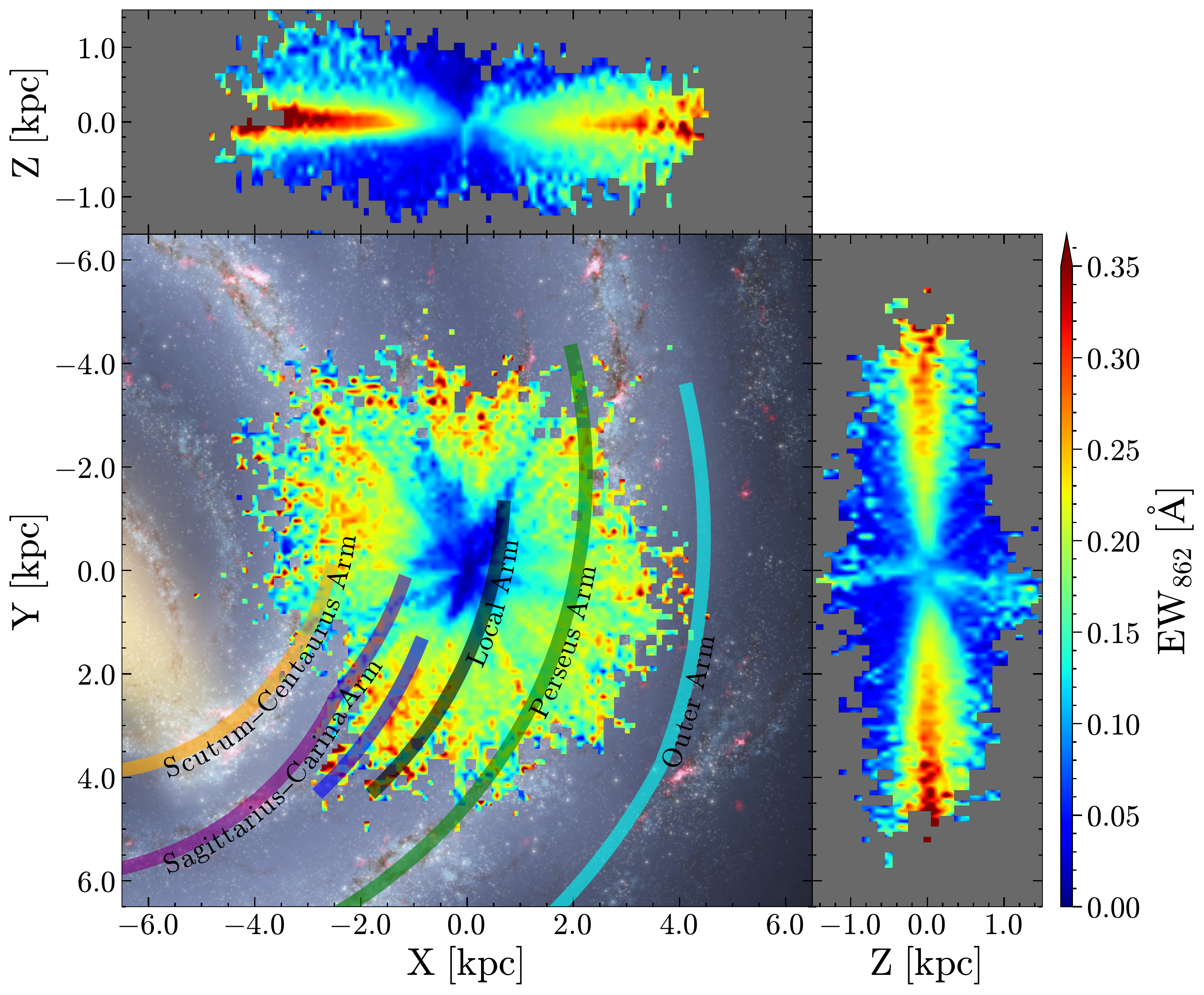}
    \caption{Face-on and side-on views of the spatial distribution of the DIB\,$\lambda$862 for the whole HQ sample  
    plotted over the Milky Way sketch created by Robert Hurt and Robert Benjamin \citep{Churchwell2009}. Median $\rm EW_{862}$ are taken from $0.1\,{\rm kpc}\,{\times}\,0.1\,{\rm kpc}$ bins in XY, XZ, and
    YZ planes, respectively. The Galactic centre is located at 
    (X,Y,Z)=(--8,0,0). The coloured lines represent the Galactic log-periodic spiral arms described by the parameters from
    \citet{Reid2019}:  Scutum--Centaurus arm, orange; Sagittarius--Carina arm, purple; Local 
    arm, black; Perseus arm, green; Outer arm, cyan. The spur between the Local and Sagittarius--Carina arms is 
    indicated by the blue line.}
    \label{fig:Spatial}
\end{figure*}

\begin{figure*}[!htbp]
    \centering
    \includegraphics[width=0.80\textwidth]{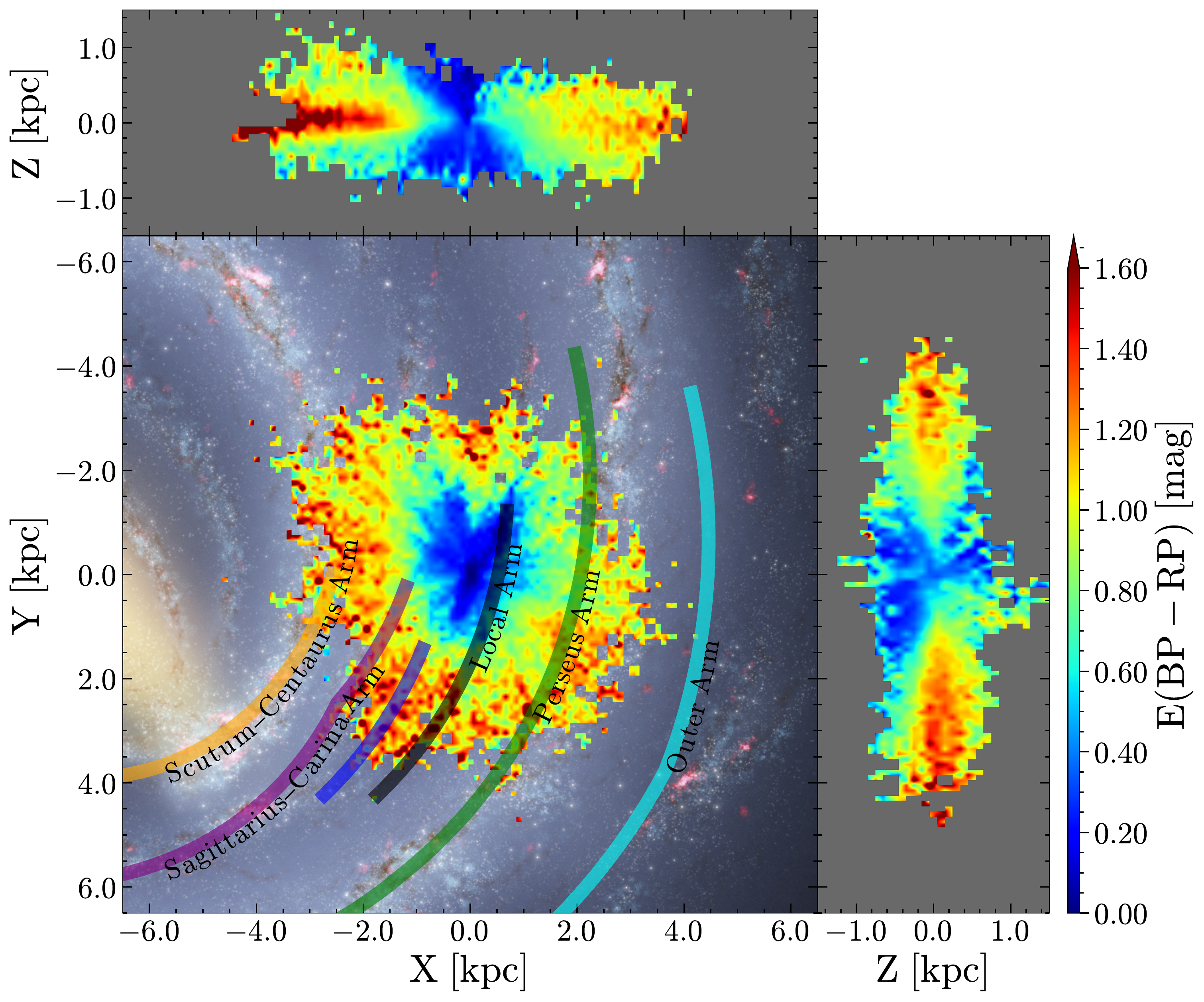}
    \includegraphics[width=0.80\textwidth]{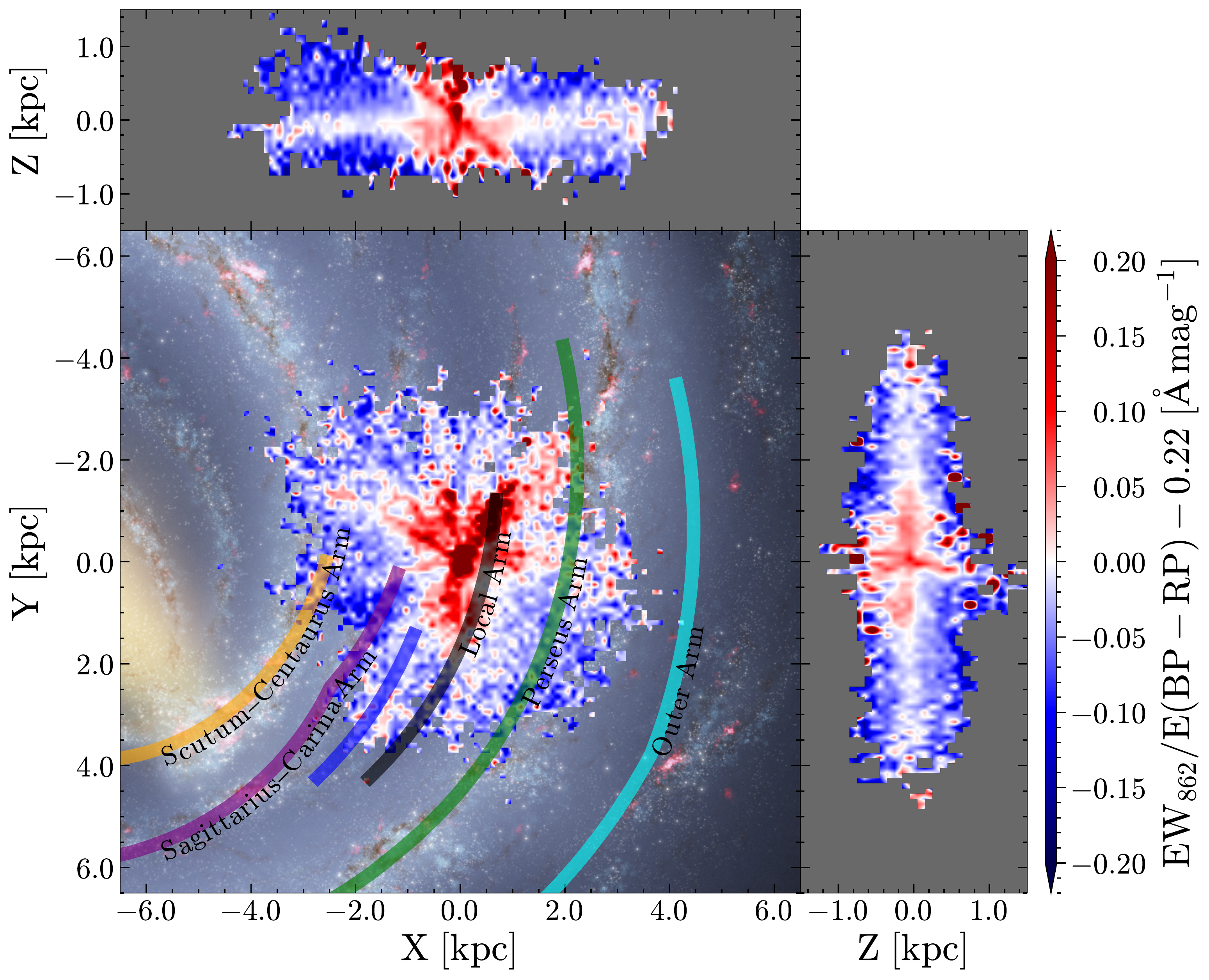}
    \caption{Same as Fig. \ref{fig:Spatial}, but for $\rm E(BP-RP)$ from GSP-Phot (upper panel),
    and the ratio of $\rm EW_{862}/E(BP-RP)$ (lower panel), subtracting 0.22,
    the inverse of the linear gradient fitted in Fig. \ref{EWvsEBPRPvsTeff}.
    Only 55 080 sources in the HQ sample with $\rm E(BP-RP)$ measurements are used.}
    \label{fig:ratio}
\end{figure*}

\begin{figure*}[!htbp]
    \centering
    \includegraphics[width=0.76\textwidth]{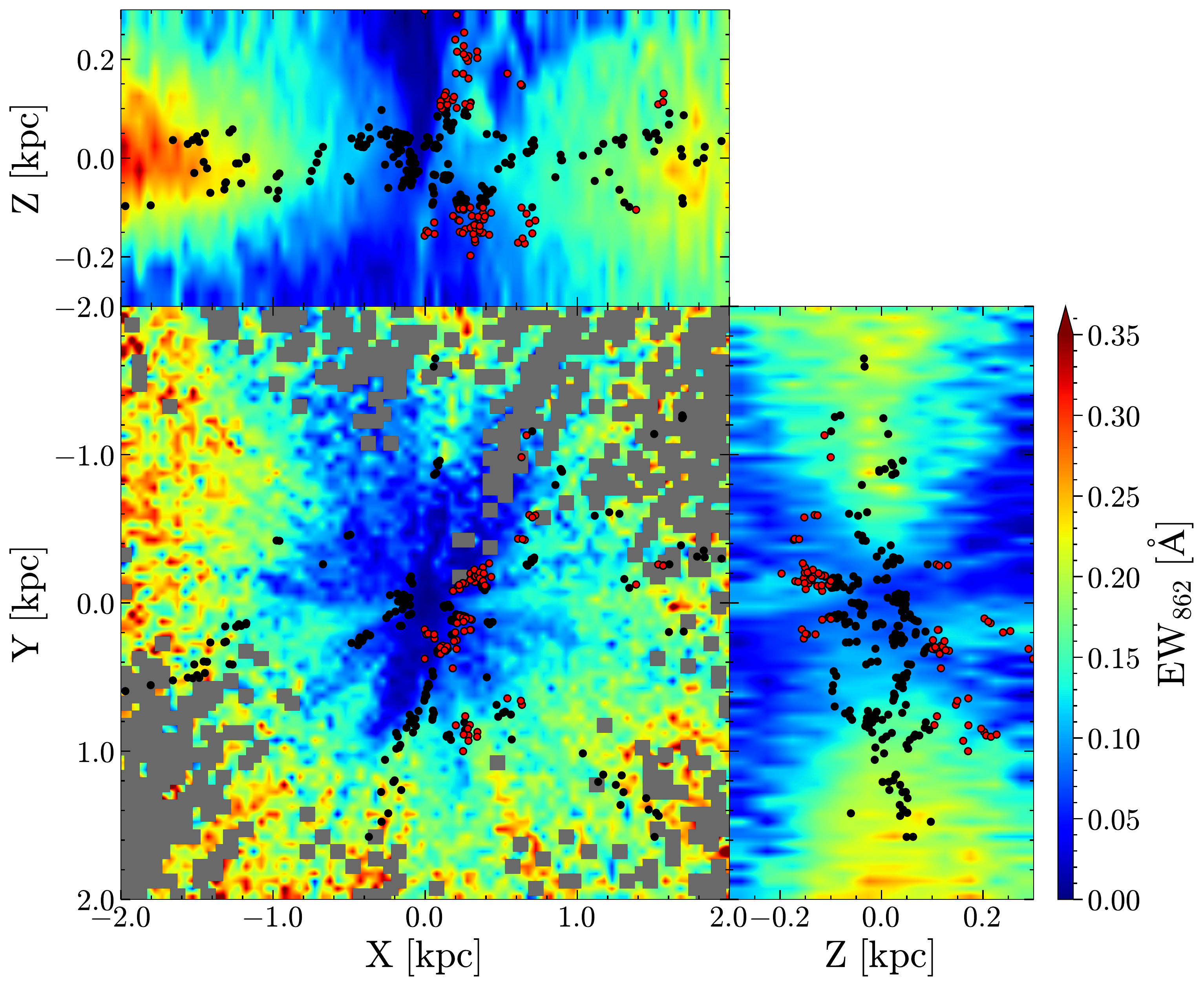}
    \includegraphics[width=0.76\textwidth]{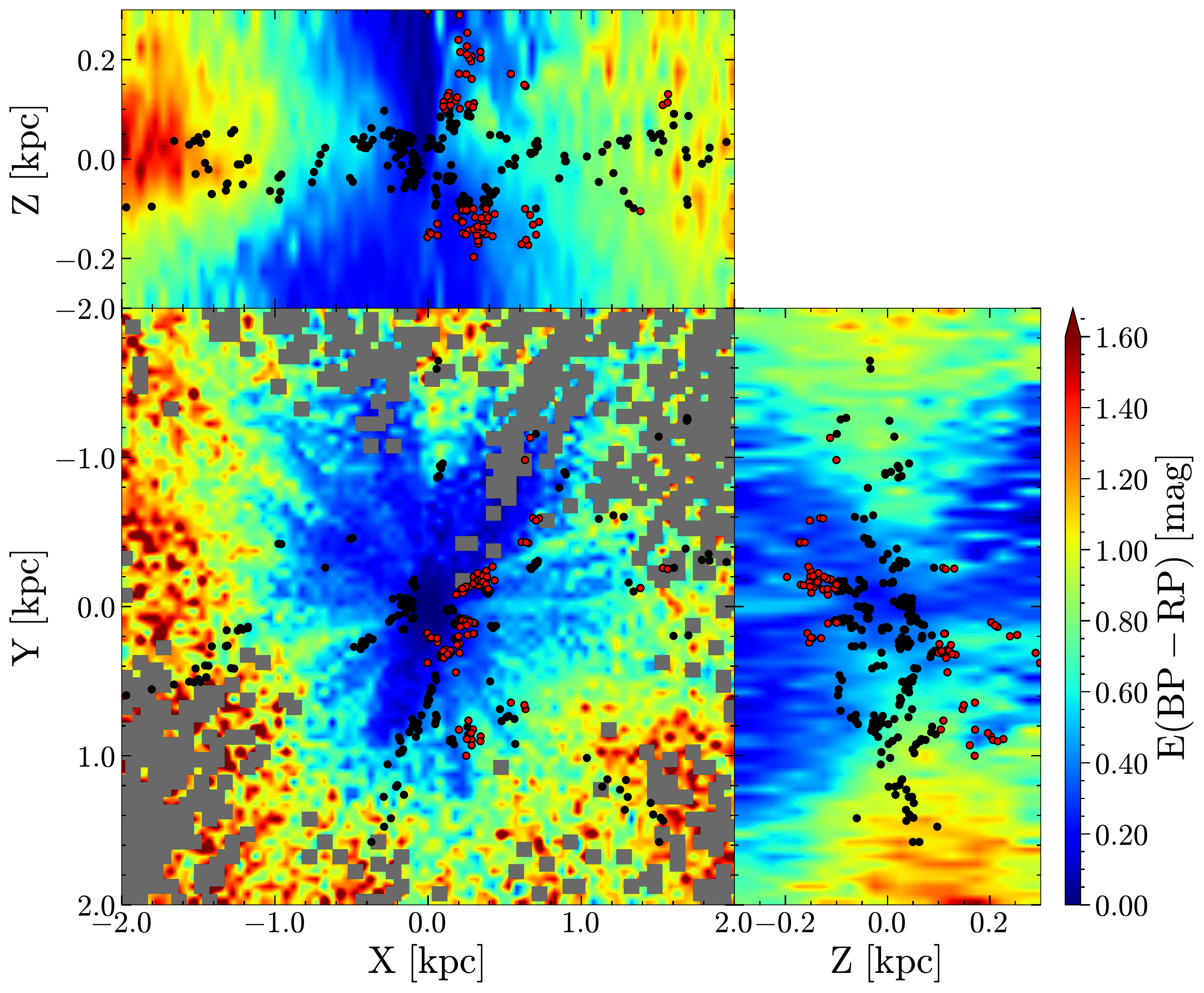}
    \caption{Same as Fig. \ref{fig:Spatial} but for a subsample containing 39 224 cases with
    $|X| \leqslant 2$\,kpc, $|Y| \leqslant 2$\,kpc, $|Z| \leqslant 0.3$\,kpc, and valid $\rm E(BP-RP)$. Median $\rm EW_{862}$ are taken from $0.05\,{\rm kpc}\,{\times}\,0.05\,{\rm kpc}$ bins in XY, XZ, and
    YZ planes, respectively.
    Overplotted are nearby MCs measured in \citet{Zucker2020}. The MCs with $\rm Z \geqslant 
    0.1$\,kpc are indicated as red dots.}
    \label{fig:Spatial-MC}
\end{figure*}

The resulting DIB $\rm EW_{862}$ HEALPix map is shown at level 5 in Fig.\ref{fig:tge_comp_level5} (top  left panel). We note that, due to our selection of DIB\,$\lambda$862 sources, this figure is not the same as the top panel of Figure \ref{comparison}.
Also shown in the top right panel of Fig.\ref{fig:tge_comp_level5} is the TGE map at level 5, where the value of a level-5 superpixel is the mean of the four level-6 pixels. Any level-5 HEALPix containing at least one level-6 HEALPix with insufficient tracers (less than three) is flagged as having no data. 
The lower left panel of Fig.\ref{fig:tge_comp_level5} shows the resulting skymap of the $\rm EW_{862}$/$A_0$ ratio, and the lower right panel shows a scatter plot of $\rm EW_{862}$ as a function of TGE $A_0$. Although the DIB\,$\lambda$862 map does not cover the entire sky (due to a lack of sufficient tracers), the two maps trace the same large-scale structures across the sky. The ratio of the two values is fairly constant from low to mid Galactic latitudes, but large fluctuations are seen at higher latitudes where the number of tracers drops considerably.
The scatter plot shows good correlation between the two values up to an $A_0$ of 1.5 mag, after which the  $\rm EW_{862}$ rises more slowly than the TGE $A_0$. This is a consequence of the fact that $A_0$ traces asymptotic values of extinction which (in the highly extinct regions) may occur beyond the distance of stars observed in DIB\,$\lambda$862 measurements.
A straight line fit to the scatter plot (broken line) below 1.5 mag results in a slope of 0.07 and an intercept of 0.03.

\section{Spatial distribution of the DIB\,$\lambda$862 } \label{Spatial}

Figure~\ref{fig:Spatial} shows a full sky map of the median values of the integrated $\rm 
EW_{862}$ of the DIB\,$\lambda$862 for the whole HQ sample, taken from $0.1\,{\rm kpc}\,{\times}\,0.1\,{\rm kpc}$ bins in XY, XZ, and YZ planes, 
respectively. Stellar  photogeometric distances are those from \citet{Bailer-Jones2021}. The overall distribution is similar to the pseudo-3D map \citep{Kos2014} from RAVE data (\citealt{Steinmetz2020b}), although a larger number of sight lines and coverage over the whole sky with \gdr{3} allow us to draw more specific conclusions. 

First, we note that $\rm EW_{862}$ increases with distance. This is expected, but it is a nice validation of our results, as this increase was not assumed when measurements of the DIB\,$\lambda$862 were made. The two cross-sections perpendicular to the Galactic plane in Fig.~\ref{fig:Spatial} show that DIB\,$\lambda$862 carriers are largely confined to the Galactic plane, as expected. We note that the regions with strong DIBs\,$\lambda$862 in two directions away from the plane (seen in the YZ cross-section) start locally and do not increase in intensity with distance. They therefore originate in clouds of DIB\,$\lambda$862 carriers which reside close to the Sun and cause DIB\,$\lambda$862 absorption in spectra of all stars located behind them.

The XY panel of Fig.~\ref{fig:Spatial} suggests  that stars within spiral arms generally show  stronger $\rm EW_{862}$ of the DIB\,$\lambda$862 carriers. This is true for the Scutum--Centaurus arm and for the Perseus arm. Our map lacks the reach needed to claim the same for the Outer arm, though an increase of DIB\,$\lambda$862 intensity at a distance of $\sim$4\,kpc in the Galactic anticentre direction agrees with this conjecture. The situation for the Local arm and the Sagittarius--Carina arm is more complicated: a region with strong DIBs\,$\lambda$862 at $\ell\,{\simeq}\,60^{\circ}$ coincides with the spur between these two arms (indicated by the blue line in Fig.~\ref{fig:Spatial}). However, there is also an indication of a region of strong DIBs\,$\lambda$862 in the opposite direction, at 
$\ell\,{\simeq}\,270^{\circ}$. This may indicate that DIBs\,$\lambda$862 fill in the region between the Sagittarius--Carina and Local arms, with the exception of a large void around the Solar position. However, we note that we do not claim the DIB carrier clouds are seen to reside within the spiral arms, as the presence of the Local Bubble around the Sun amplifies a general rise of EW with distance in any direction along the Galactic plane. A detailed investigation of the spatial distribution of DIB carriers is beyond the scope of this paper and will be discussed in Zhao et al. (in preparation).  

Figure~\ref{fig:ratio} compares the spatial distribution of DIB\,$\lambda$862 and dust absorptions. We note that only 40\%\ of the DIB\,$\lambda$862 sample has valid $\mathrm{E(BP-RP)}$ measurements due to a strong quality filtering in GSP-Phot. The comparison therefore only refers to 55\,080 sources in common and not to the whole DIB\,$\lambda$862 HQ sample shown in Fig.~\ref{fig:Spatial}. The top panels show the distribution of the colour excess, and the bottom panel is the ratio between $\rm EW_{862}$ and $\rm E(BP-RP)$ with a subtracted linear fit from Fig. \ref{EWvsEBPRPvsTeff}. 

\begin{figure*}[!htbp]
    \centering
    \includegraphics[width=0.9\textwidth]{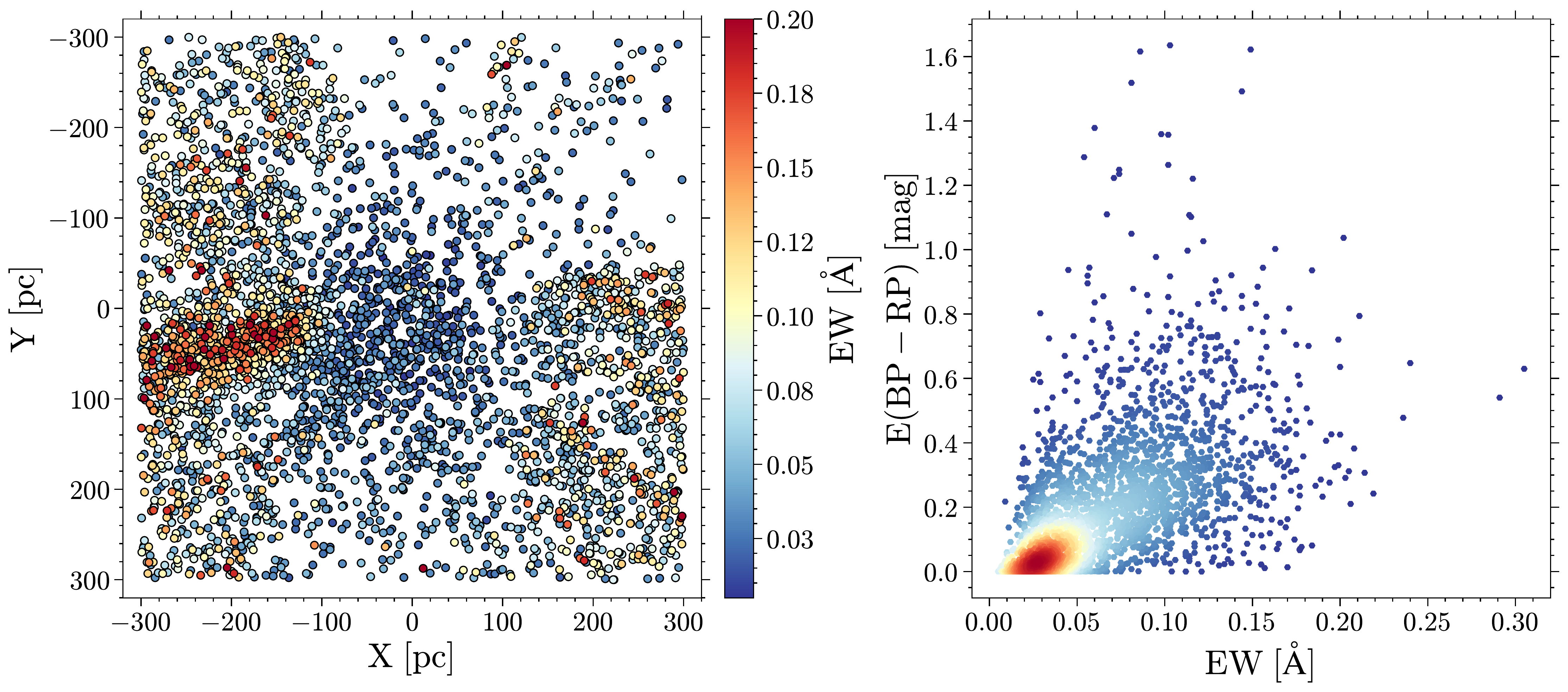}
    \caption{The local Bubble:  {\it Left panel:} Face-on view of the $\rm EW_{862}$ distribution of 3861 stars with $\rm |X| \leqslant 300$\,pc, $\rm |Y| \leqslant 300$\,pc, and $\rm |Z| \leqslant 100$\,pc.
    The Galactic centre is located at $\rm (X,Y)=(-8,0)$. {\it Right panel:} Density plot of the correlation between $\rm EW_{862}$ and $\rm E(BP-RP)$ for 2746  cases with valid
    $\rm E(BP-RP)$ measurements.}
    \label{fig:LB}
\end{figure*}

\begin{figure}[!htbp]
    \centering
    \includegraphics[width=0.48\textwidth]{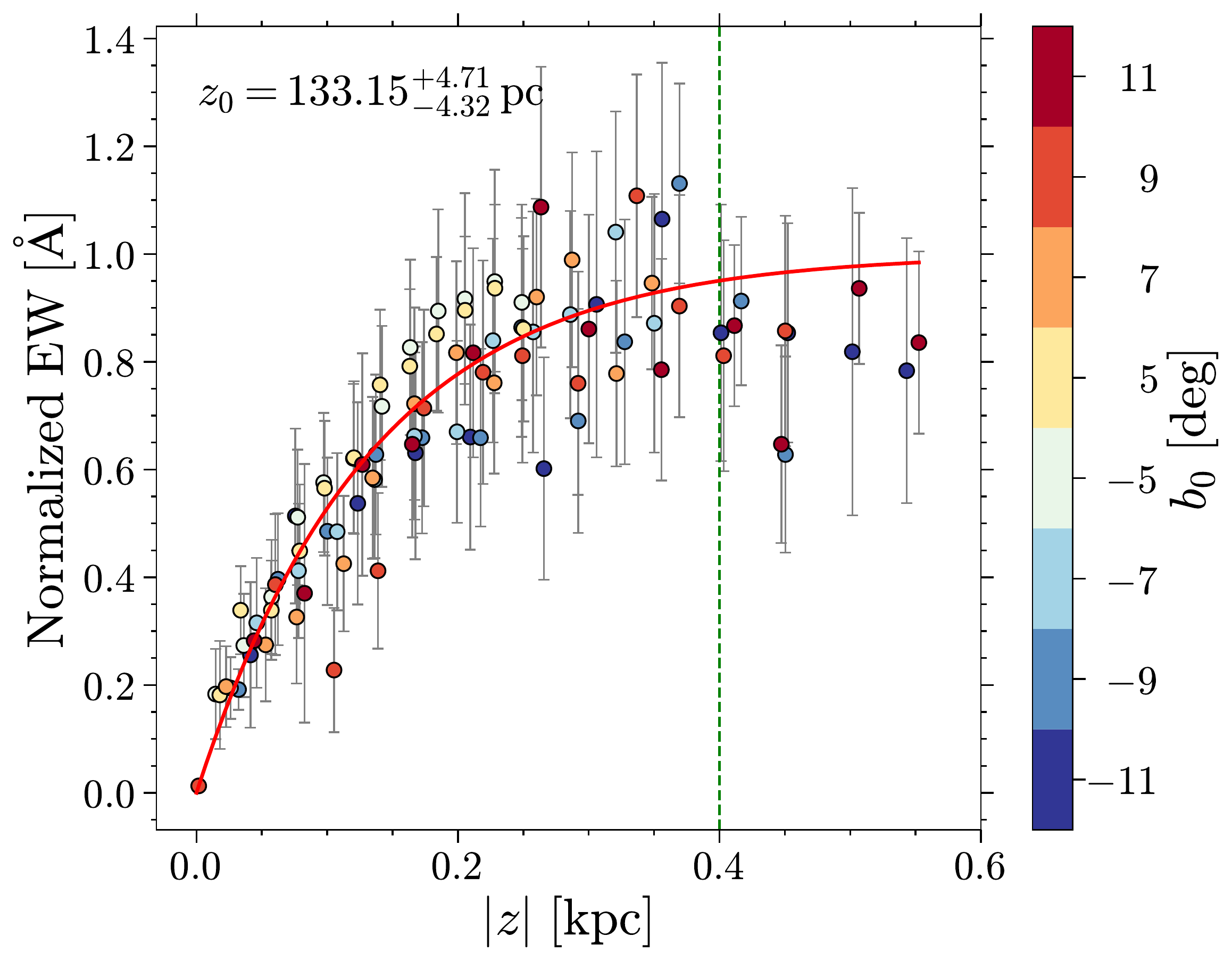}
    \includegraphics[width=0.48\textwidth]{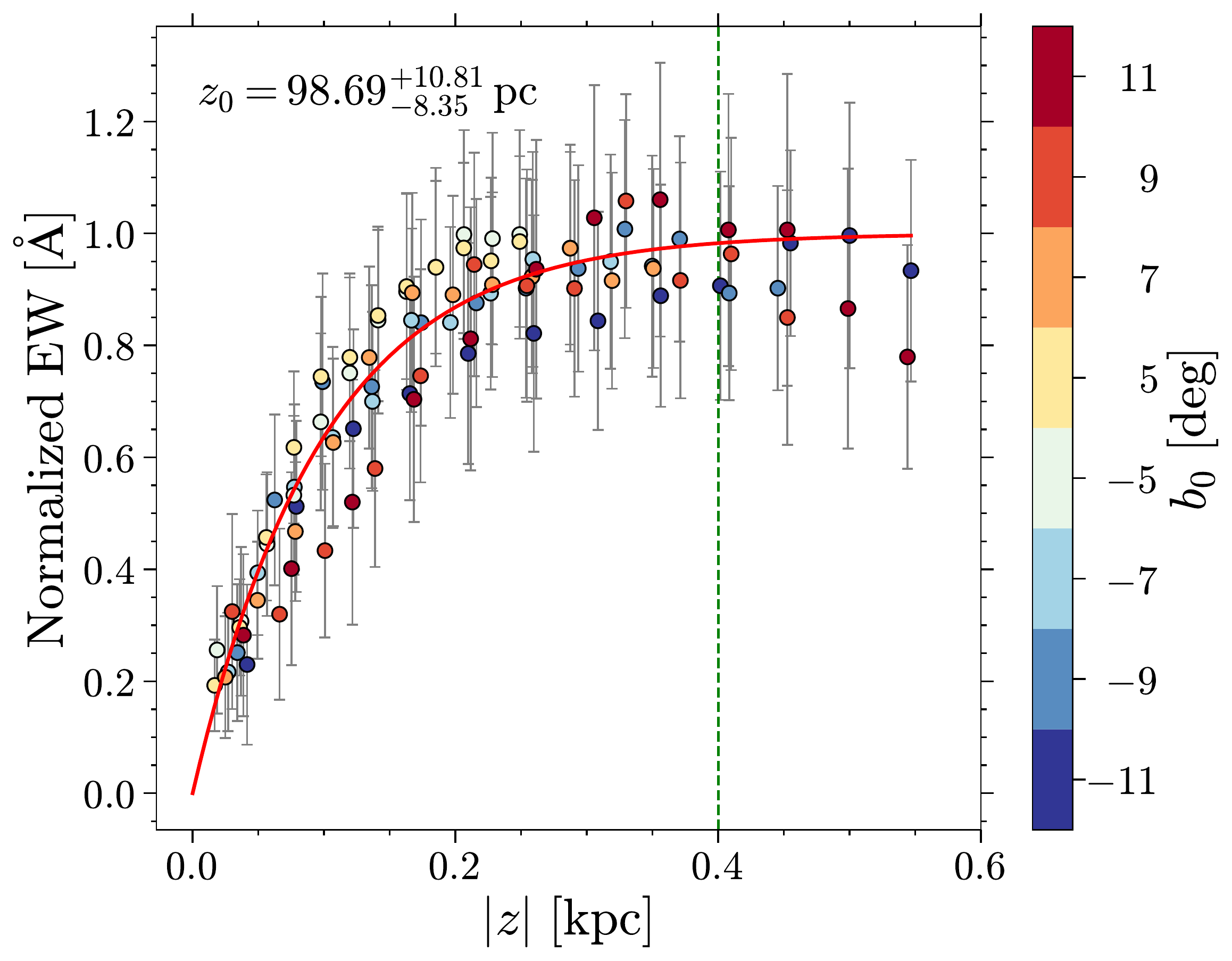}
    \caption{Determination of the scale height of the $\lambda$862 carrier by the DIB measurements with
    $4^{\circ} \leqslant |b| \leqslant 12^{\circ}$, and {\it upper panel:} $240^{\circ} \leqslant \ell \leqslant 330^{\circ}$; {\it lower panel:} toward all available longitude directions, respectively. The data points at different latitude slabs are
    coloured according to  the central latitude values ($b_0$). The dashed green line indicates $z=0.4$\,kpc. The red
    curve in the upper panel is the fit to data points with $z \leqslant 0.4$\,kpc, while in the lower panel, the red
    curve is the fit to all the data points.}
    \label{fig:sh-fit}
\end{figure}

\begin{figure*}[!htbp]
    \centering
    \includegraphics[width=0.9\textwidth]{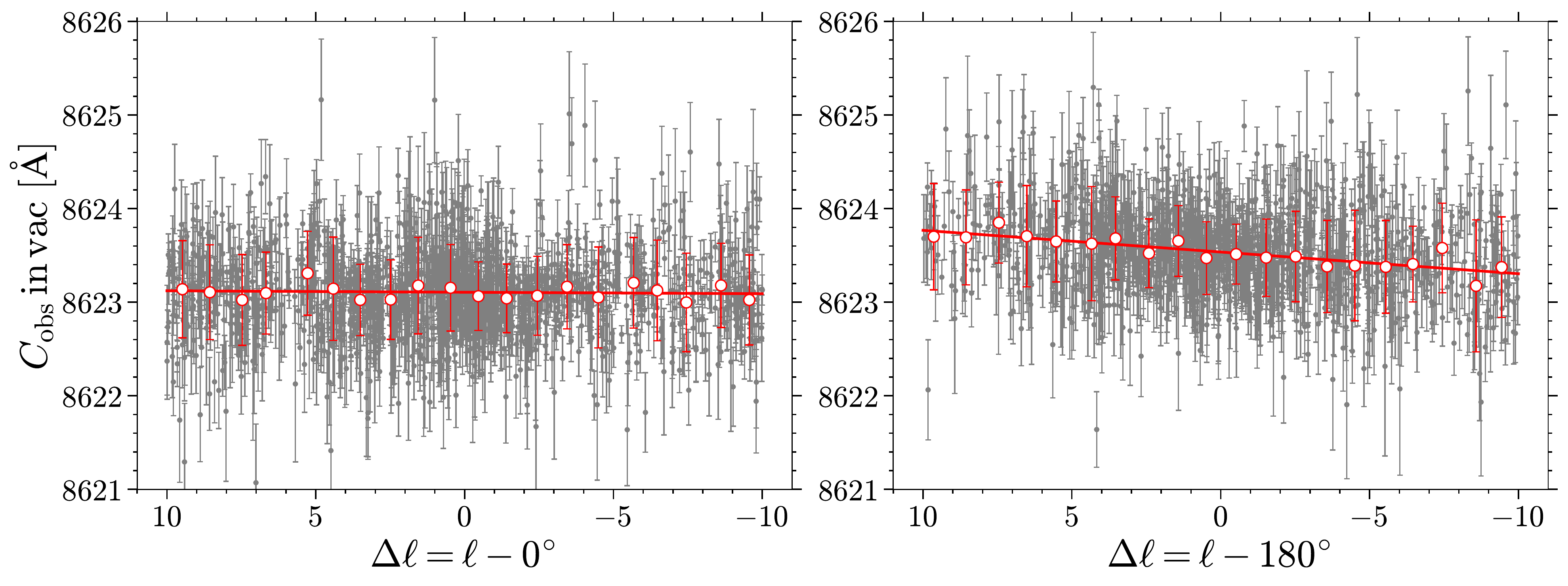}
    \caption{Observed central wavelengths ($C_{\rm obs}$, in vacuum) of DIB\,$\lambda$862 in the heliocentric 
    frame as a function of the angular distance from the longitude centre ($\Delta \ell$) for the 
    Galactic centre ({\it left panel}) and the Galactic anti-centre ({\it right panel}), respectively. 
    The grey points are the individual measurements with the fitted uncertainties. The red dots are the median 
    values taken in each $\Delta \ell\,{=}\,1^{\circ}$ bin with the standard deviation. The red lines are the 
    linear fit to the red dots.}
    \label{fig:lambda0}
\end{figure*}

Two important results of Figs.~\ref{fig:Spatial} and \ref{fig:ratio} are that the spatial distribution of DIB\,$\lambda$862 carriers and dust are qualitatively similar, but their ratio shows a pronounced lack of dust absorption for nearby sight lines. The red regions in the bottom panels of Fig.~\ref{fig:ratio} demonstrate that the local bubble around the Sun which contains very little dust does not have a similar low density of DIB\,$\lambda$862 carriers. This is confirmed with a median $\rm EW_{862}\,{\sim}\,0.1\,\AA $ within the inner 150\,pc from the Sun.  To investigate the situation further,  Fig.~\ref{fig:Spatial-MC} shows a zoom into the $4 \times 4 \times 0.6$~kpc rectangular box centred on the Sun for stars that have valid $\rm EW_{862}$ and $\rm E(BP-RP)$ measurements. In addition, the positions of the nearby molecular clouds from \citet{Zucker2020} are indicated by dots: black for clouds within 100~pc from the plane and red for those at heights between 100 and 300~pc. It is encouraging to see that molecular clouds at low Galactic heights are indeed at the head of strong DIB\,$\lambda$862 directions  and dust absorptions in the XY plane. This suggests that the light from behind stars passes through these clouds of simple  molecules, dust, and DIB\,$\lambda$862 carriers and so their volume-filling factor is large enough for this to happen. Similarly, molecular clouds at larger distances from the Galactic plane (red dots) seem to correspond to directions of enhanced dust absorption and DIB\,$\lambda$862 presence away from the plane. 

 We note that Figs.~\ref{fig:Spatial} and \ref{fig:ratio} are based on the assumption of a Gaussian profile for the DIB carrier. The profile of the DIB may be more complicated or may vary in shape; in some cases one may expect a superposition of absorptions originating in multiple clouds along the line of sight, but the $\rm EW_{862}$  values we derive are not affected significantly, as long as the radial velocities of the DIB carriers and profile variations are small compared to the width of the profile in our spectra with a moderate resolving power. The EWs we derive are always small, and so we are in a linear regime where the total value is a simple sum of individual absorptions.
In addition, the departures from the Gaussian profile caused by the superposition effect have been shown to be insignificant for DIB\,$\lambda862$ by comparing the fitted
EW with the integrated EW \citep{Kos2013} and EW calculated from an asymmetric Gaussian function \citep{paperI}.

Due to the large catalogue of DIB\,$\lambda862$, and the better sampling for different sightlines, we can trace the spatial variation  of $\rm EW_{862}/E(BP-RP)$
(bottom in Fig. \ref{fig:ratio}) which  can be used as a tracer to reveal the local physical conditions; as in the work of  \citet{Vos2011}    for the Scorpius OB2 association.
The ultimate goal would be to compare the densities of dust and DIB\,$\lambda862$ carrier derived by extinction and EW, respectively. A series of works carried out such a comparison for the dust \citep[e.g.][]{Capitanio2017,Rezaei2018,Lallement2014,Lallement2019,Rezaei2020}. No attempt has been made so far for DIB\,$\lambda862$.

%\hzhao{In the time of DIB research based on several to hundreds of spectra of early-type stars, the superposition effect could make the EW--extinction correlation
%tighter when the range of the observed quantities are broader. Thus we cannot treat their ratio as a description of the local environment near the background star.
%While with a large catalog such as DIB\,$\lambda862$, different sightlines are better sampled, which allows us to see the variation of the ratio of $\rm EW_{862}/E(BP-RP)$
%(bottom in Fig. \ref{fig:ratio}) at different positions. Therefore, it breaks the superposition of multiple cloud to some extent and explores the physical connection and
%disconnection between the dust and DIB\,$\lambda862$ carrier, although the ratio is not a local value but an integrated result. A more sophisticated method is to
%compare the densities of dust and DIB\,$\lambda862$ carrier derived by extinction and EW, respectively. A series works have completed this for the dust
%citep[e.g.][]{Capitanio2017,Rezaei2018, Lallement2014, Lallement2019,Rezaei2020}. But no attempts have been done for DIB\,$\lambda862$.}
A detailed analysis of the spatial co-location of molecular clouds and clouds of DIB\,$\lambda$862 carriers and interstellar dust, together with a study of their spatial filling factors, is beyond the scope of this paper and will be explored in the future. 

\subsection{The Local Bubble}

\citet{Farhang2019} studied the low-density cavity known as the Local
Bubble and found the presence of the DIB carriers at $\lambda 5797$
and $\lambda 5780$ in the bubble. Other detailed studies of the local
ISM were obtained from  \citet{Vergely2001,Vergely2010,Welsh2010}.
 Figure~\ref{fig:LB}  shows the distribution of the DIB\,$\lambda$862 carrier in the inner 300 pc volume with respect to the Sun within 100\,pc from the Galactic plane.  In the left-hand panel, a  clear asymmetry can be seen in the distribution of the DIB\,$\lambda$862, which is also seen in other DIB maps in the Local Bubble (see e.g. \citealt{Farhang2019}, \citealt{Bailey2016}),
while in the inner 100\,pc we see a homogeneous distribution of weak DIBs ($\rm EW < 0.05\,\AA$). 

Figure  \ref{fig:LB} shows the correlation of the  DIB\,$\lambda$862 of our sample  with the dust extinction derived from E(BP--RP).  Here, we see a clear linear relation in this extreme  low-extinction region  even for very small EW ($\rm < 0.05\,\AA$). However,  a more detailed discussion of the behaviour of the DIB\,$\lambda$862 in the Local Bubble is beyond the scope of this paper.

\subsection{Scale height}
\label{Sec_scale_height}

To characterise the vertical distribution of the carrier of the DIB\,$\lambda$862, we assume an exponential model
and follow the straightforward method used in \citet{Kos2014}. Following this approach, the
DIB strength $\rm EW_{862}$ and the stellar distance ($d$) in a narrow latitude slab can be derived as  
\begin{equation}
    {\rm EW_{862}} = \int_0^d \rho_0 \exp\left(\frac{- s \sin (|b|)}{z_0}\right) ds +B = A\,[1-{\rm exp}(-d/d_0)] + B, \label{eq:scale-height}
\end{equation} 
where  $z_0$ is the scale height, $b$ is the galactic latitude, $d$ is the heliocentric distance, $d_0 = z_0/{\sin}(|b|)$,  $A = \rho_0 z_0 / {\sin}(|b|)$, and $B$ is a small offset of our $\rm EW_{862}$ values due to the fact that only sufficiently strong DIBs\,$\lambda$862 pass the selection criteria for the HQ sample.  So that we can compare the data points at different latitudes, we follow \citet{Kos2014} and first normalise the curves in different latitude bins by fitting parameters ($({\rm EW_{862}}-B)/A$). This normalised $\rm EW_{862}$ is then fitted again by Eq. \ref{eq:scale-height} in order to
to get the scale height $z_0$. We refer to \citet{Kos2014} for more details,  especially their
Fig. 2.

\citet{Kos2014} applied this method for 20 latitude slabs from $b=-20^{\circ}$ to $b=20^{\circ}$ with a bin size of $2^{\circ}$ and obtained $z_0=209.0\,{\pm}\,11.9$\,pc. We only use eight slabs with moderate latitudes
($-12^{\circ} \leqslant b \leqslant -4^{\circ}$ and $4^{\circ} \leqslant b \leqslant 12^{\circ}$) which show exponential saturation, and take median $\rm EW_{862}$ in each 0.25\,kpc bin from 0 to $d=3$\,kpc. To compare with the result of \citet{Kos2014}, we first consider measurements with $240^{\circ} \leqslant \ell \leqslant 330^{\circ}$ (upper panel in Fig.~\ref{fig:sh-fit}). The normalised $\rm EW_{862}$ with $z>0.4$\,kpc show an apparent offset due to the low quality of the fitting at large distances from the Galactic plane. Therefore, we only fit the data points with $|z| \leqslant 0.4$\,kpc by Eq. \ref{eq:scale-height} and
get $z_0=133.15_{-4.32}^{+4.71}$\,pc, which is a  smaller value than that derived by \citet{Kos2014}. We note that we do not survey the same sample here and that \citet{Kos2014} had to resort to averaging of DIB\,$\lambda$862 measurements from different stars, meaning that their sample may be influenced by systematic errors in distance measurements available in the pre-Gaia era. 

Gaia makes an all sky survey of DIBs\,$\lambda$862 which is not restricted to the Southern hemisphere and equatorial region, as is the case for RAVE. Using all available lines of sight (lower panel in Fig.~\ref{fig:sh-fit}), the fitted $\rm z_{0}$ decreases to $\rm 98.69_{-8.35}^{+10.81}$\,pc. The uncertainties are small and may indicate a variation of the DIB\,$\lambda$862 scale height on the line of sight. This is consistent with the spatial distribution of the DIBs\,$\lambda$862 (see Fig.~\ref{comparison}) where we notice, for example, a larger $\rm z_{0}$  for the inner disc ($|\ell| < 30^{\circ}$). Our derived $z_0$ of the DIB\,$\lambda862$ carrier towards all available lines of sight with $4^{\circ} \leqslant |b| \leqslant 12^{\circ}$ is close to the scale height of the carrier of the DIB at $1.527\,\mu$m derived by \citet{Zasowski2015c} with $\rm z_{0} = 108 \pm 8\,pc$ but is slightly smaller than the scale height of the dust grains as measured by various authors, such as $134.4 \pm 8.5$\,pc by \citet{DS2001}, $125_{-7}^{+17}$\,pc by \citet{Marshall2006}, $119 \pm 15$\,pc by \citet{Jones2011}.
On the other hand,  \citet{Li2018} reported a smaller value of 103\,pc while  \citet{Guo2021} obtained two 
$z_0$, $72.7 \pm 2.4$\,pc and $224.6 \pm 0.7$\,pc, for a two-disk model. For comparison, dense molecular gas such as $\rm ^{12}CO$  has a smaller scale height of $\sim$50--70\,pc (\citealt{Sanders1984}).  For stars with $4^{\circ} \leqslant |b| \leqslant 12^{\circ}$, we derive $\rho_0 = 0.19 \pm 0.04$\,{\AA}/kpc and $B = 0.05 \pm 0.01$\,{\AA}. This allows the reader to use Eq.\ \ref{eq:scale-height} as an estimate of the expected DIB\,$\lambda 8620$ carrier strength towards any star in the solar neighbourhood with $4^{\circ} \leqslant |b| \leqslant 12^{\circ}$. The ratio of the measured $\rm EW_{862}$ over the expected $\rm EW_{862}$ has 16 and 84 percentile values of 0.66 and 1.30. A detailed characterisation of the DIB\,$\lambda$862  carrier extending beyond the symmetric models is needed to study local substructures in and out of the Galactic plane.

\section{Rest-frame wavelength} \label{sect:lambda0}

One of the most important observational properties of the DIB\,$\lambda$862 is its central rest-frame wavelength ($\rm \lambda_{0}$), 
which is necessary to identify the DIB\,$\lambda$862 carrier through comparison to laboratory measurements. A frequently 
used method is to use the well-identified interstellar atomic or molecular lines to shift the whole spectrum to 
the rest velocity frame assuming a tight correlation between the DIB\,$\lambda$862 and the interstellar lines
\citep[e.g.][]{JD1994,Galazutdinov2000a}. Without the interstellar counterpart, $\rm \lambda_{0}$ can also be 
statistically determined with the empirical assumption that the radial velocity in the Local Standard of
Rest (LSR) towards the Galactic centre  (GC) or the Galactic
anti-centre (GAC) is almost  null 
\citep[e.g.][]{Munari2008,Zasowski2015c,paperII}.

\begin{figure*}[!htbp]
    \centering
    \includegraphics[width=0.9\textwidth]{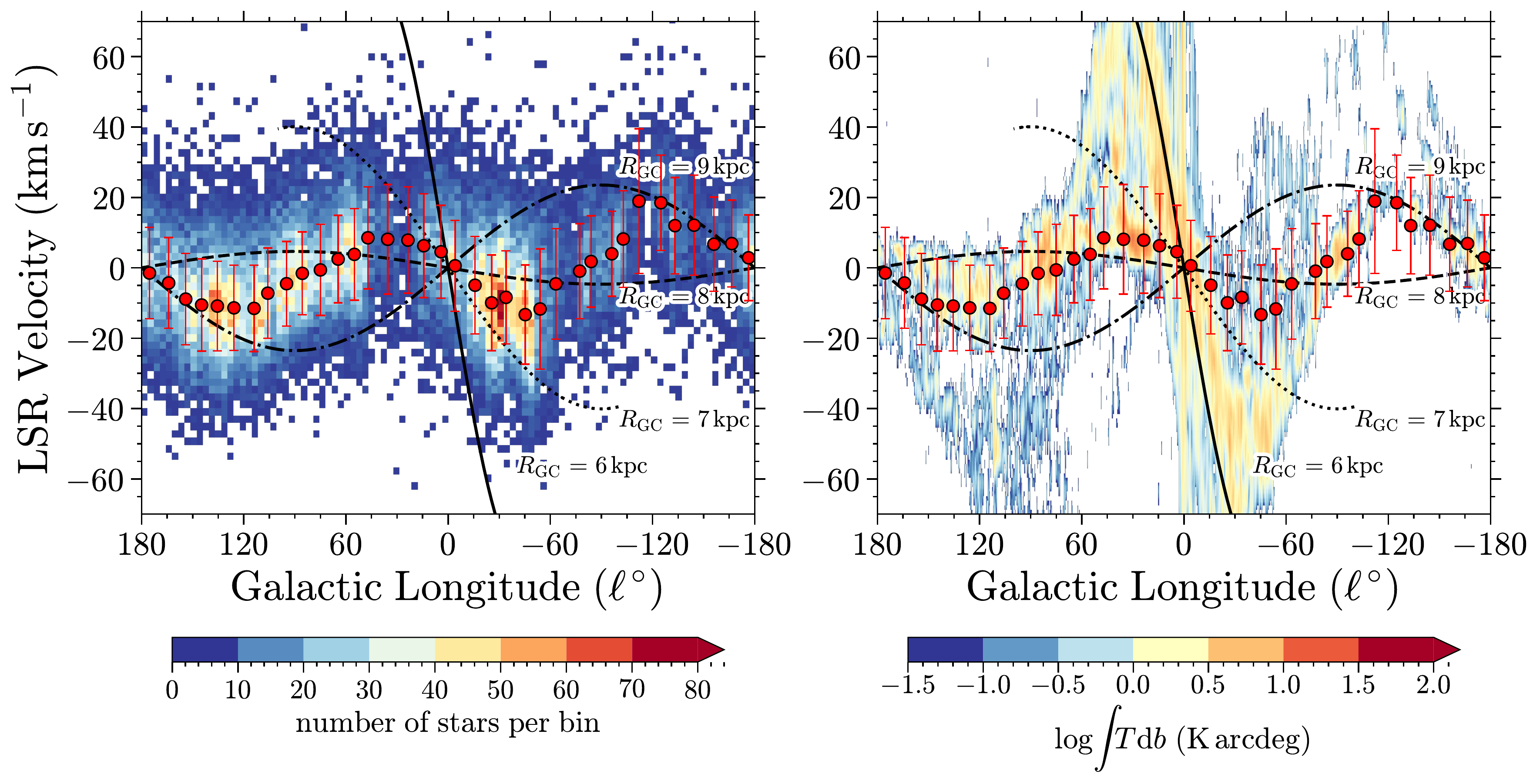}
    \caption{(Left panel): Longitude--velocity diagram for the Gaia HQ DIB\,$\lambda$862  sample. The circles indicate the median $\Vlsr$ and standard uncertainty of the mean for each field. Velocity curves calculated by Model A5 in \citet{Reid2019} for different galactocentric distances ($\Rgc$) are overplotted.  (Right panel): Same as left panel but superimposed on the $\rm ^{12}CO$ data from \citet{Dame2001}. The colour-scale displays the $\rm ^{12}CO$ brightness temperature in a logarithmic scale integrated over the velocity range.  }
    \label{rotation}
\end{figure*}

We apply this statistical method for both GC and GAC by selecting targets with $\Delta \ell \leqslant
10^{\circ}$, $|b| \leqslant 2^{\circ}$, $d \leqslant 4$\,kpc, $\rm QF=0$, ${\rm err}(\lambda_C) < 1.0$\,{\AA},
and valid stellar radial velocities. This provides 1405 stars for GC and 1106 cases for GAC. Figure 
\ref{fig:lambda0} shows their measured central wavelengths in the heliocentric frame ($C_{\rm obs}$) as a function of the angular distance from GC and GAC, respectively.
By the linear fit to the median values in each $\Delta \ell = 1^{\circ}$ bin, we get $C_{\rm obs}=
8623.10 \pm 0.018$\,{\AA} at $\ell=0^{\circ}$ and $C_{\rm obs}=8623.54 \pm 0.019$\,{\AA} at $\ell=180^{\circ}$. We stress that these are 
vacuum wavelengths, which means they are\ appropriate for Gaia observations.
For GAC, $C_{\rm obs}$ increases with Galactic longitude, having a slope of $23 \pm 3.4$\,m{\AA}\,deg$^{-1}$,
while the longitude trend is flatter toward the GC, with a slope of $1.2 \pm 
3.1$\,m{\AA}\,deg$^{-1}$. Fitting with a more constrained longitude region, such as $\Delta \ell \leqslant 
2^{\circ}$, yields very similar intercepts, that is $C_{\rm obs}= 8623.10 \pm 0.016$\,{\AA} at $\ell=0^{\circ}$ 
and $C_{\rm obs}=8623.52 \pm 0.023$\,{\AA} at $\ell=180^{\circ}$. Nevertheless, both of the slopes toward GC and
GAC become larger and much closer to each other: $47 \pm 14$\,m{\AA}\,deg$^{-1}$ for GC and 
$45 \pm 20$\,m{\AA}\,deg$^{-1}$ for GAC.
These slopes are also consistent with the values of $57 \pm 8$\,m{\AA}\,deg$^{-1}$ derived by
\citet{Zasowski2015c} for the DIB at 1.5273\,$\mu m$, 47\,m{\AA}\,deg$^{-1}$ derived from the CO rotation
curve (\citealt{Clemens1985}), and 40\,m{\AA}\,${\rm deg}^{-1}$ derived from 
the stellar rotation curve (\citealt{Bovy2012a}).

Considering the effect of solar motion, $\lambda_0$ in vacuum is derived as $c/(c-\Usun) \times 
C_{\rm obs}\,{=}\,8623.41$\,{\AA} for GC, and $c/(c+\Usun) \times C_{obs}\,{=}\,8623.23$\,{\AA} for GAC, where $c$ is the 
speed of the light and $\Usun\,{=}\,10.6\,\kms$ \citep{Reid2019} is the radial solar motion. The difference between them
may be caused by non-circular motion of the DIB\,$\lambda$862 carrier about the Galactic centre, which makes the LSR velocity non-zero. We believe this 
systematic effect is less pronounced in the  direction of the GAC, and so we use this value to derive its counterpart wavelength in the air of 8620.86\,{\AA}. 
This number agrees well with our previous result from the Giraffe Inner Bulge Survey \citep{Zoccali14} towards the GC \citep[8620.83\,{\AA};][]{paperII}. 
The obtained value in this work is slightly larger than the values of $\rm 8620.70 \pm \,0.3$\,{\AA} \citep{Sanner1978},
8620.75\,{\AA} \citep{HL1991}, and 8620.79\,{\AA} \citep{Galazutdinov2000b}. The result of \citet{JD1994}, namely
$8621.11 \pm 0.34$\,{\AA}, is very close to our result towards GC ($8621.03$\,{\AA} in air). Based on 68
hot stars from RAVE, \citet{Munari2008} measured a mean $C_{\rm obs}$ toward GC as $8620.4 \pm 0.1$\,{\AA},
corresponding to a $\lambda_0=8620.70$\,{\AA} after the solar-motion correction, which is also smaller than
our result. \citet{Fan2019} obtained a much smaller $\rm \lambda_0 = 8620.18 \pm 0.25$\,{\AA}, an average
value of 17 for their program spectra, which was measured in the averaged optical-depth profiles and corrected by the
interstellar $\KI$ line at 7699\,{\AA}. The lower quality of their spectra at longer wavelengths  and the complex velocity structure of the atomic species could be the cause of the large difference between their results and others (Haoyu, priv. communication).

\section{Kinematics of the DIB carrier} \label{Kinematics}
Although most of the DIB carriers are unknown, they have been proven to be a powerful tool for ISM tomography and consequently can probe the Galactic structure and interstellar environments. The most comprehensive kinematic study to date was  performed by \citet{Zasowski2015b} using APOGEE (SDSS-III) data, and allowed the authors to reveal the average Galactic rotation curve of the $\lambda1527$ DIB carriers spanning several kiloparsecs (kpc) from the Sun. They probed the DIB\,$\lambda$1527 carrier distribution in 3D and showed that DIBs\,$\lambda$1527 can be used to trace large-scale Galactic structures, such as the Galactic long bar and the warp of the outer disk. \citet{paperII} studied the kinematics of the DIB\,$\lambda$862 in the Galactic Bulge using Gaia-ESO (\citealt{Gilmore2012}) and GIBS  data (\citealt{Zoccali14}). These authors concluded that the DIB\,$\lambda$862 carrier  is located in the inner few kpc of the Galactic disk  based on their rotation velocities and radial velocity dispersion. However, these studies are based on specific pencil beams with a limited number of  objects.  Figure~\ref{rotation} demonstrates the enormous potential of Gaia for studying the kinematic behaviour of the DIBs\,$\lambda$862; it shows the Galactic rotation curve of the DIB\,$\lambda$862 carrier for $\rm |b| < 5^\mathrm{o}$ and in bins of 10 degrees in galactic longitude. Indicated are Galactic rotation curves computed by Model A5 in \citet{Reid2019} with different galactocentric  radii ($\Rgc$). For sightlines with $\ell \gtrsim 150^{\circ}$, the DIB\,$\lambda$862 velocities are consistent with the model rotation curves for $\rm R_{GC} \sim 9\,kpc$. On the other hand, for the inner disc with $\ell \lesssim 30^{\circ}$ the DIB\,$\lambda$862 carrier is best represented by  $\rm R_{GC} \sim 7.5\,kpc$, thus closer to the Sun. This is different from the findings of \citet{Zasowski2015c}, namely that the DIB\,$\lambda$1527 carrier in the inner Galaxy is farther from the Sun. Indeed, the inner disc sample of these latter authors shows   higher velocities compared to our sample by a
factor of almost two. This is most likely due to the fact that APOGEE observes in the infrared and so probes the  DIB\,$\lambda$1527 in the inner Galaxy up to larger distances compared to Gaia. The majority of stars in APOGEE are within $\sim$ 6\,kpc from the Sun while our sample is mostly confined to $\sim$ 2--3\,kpc.

Assuming a galactic rotation model, \citet{paperII} demonstrated that  kinematic distances of the DIB\,$\lambda$862 can be obtained, allowing the real 3D distribution of the DIB carrier to be traced. We plan to present this in a forthcoming paper. 

Correlations between the DIB\,$\lambda$862 carrier and gas kinematics
using different tracers  such as CO and HI can provide  additional
clues as to the origin of the DIB\,$\lambda862$
carrier. Figure~\ref{rotation} shows one example with  the comparison
of the  $\rm ^{12}CO$ data from \citet{Dame2001}. In the present study
we use the momentum-masked cube restricted to the latitude range $\rm
\pm 5^{o}$
\footnote{\href{https://lweb.cfa.harvard.edu/rtdc/CO/CompositeSurveys/}{https://lweb.cfa.harvard.edu/rtdc/CO/CompositeSurveys/}}. We
see that,  in general, the DIB\,$\lambda$862  closely follows the CO gas pattern, especially in the Galactic anticentre region, while higher velocities are seen in CO for $\rm |\ell| < 50^{o}$. This close relation between the DIB\,$\lambda$862 and the gas reinforces the suggestion that the DIB\,$\lambda$862 carrier could be related to macro molecules.
We want to stress again that Gaia data allow us to discuss such a large-scale picture for the first time.

\section{Conclusions}  \label{Discussions}

We present the largest sample of individual DIBs at 862nm published to date, as obtained by the Gaia  RVS spectrometer. This is the first homogeneous and all-sky survey of the DIB\,$\lambda$862, and allows us to study the global properties of this DIB\,$\lambda$862 carrier  in detail. Defining a high-quality sample, we demonstrate that  DIBs at 862nm show a  tight relation with interstellar reddening such  as $\rm E(BP-RP)$ or $E(B-V)$. Despite the use of  different algorithms in the measurement of DIBs at 862nm between hot stars ($\teff\,{>}\,7000$\,K) and cool stars ($\teff\,{\leqslant}\,7000$\,K), we see very similar relations between $\rm EW_{862}$ and $\rm E(BP-RP)$, demonstrating the robustness of the DIB\,$\lambda$862 measurement. While we see  similarities in the spatial distributions between the DIB\,$\lambda$862 carrier and the interstellar reddening, we also notice some differences, in  particular that the scale height of the DIB\,$\lambda$862 carrier is smaller compared to the dust and that the DIB\,$\lambda$862 carrier is concentrated within the inner kpc from the Sun.
A similar conclusion can be drawn from the comparison with the total Galactic extinction map. The main and most striking difference between the DIB\,$\lambda$862 carrier and dust distributions is that DIB\,$\lambda$862 carriers are present in the Local Bubble around the Sun, while this region is known to contain almost no dust. To first order, the spatial distribution of DIB\,$\lambda$862 carriers follows a simple slab model. We derive its local density and scale height, which can be used to predict the expected EW of the DIB\,$\lambda$862 towards any star up to $\sim$3\,kpc from the Sun. 

Taking advantage of the full sky coverage of the DIB\,$\lambda$862, we determined  the rest-frame wavelength of the DIB\,$\lambda$862 in the Galactic anticentre  with an estimated $\rm \lambda_{0} = 8620.86\, \pm 0.019$\,{\AA} in air. This is the most precise determination of $\rm \lambda_{0}$ to date. We note that using a large number of sources diminishes the formal measurement errors and, more importantly, largely negates the systematic errors of unknown radial velocities of clouds of DIB carriers which may influence any studies based on a small number of sources.  For the first time, we demonstrate here the Galactic rotation curve traced by the DIB\,$\lambda$862 carrier within 1--2\,kpc from the Sun and reveal the remarkable correspondence between the DIB\,$\lambda$862 velocities and the CO gas velocities, reinforcing  the suggestion that DIB\,$\lambda$862 carriers could be related to gaseous macromolecules.

 \section*{Acknowledgements\label{sec:acknowl}}
\addcontentsline{toc}{chapter}{Acknowledgements}
This work presents results from the European Space Agency (ESA) space
mission \gaia. \gaia\ data are being processed by the \gaia\ Data
Processing and Analysis Consortium (DPAC). Funding for the DPAC is
provided by national institutions, in particular the institutions
participating in the \gaia\ MultiLateral Agreement (MLA). The \gaia\
mission website is \url{https://www.cosmos.esa.int/gaia}. The \gaia\
archive website is \url{https://archives.esac.esa.int/gaia}. Acknowledgements are given in Appendix~\ref{ssec:appendixA}. 
%  This work has made use of data from the European Space Agency (ESA)
 % mission Gaia (https://www.cosmos.esa.int/gaia), processed by the
 % Gaia Data Processing and Analysis Consortium (DPAC,
 % https://www.cosmos.esa.int/web/gaia/dpac/consortium). Funding for
  %the DPAC has been provided by national institutions, in particular
  %the institutions participating in the Gaia Multilateral Agreement.

    T. Z. acknowledges financial support of the Slovenian Research Agency (research core funding No. P1-0188) and the European Space Agency (Prodex Experiment Arrangement No. C4000127986).  Part of the calculations have been performed with the high-performance computing facility SIGAMM, hosted by the Observatoire de la Côte d'Azur.
The GSP-spec group acknowledges financial supports from the french space agency (CNES), Agence National de la Recherche (ANR 14-CE33-014-01) and  Programmes Nationaux de Physique Stellaire \& Cosmologie et Galaxies (PNPS \& PNCG) of CNRS/INSU. H.Z. is funded by the China Scholarship Council (No.201806040200).
YF acknowledges the BELgian federal Science Policy Office (BELSPO) through various PROgramme de D\'eveloppement d'Exp\'eriences scientifiques (PRODEX) grants.\\

\bibliographystyle{aa}
%\vspace*{-0.4cm}
\bibliography{43283corr}

%\onecolumn
\begin{appendix}

\section{}\label{ssec:appendixA}

%This work presents results from the European Space Agency (ESA) space mission \gaia. \gaia\ data are being processed by the \gaia\ Data Processing and Analysis Consortium (DPAC). Funding for the DPAC is provided by national institutions, in particular the institutions participating in the \gaia\ MultiLateral Agreement (MLA). The \gaia\ mission website is \url{https://www.cosmos.esa.int/gaia}. The \gaia\ archive website is \url{https://archives.esac.esa.int/gaia}.

The \gaia\ mission and data processing have financially been supported by, in alphabetical order by country:
\begin{itemize}
\item the Algerian Centre de Recherche en Astronomie, Astrophysique et G\'{e}ophysique of Bouzareah Observatory;
\item the Austrian Fonds zur F\"{o}rderung der wissenschaftlichen Forschung (FWF) Hertha Firnberg Programme through grants T359, P20046, and P23737;
\item the BELgian federal Science Policy Office (BELSPO) through various PROgramme de D\'{e}veloppement d'Exp\'{e}riences scientifiques (PRODEX)
   grants, the Research Foundation Flanders (Fonds Wetenschappelijk Onderzoek) through grant VS.091.16N, the Fonds de la Recherche Scientifique (FNRS), and the Research Council of Katholieke Universiteit (KU) Leuven through grant C16/18/005 (Pushing AsteRoseismology to the next level with TESS, GaiA, and the Sloan DIgital Sky SurvEy -- PARADISE);  
\item the Brazil-France exchange programmes Funda\c{c}\~{a}o de Amparo \`{a} Pesquisa do Estado de S\~{a}o Paulo (FAPESP) and Coordena\c{c}\~{a}o de Aperfeicoamento de Pessoal de N\'{\i}vel Superior (CAPES) - Comit\'{e} Fran\c{c}ais d'Evaluation de la Coop\'{e}ration Universitaire et Scientifique avec le Br\'{e}sil (COFECUB);
\item the Chilean Agencia Nacional de Investigaci\'{o}n y Desarrollo (ANID) through Fondo Nacional de Desarrollo Cient\'{\i}fico y Tecnol\'{o}gico (FONDECYT) Regular Project 1210992 (L.~Chemin);
\item the National Natural Science Foundation of China (NSFC) through grants 11573054, 11703065, and 12173069, the China Scholarship Council through grant 201806040200, and the Natural Science Foundation of Shanghai through grant 21ZR1474100;  
\item the Tenure Track Pilot Programme of the Croatian Science Foundation and the \'{E}cole Polytechnique F\'{e}d\'{e}rale de Lausanne and the project TTP-2018-07-1171 `Mining the Variable Sky', with the funds of the Croatian-Swiss Research Programme;
\item the Czech-Republic Ministry of Education, Youth, and Sports through grant LG 15010 and INTER-EXCELLENCE grant LTAUSA18093, and the Czech Space Office through ESA PECS contract 98058;
\item the Danish Ministry of Science;
\item the Estonian Ministry of Education and Research through grant IUT40-1;
\item the European Commission’s Sixth Framework Programme through the European Leadership in Space Astrometry (\href{https://www.cosmos.esa.int/web/gaia/elsa-rtn-programme}{ELSA}) Marie Curie Research Training Network (MRTN-CT-2006-033481), through Marie Curie project PIOF-GA-2009-255267 (Space AsteroSeismology \& RR Lyrae stars, SAS-RRL), and through a Marie Curie Transfer-of-Knowledge (ToK) fellowship (MTKD-CT-2004-014188); the European Commission's Seventh Framework Programme through grant FP7-606740 (FP7-SPACE-2013-1) for the \gaia\ European Network for Improved data User Services (\href{https://gaia.ub.edu/twiki/do/view/GENIUS/}{GENIUS}) and through grant 264895 for the \gaia\ Research for European Astronomy Training (\href{https://www.cosmos.esa.int/web/gaia/great-programme}{GREAT-ITN}) network;
\item the European Cooperation in Science and Technology (COST) through COST Action CA18104 `Revealing the Milky Way with \gaia (MW-Gaia)';
\item the European Research Council (ERC) through grants 320360, 647208, and 834148 and through the European Union’s Horizon 2020 research and innovation and excellent science programmes through Marie Sk{\l}odowska-Curie grant 745617 (Our Galaxy at full HD -- Gal-HD) and 895174 (The build-up and fate of self-gravitating systems in the Universe) as well as grants 687378 (Small Bodies: Near and Far), 682115 (Using the Magellanic Clouds to Understand the Interaction of Galaxies), 695099 (A sub-percent distance scale from binaries and Cepheids -- CepBin), 716155 (Structured ACCREtion Disks -- SACCRED), 951549 (Sub-percent calibration of the extragalactic distance scale in the era of big surveys -- UniverScale), and 101004214 (Innovative Scientific Data Exploration and Exploitation Applications for Space Sciences -- EXPLORE);
\item the European Science Foundation (ESF), in the framework of the \gaia\ Research for European Astronomy Training Research Network Programme (\href{https://www.cosmos.esa.int/web/gaia/great-programme}{GREAT-ESF});
\item the European Space Agency (ESA) in the framework of the \gaia\ project, through the Plan for European Cooperating States (PECS) programme through contracts C98090 and 4000106398/12/NL/KML for Hungary, through contract 4000115263/15/NL/IB for Germany, and through PROgramme de D\'{e}veloppement d'Exp\'{e}riences scientifiques (PRODEX) grant 4000127986 for Slovenia;  
\item the Academy of Finland through grants 299543, 307157, 325805, 328654, 336546, and 345115 and the Magnus Ehrnrooth Foundation;
\item the French Centre National d’\'{E}tudes Spatiales (CNES), the Agence Nationale de la Recherche (ANR) through grant ANR-10-IDEX-0001-02 for the `Investissements d'avenir' programme, through grant ANR-15-CE31-0007 for project `Modelling the Milky Way in the \gaia era’ (MOD4Gaia), through grant ANR-14-CE33-0014-01 for project `The Milky Way disc formation in the \gaia era’ (ARCHEOGAL), through grant ANR-15-CE31-0012-01 for project `Unlocking the potential of Cepheids as primary distance calibrators’ (UnlockCepheids), through grant ANR-19-CE31-0017 for project `Secular evolution of galxies' (SEGAL), and through grant ANR-18-CE31-0006 for project `Galactic Dark Matter' (GaDaMa), the Centre National de la Recherche Scientifique (CNRS) and its SNO \gaia of the Institut des Sciences de l’Univers (INSU), its Programmes Nationaux: Cosmologie et Galaxies (PNCG), Gravitation R\'{e}f\'{e}rences Astronomie M\'{e}trologie (PNGRAM), Plan\'{e}tologie (PNP), Physique et Chimie du Milieu Interstellaire (PCMI), and Physique Stellaire (PNPS), the `Action F\'{e}d\'{e}ratrice \gaia' of the Observatoire de Paris, the R\'{e}gion de Franche-Comt\'{e}, the Institut National Polytechnique (INP) and the Institut National de Physique nucl\'{e}aire et de Physique des Particules (IN2P3) co-funded by CNES;
\item the German Aerospace Agency (Deutsches Zentrum f\"{u}r Luft- und Raumfahrt e.V., DLR) through grants 50QG0501, 50QG0601, 50QG0602, 50QG0701, 50QG0901, 50QG1001, 50QG1101, 50\-QG1401, 50QG1402, 50QG1403, 50QG1404, 50QG1904, 50QG2101, 50QG2102, and 50QG2202, and the Centre for Information Services and High Performance Computing (ZIH) at the Technische Universit\"{a}t Dresden for generous allocations of computer time;
\item the Hungarian Academy of Sciences through the Lend\"{u}let Programme grants LP2014-17 and LP2018-7 and the Hungarian National Research, Development, and Innovation Office (NKFIH) through grant KKP-137523 (`SeismoLab');
\item the Science Foundation Ireland (SFI) through a Royal Society - SFI University Research Fellowship (M.~Fraser);
\item the Israel Ministry of Science and Technology through grant 3-18143 and the Tel Aviv University Center for Artificial Intelligence and Data Science (TAD) through a grant;
\item the Agenzia Spaziale Italiana (ASI) through contracts I/037/08/0, I/058/10/0, 2014-025-R.0, 2014-025-R.1.2015, and 2018-24-HH.0 to the Italian Istituto Nazionale di Astrofisica (INAF), contract 2014-049-R.0/1/2 to INAF for the Space Science Data Centre (SSDC, formerly known as the ASI Science Data Center, ASDC), contracts I/008/10/0, 2013/030/I.0, 2013-030-I.0.1-2015, and 2016-17-I.0 to the Aerospace Logistics Technology Engineering Company (ALTEC S.p.A.), INAF, and the Italian Ministry of Education, University, and Research (Ministero dell'Istruzione, dell'Universit\`{a} e della Ricerca) through the Premiale project `MIning The Cosmos Big Data and Innovative Italian Technology for Frontier Astrophysics and Cosmology' (MITiC);
\item the Netherlands Organisation for Scientific Research (NWO) through grant NWO-M-614.061.414, through a VICI grant (A.~Helmi), and through a Spinoza prize (A.~Helmi), and the Netherlands Research School for Astronomy (NOVA);
\item the Polish National Science Centre through HARMONIA grant 2018/30/M/ST9/00311 and DAINA grant 2017/27/L/ST9/03221 and the Ministry of Science and Higher Education (MNiSW) through grant DIR/WK/2018/12;
\item the Portuguese Funda\c{c}\~{a}o para a Ci\^{e}ncia e a Tecnologia (FCT) through national funds, grants SFRH/\-BD/128840/2017 and PTDC/FIS-AST/30389/2017, and work contract DL 57/2016/CP1364/CT0006, the Fundo Europeu de Desenvolvimento Regional (FEDER) through grant POCI-01-0145-FEDER-030389 and its Programa Operacional Competitividade e Internacionaliza\c{c}\~{a}o (COMPETE2020) through grants UIDB/04434/2020 and UIDP/04434/2020, and the Strategic Programme UIDB/\-00099/2020 for the Centro de Astrof\'{\i}sica e Gravita\c{c}\~{a}o (CENTRA);  
\item the Slovenian Research Agency through grant P1-0188;
\item the Spanish Ministry of Economy (MINECO/FEDER, UE), the Spanish Ministry of Science and Innovation (MICIN), the Spanish Ministry of Education, Culture, and Sports, and the Spanish Government through grants BES-2016-078499, BES-2017-083126, BES-C-2017-0085, ESP2016-80079-C2-1-R, ESP2016-80079-C2-2-R, FPU16/03827, PDC2021-121059-C22, RTI2018-095076-B-C22, and TIN2015-65316-P (`Computaci\'{o}n de Altas Prestaciones VII'), the Juan de la Cierva Incorporaci\'{o}n Programme (FJCI-2015-2671 and IJC2019-04862-I for F.~Anders), the Severo Ochoa Centre of Excellence Programme (SEV2015-0493), and MICIN/AEI/10.13039/501100011033 (and the European Union through European Regional Development Fund `A way of making Europe') through grant RTI2018-095076-B-C21, the Institute of Cosmos Sciences University of Barcelona (ICCUB, Unidad de Excelencia `Mar\'{\i}a de Maeztu’) through grant CEX2019-000918-M, the University of Barcelona's official doctoral programme for the development of an R+D+i project through an Ajuts de Personal Investigador en Formaci\'{o} (APIF) grant, the Spanish Virtual Observatory through project AyA2017-84089, the Galician Regional Government, Xunta de Galicia, through grants ED431B-2021/36, ED481A-2019/155, and ED481A-2021/296, the Centro de Investigaci\'{o}n en Tecnolog\'{\i}as de la Informaci\'{o}n y las Comunicaciones (CITIC), funded by the Xunta de Galicia and the European Union (European Regional Development Fund -- Galicia 2014-2020 Programme), through grant ED431G-2019/01, the Red Espa\~{n}ola de Supercomputaci\'{o}n (RES) computer resources at MareNostrum, the Barcelona Supercomputing Centre - Centro Nacional de Supercomputaci\'{o}n (BSC-CNS) through activities AECT-2017-2-0002, AECT-2017-3-0006, AECT-2018-1-0017, AECT-2018-2-0013, AECT-2018-3-0011, AECT-2019-1-0010, AECT-2019-2-0014, AECT-2019-3-0003, AECT-2020-1-0004, and DATA-2020-1-0010, the Departament d'Innovaci\'{o}, Universitats i Empresa de la Generalitat de Catalunya through grant 2014-SGR-1051 for project `Models de Programaci\'{o} i Entorns d'Execuci\'{o} Parallels' (MPEXPAR), and Ramon y Cajal Fellowship RYC2018-025968-I funded by MICIN/AEI/10.13039/501100011033 and the European Science Foundation (`Investing in your future');
\item the Swedish National Space Agency (SNSA/Rymdstyrelsen);
\item the Swiss State Secretariat for Education, Research, and Innovation through the Swiss Activit\'{e}s Nationales Compl\'{e}mentaires and the Swiss National Science Foundation through an Eccellenza Professorial Fellowship (award PCEFP2\_194638 for R.~Anderson);
\item the United Kingdom Particle Physics and Astronomy Research Council (PPARC), the United Kingdom Science and Technology Facilities Council (STFC), and the United Kingdom Space Agency (UKSA) through the following grants to the University of Bristol, the University of Cambridge, the University of Edinburgh, the University of Leicester, the Mullard Space Sciences Laboratory of University College London, and the United Kingdom Rutherford Appleton Laboratory (RAL): PP/D006511/1, PP/D006546/1, PP/D006570/1, ST/I000852/1, ST/J005045/1, ST/K00056X/1, ST/\-K000209/1, ST/K000756/1, ST/L006561/1, ST/N000595/1, ST/N000641/1, ST/N000978/1, ST/\-N001117/1, ST/S000089/1, ST/S000976/1, ST/S000984/1, ST/S001123/1, ST/S001948/1, ST/\-S001980/1, ST/S002103/1, ST/V000969/1, ST/W002469/1, ST/W002493/1, ST/W002671/1, ST/W002809/1, and EP/V520342/1.
\end{itemize}

We made use of the following tools in the preparation of this paper:

\citep[SIMBAD,][]{2000AAS..143....9W} and VizieR \citep{2000AAS..143...23O} operated at (\href{http://cds.u-strasbg.fr/}{CDS}) Strasbourg; 
NASA \href{http://adsabs.harvard.edu/abstract_service.html}{ADS}; 
\href{http://www.starlink.ac.uk/topcat/}{TOPCAT} \citep{2005ASPC..347...29T};
Matplotlib \citep{Hunter:2007};
IPython \citep{PER-GRA:2007};  
Astropy, a community-developed core Python package for Astronomy \citep{2018AJ....156..123A};

Funding for SDSS-III has been provided by the Alfred P. Sloan Foundation, the Participating Institutions, the National Science Foundation, and the U.S. Department of Energy Office of Science. The SDSS-III web site is http://www.sdss3.org/. 
SDSS-III is managed by the Astrophysical Research Consortium for the Participating Institutions of the SDSS-III Collaboration including the University of Arizona, the Brazilian Participation Group, Brookhaven National Laboratory, Carnegie Mellon University, University of Florida, the French Participation Group, the German Participation Group, Harvard University, the Instituto de Astrofisica de Canarias, the Michigan State/Notre Dame/JINA Participation Group, Johns Hopkins University, Lawrence Berkeley National Laboratory, Max Planck Institute for Astrophysics, Max Planck Institute for Extraterrestrial Physics, New Mexico State University, New York University, Ohio State University, Pennsylvania State University, University of Portsmouth, Princeton University, the Spanish Participation Group, University of Tokyo, University of Utah, Vanderbilt University, University of Virginia, University of Washington, and Yale University.

Funding for the Sloan Digital Sky Survey IV has been provided by the Alfred P. Sloan Foundation, the U.S.  Department of Energy Office of Science, and the Participating Institutions.  SDSS-IV acknowledges support and resources from the Center for High Performance Computing at the University of Utah. The SDSS website is www.sdss.org.  SDSS-IV is managed by the Astrophysical Research Consortium for the Participating Institutions of the SDSS Collaboration including the Brazilian Participation Group, the Carnegie Institution for Science, Carnegie Mellon University, Center for Astrophysics | Harvard \& Smithsonian, the Chilean Participation Group, the French Participation Group, Instituto de Astrof\'isica de Canarias, The Johns Hopkins University, Kavli Institute for the Physics and Mathematics of the Universe (IPMU) / University of Tokyo, the Korean Participation Group, Lawrence Berkeley National Laboratory, Leibniz Institut f\"ur Astrophysik Potsdam (AIP), Max-Planck-Institut f\"ur Astronomie (MPIA Heidelberg), Max-Planck-Institut f\"ur Astrophysik (MPA Garching), Max-Planck-Institut f\"ur Extraterrestrische Physik (MPE), National Astronomical Observatories of China, New Mexico State University, New York University, University of Notre Dame, Observat\'ario Nacional / MCTI, The Ohio State University, Pennsylvania State University, Shanghai Astronomical Observatory, United Kingdom Participation Group, Universidad Nacional Aut\'onoma de M\'exico, University of Arizona, University of Colorado Boulder, University of Oxford, University of Portsmouth, University of Utah, University of Virginia, University of Washington, University of Wisconsin, Vanderbilt University, and Yale University.

%
% From Martin Altmann, 13 March 2019:
%  092.B-0165   01.10.13 - 31.03.14
%  093.B-0236   01.04.14 - 30.09.14
%  094.B-0181   01.10.14 - 31.03.15
%  095.B-0046   01.04.15 - 30.09.15
%  096.B-0162   01.10.15 - 31.03.16
%  097.B-0304   01.04.16 - 30.09.16
%  098.B-0030   01.10.16 - 31.03.17
%  099.B-0034   01.04.17 - 30.09.17
% 0100.B-0131   01.10.17 - 31.03.18
% 0101.B-0156   01.04.18 - 30.09.18
% 0102.B-0174   01.10.18 - 31.03.19
% 0103.B-0165   01.04.19 - 30.09.19
%

%In case of errors or omissions, please contact the \href{https://www.cosmos.esa.int/web/gaia/gaia-helpdesk}{\gaia\ Helpdesk}.

\section*{ADQL Queries} \label{queries}
\small

\noindent
{\bf Use Case}: Retrieve full DIB  sample 
\begin{verbatim}
SELECT *
FROM user_dr3int5.astrophysical_parameters AS gaia 
INNER JOIN user_dr3int5.astrophysical_parameters_supp AS m
ON gaia.source_id = m.source_id 
WHERE gaia.dibqf_gspspec >= 0 
\end{verbatim}

\noindent
{\bf Use Case}: Retrieve DIB  results for HQ sample

\noindent
SELECT *\\
FROM user\_dr3int5.astrophysical\_parameters\\
WHERE ((flags\_gspspec LIKE  '0\%') OR (flags\_gspspec LIKE '1\%')) AND
((flags\_gspspec LIKE '\_0\%') OR (flags\_gspspec LIKE '\_1\%')) AND
((flags\_gspspec LIKE '\_\_0\%') OR (flags\_gspspec LIKE '\_\_1\%')) AND
((flags\_gspspec LIKE '\_\_\_0\%') OR (flags\_gspspec LIKE '\_\_\_1\%')) AND
((flags\_gspspec LIKE '\_\_\_\_0\%') OR (flags\_gspspec LIKE '\_\_\_\_1\%')) AND
((flags\_gspspec LIKE '\_\_\_\_\_0\%') OR (flags\_gspspec LIKE '\_\_\_\_\_1\%')) AND
((flags\_gspspec LIKE '\_\_\_\_\_\_0\%') OR (flags\_gspspec LIKE '\_\_\_\_\_\_1\%')) AND
((flags\_gspspec LIKE '\_\_\_\_\_\_\_0\%') OR (flags\_gspspec LIKE '\_\_\_\_\_\_\_1\%')) AND
((flags\_gspspec LIKE '\_\_\_\_\_\_\_\_0\%') OR (flags\_gspspec LIKE '\_\_\_\_\_\_\_\_1\%')) AND
((flags\_gspspec LIKE '\_\_\_\_\_\_\_\_\_0\%') OR (flags\_gspspec LIKE '\_\_\_\_\_\_\_\_\_1\%')) AND
((flags\_gspspec LIKE '\_\_\_\_\_\_\_\_\_\_0\%') OR (flags\_gspspec LIKE '\_\_\_\_\_\_\_\_\_\_1\%')) AND
((flags\_gspspec LIKE '\_\_\_\_\_\_\_\_\_\_\_0\%') OR (flags\_gspspec LIKE '\_\_\_\_\_\_\_\_\_\_\_1\%')) AND
((flags\_gspspec LIKE '\_\_\_\_\_\_\_\_\_\_\_\_0\%') OR (flags\_gspspec LIKE '\_\_\_\_\_\_\_\_\_\_\_\_1\%')) AND
(dibqf\_gspspec $<=$ 2) AND (dibqf\_gspspec $>=$ 0 ) AND
(dib\_gspspec\_lambda$>$ 862.0) AND (dib\_gspspec\_lambda $<$ 862.6) AND
((dibew\_gspspec\_uncertainty/dibew\_gspspec) $<$ 0.35)

%\end{verbatim}

\section*{Hot star outliers} \label{outliers}

\begin{table*}[!htbp]
    \caption{Outliers found in the hot stars sample (Fig.\,\ref{EW_ESPHS}). Description of the table columns: number (col.1), GDR3 ID (col.2), Simbad ID and spectral/object type between brackets when available (col.3), effective temperatures from ESP-HS and spectral type found in then field {\tt spectraltype\_esphs}  (col.4), GSP-Spec (col.5), and GSP-Phot (col.6). } 
    \label{tab:hot.outliers}
    \centering
    \begin{tabular}{r|rcrrr}
n & DR3 ID & ID Simbad & T$\mathrm{eff}$ & T$\mathrm{eff}$ & T$\mathrm{eff}$ \\
 & & & ESP-HS & GSP-Spec & GSP-Phot \\
 \hline
&\multicolumn{5}{c}{ESP-HS, upper panel of Fig.\,\ref{EW_ESPHS}}\\
\cline{2-6}\\
1 & 2066753415480268800 & 2MASS J20500395+4300117 & 54\,705 (O) & 6\,904 & 7\,690 \\
 2 & 444955867385000832 & - & 54\,650 (O) & 6\,248 & 7\,023  \\ 
 3 & 4054566946966876288 & CD-32 12958 & 54\,706 (B) & 7\,115 & -  \\
 4 & 2164089679515345280 & TYC 3589-1199-1 & 13\,541 (B) & 8\,404 & -  \\
 5 & 5537927056196905984 & TYC 7659-1313-1 & 16\,054 (B) & 8\,339 & 12\,109  \\
 6 & 1730824030187372416 & FQ Aqr & 17\,364 (B) & 8\,289 & -  \\
 7 & 3455454953759211264 & LS V +35 26 (OB-e) & 20\,000 (B) & 8\,068 & -  \\
 8 & 2005574977916673792 & BD+53 2784 (B3 III) & 17\,695 (B) & 7\,869 & -      \\
 \cline{2-6}
&\multicolumn{5}{c}{GSP-Phot, middle panel of Fig.\,\ref{EW_ESPHS}}\\
\cline{2-6}\\
1 & 5999123049637219072 & IRAS 15212-4624 & - (M) & 7\,316 & 3\,641 \\
 2 & 5843278232842959872 & IRAS 12365-6959 & - (M) & 7\,312 & 3\,523\\ 
 3 & 5878260883212542208 & - & - (O) & 7\,843 & 6\,495\\ 
 4 & 5854026787978149760 & IRAS 14112-6224 & - (M) & 8\,000 & 3\,712\\ %  - Gaia DR2 5854026787970760704
 5 & 5889006272967734144 &IRAS 15230-5132 (LP?) & - (M) & 7\,348 & 3\,584 \\ % - Gaia DR2 5889006272948945920
 6 & 4152556797623844608 & TYC 5702-740-1 & - (O) & 7\,679 & 8\,061 \\
 7 & 4134885451076972416 & IRAS 17170-1756 & - (M) & 7\,900 & 3\,707 \\
 8 & 5341747587387330432 & IRAS 11464-5753 (M7) & - (M) & 7\,900 & 3\,500 \\
 9 & 1931994246725494400 & V608 And (M7/M8) & - (M) & 7\,298 & 3\,623 \\
 10 & 4478836843125274496 & IRAS 18313+0720 & - (M) & 8\,000 & 3\,637 \\
 \hline
\end{tabular}

\end{table*}

%\section{}\label{ssec:appendixA}
%\input{acknow-nonames}

\end{appendix}

%\section{The second peak of EW error}

%\begin{figure*}
%    \centering
%    \includegraphics[width=0.9\textwidth]{figures/EWerror.pdf}
%    \caption{The origin of the second peak of EW error.}
%    \label{EWerr-origin}
%\end{figure*}

% \section{Different EW-$\rm E(BP-RP)$ relation}

% \begin{figure*}
%     \centering
%     \includegraphics[width=0.48\textwidth]{figures/EW-Ext-discussion.png}
%     \includegraphics[width=0.48\textwidth]{figures/EW-Ext-discussion2.png}
%     \caption{{\it Left:} HQ sample. {\it Right:} median values taken in $3.7^{\circ}$ ($\ell,b$) bins
%     and 0.5\,kpc distance bins. Red dots are from EW bins and blue dots are from $\rm E(BP-RP)$ bins. The black
%     dots are calculated by the red points in Fig. 8.}
%     \label{EWEBPRP}
% \end{figure*}

\end{document}